\documentclass[11pt]{article}
\pdfoutput=1
\usepackage{cite}
\usepackage{graphicx}
\usepackage{multicol}
\usepackage{multirow}
\usepackage{amsfonts}
\usepackage{amssymb}
\usepackage{array}
\usepackage{amsmath}
\usepackage{heck}
\usepackage{setspace}
\usepackage{verbatim}
\usepackage{color}
\usepackage{float}
\usepackage{epsfig}
\usepackage[font=footnotesize,labelfont=bf]{caption}
\usepackage{subcaption}
\usepackage[titletoc]{appendix}

\newcommand{\ba}{\begin{eqnarray}}
\newcommand{\ea}{\end{eqnarray}}

\begin{document}

\def\simgt{\mathrel{\lower2.5pt\vbox{\lineskip=0pt\baselineskip=0pt
           \hbox{$>$}\hbox{$\sim$}}}}
\def\simlt{\mathrel{\lower2.5pt\vbox{\lineskip=0pt\baselineskip=0pt
           \hbox{$<$}\hbox{$\sim$}}}}

\begin{titlepage}
\begin{flushright}\small{MCTP-15-06} 
\end{flushright}

\begin{center}

{\LARGE \bf Dark Matter Production Mechanisms with a Non-Thermal \\ \vspace{1cm} Cosmological History - A Classification}

\vspace{0.8cm}

\small
\bf{Gordon L. Kane$^{1}$, Piyush Kumar$^{2}$, Brent D. Nelson$^{3}$, Bob Zheng$^{1}$}
\normalsize

\vspace{.5cm}
{\it $^1$ Michigan Center for Theoretical Physics, University of Michigan, Ann Arbor, MI 48109} \\
{\it $^2$ Department of Physics, Yale University, New Haven, CT 06520} \\
{\it $^3$ Department of Physics, Northeastern University, Boston, MA 02115, USA}\\

\abstract{We perform a comprehensive study of models of dark matter (DM) in a Universe with a non-thermal cosmological history, i.e with a phase of pressure-less matter domination before the onset of big-bang nucleosynethesis (BBN). Such cosmological histories are generically predicted by UV completions that contain gravitationally coupled scalar fields (moduli). We classify the different production mechanisms for DM in this framework, generalizing previous works by considering a wide range of DM masses/couplings and allowing for DM to be in equilibrium with a ``dark" sector. We identify four distinct parametric regimes for the production of relic DM, and derive accurate semi-analytic approximations for the DM relic abundance. Our results are particularly relevant for supersymmetric theories, in which the standard non-thermally produced DM candidates are disfavored by indirect detection constraints. We also comment on experimental signals  in this framework, focusing on novel effects involving the power spectrum of DM density perturbations. In particular, we identify a class of models where the spectrum of DM density perturbations is sensitive to the pressure-less matter dominated era before BBN, giving rise to interesting astrophysical signatures to be looked for in the future. A worthwhile future direction would be to study well-motivated theoretical models within this framework and carry out detailed studies of the pattern of expected experimental signals.}

\end{center}
\maketitle

\end{titlepage}
\tableofcontents

\section{Motivation and Introduction}

Apart from its existence, the nature and non-gravitational interactions of dark matter (DM) are still very uncertain. The most popular class of dark matter models - Weakly Interacting Massive Particles (WIMPs) - rely on two key assumptions to  reproduce the observed relic abundance. First, WIMPs are assumed to annihilate into Standard Model (SM) particles with an electroweak-scale cross section. Second, the universe is usually assumed to be radiation dominated between the end of inflation and matter-radiation equality. However, there are no clear indications that either of these assumptions are valid. With regards to the former, large regions of WIMP parameter space have been ruled out by various direct and indirect detection experiments. With regards to the latter, the earliest evidence for a radiation dominated universe arises during BBN, which occurs at temperatures of order an MeV. The energy budget of the Universe has not been probed for temperatures above that at the time of BBN. Of course, it is still possible that dark matter is a simple WIMP, but because of the above reasons it is well-motivated to go \emph{beyond} the traditional WIMP paradigm, both in terms of DM candidates as well as the production mechanisms for DM.

A well-motivated alternative to the standard ``thermal" cosmological history mentioned above is that of a non-thermal cosmological history, in which BBN is preceeded by a phase of pressureless matter domination. Such a situation is predicted in many top-down theories for new physics e.g.  low-energy limits of supergravity and string/M-theory compactifications. These theories, under some very mild assumptions, contain gravitationally coupled scalars called moduli. When the Hubble parameter drops below moduli masses, moduli begin coherent oscillations and behave as pressure-less matter, dominating the energy density of the universe until the longest-lived one ($\phi$) decays to reheat the universe. In these cosmological histories, an electroweak-scale Wino provides a natural candidate for supersymmetric (SUSY) DM, provided that the modulus dominated phase ends at temperatures below a GeV or so~\cite{Moroi:1999zb, Acharya:2009zt}. However, recent FERMI-LAT and HESS observations of Galactic Center photons have placed severe limits on Wino DM~\cite{Fan:2013faa, Cohen:2013ama}. If the Wino is stable, satisfying these constraints in the cosmological histories mentioned above requires a large hierarchy between the modulus and gravitino masses~\cite{Fan:2013faa}. Such a hierarchy is quite unnatural for a broad class of models in which moduli stabilization sets the scale of supersymmetry breaking~\cite{Acharya:2012tw, Acharya:2010af, Denef:2004cf, GomezReino:2006dk}. This conclusion also holds if the lightest superpartner is some more general admixture of MSSM particles~\cite{Blinov:2014nla}. A simple way to avoid these constraints is to assume that the lightest visible sector superpartner, hereafter referred to as the LOSP, is unstable.

Motivated by the above statements, this work provides a comprehensive study of relic DM production in cosmological histories with a late phase of modulus domination. To perform as general an analysis as possible, we go beyond the standard WIMP picture by \emph{i}). allowing for a wide range of DM masses and annihilation cross sections and \emph{ii}). allowing for the possibility that DM is in kinetic equilibrium with some sector other than the visible sector. These two assumptions are well motivated in SUSY theories with an unstable LOSP, but can also be true in general. If the LOSP decays, DM is not a visible sector particle; a priori there is no reason to expect its DM mass or annihilation cross section to be near the electroweak scale. Moreover, if the DM resides in a sector that couples weakly to the visible sector, DM could be in kinetic equilibrium with a ``dark sector" instead of the thermal bath of visible sector particles. As we will see, our results can be straightforwardly reduced to that of the single-sector case despite assumption \emph{ii}).

The analysis of DM models in this framework can be effectively separated into three questions.
\begin{itemize}
\item How does one classify production mechanisms for relic DM and accurately compute $\Omega_{DM} h^2$?
\item What is the pattern of experimental and observational signals arising within such a framework?
\item What kind of DM candidates and interactions naturally arise in well-motivated theories? 
\end{itemize}

In this paper, we focus primarily on the first question by solving the Boltzmann equations for a two-sector system with a late phase of modulus domination. A brief overview of this framework, along with the Boltzmann equations describing its cosmological evolution, is presented in Section \ref{overview}. We then classify all potential mechanisms for the production of relic DM, and compute $\Omega_{DM} h^2$ for a large range of DM masses and annihilation cross sections. The entire parameter space of these DM models can be classified into four different parametric regimes, each with distinct production mechanisms for relic DM. We derive (semi) analytic approximations for $\Omega_{DM} h^2$ in these different regimes, and confirm their validity by comparison with the numerical solution. This is the main new result of our work, and is discussed in detail in Section~\ref{solutions}. Readers may go directly to Section \ref{summary}, which contains a self-contained summary of the above results. In Section~\ref{viable}, we discuss the implications of our results for UV-motivated SUSY theories where the modulus mass is of order the gravitino mass, and show that this framework provides several viable alternatives to MSSM dark matter.

In the remainder of the paper, we briefly discuss the latter two questions listed above. In Section \ref{conseq}, we discuss potential experimental signatures of the DM models considered here. A significant portion of the parameter space predicts free-streaming lengths characteristic of warm dark matter. Furthermore, we identify a class of DM models in which the DM power spectrum is sensitive to the linear growth of subhorizon DM density perturbations during the modulus dominated era. This can lead to interesting astrophysical signatures, such as an abundance of earth-mass (or smaller) DM microhalos which are far denser than their counterparts in standard cosmologies~\cite{Erickcek:2011us}. Finally, Section \ref{models} briefly addresses the third question, and describes work done in string theory models that have dark sectors. In a companion paper, we will elaborate further on some classes of these models. We present our conclusions in Section \ref{conclude}. The appendices contain technical results which will be referred to in the main text.

\section{Overview of Two-Sectors - Models and Cosmology}\label{overview}

\begin{figure}[t]
  \begin{center}
    \includegraphics[scale=0.5]{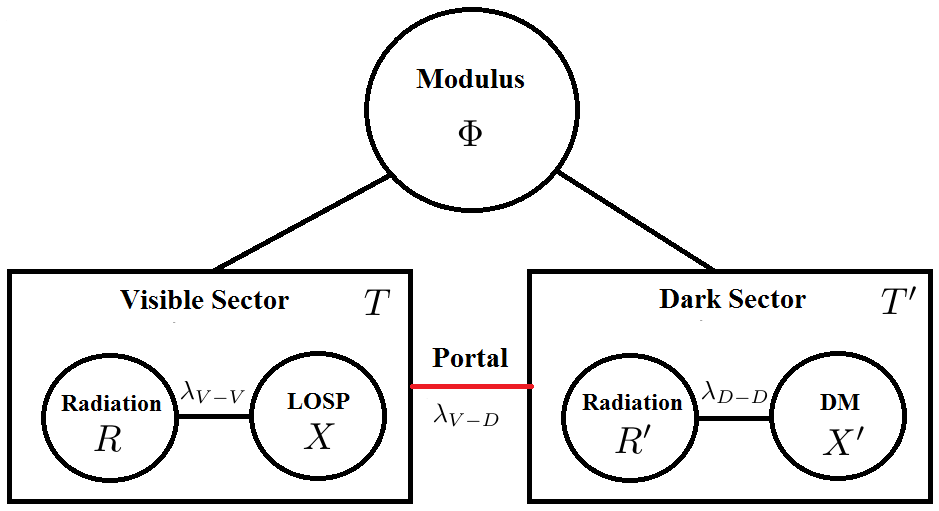}
  \end{center}
  \caption{\footnotesize{Schematic representation of the Two-sector Framework under consideration.}}\label{twosector}
\end{figure}

The framework considered here consists of two sectors: a visible sector containing SM (and perhaps MSSM) particles and a dark sector containing the DM. Both the visible and dark sectors are assumed to have sufficient interactions such that thermal equilibrium is separately maintained within the two sectors, whose temperatures are $T$ and $T'$ respectively. We assume that there exist very weak portal interactions between the two sectors, so that $T$ and $T'$ may not be equal to each other. Finally, we assume that the Universe is dominated by the coherent oscillations of a modulus field $\phi$ at some time which is much earlier than when BBN occurs\footnote{In general, there could be many moduli present in the early Universe. In this case, $\phi$ should be thought of as the longest-lived modulus. DM produced from shorter-lived  moduli will be diluted by entropy production~\cite{Acharya:2008bk}.}. The cosmological framework described above is depicted schematically in Figure~\ref{twosector}. The results of our work will be straightforward to reduce to the single sector case, see the discussion in Section \ref{summary}.

As denoted in Figure \ref{twosector}, the visible sector contains radiation degrees of freedom $R$, comprised of relativistic particles in equilibrium with the SM bath at temperature $T$. We also track the abundance of an unstable WIMP-like particle $X$ which is in equilibrium with the visible sector. $X$ corresponds to the LOSP in the SUSY theories discussed in the introduction. The dark sector is assumed to contain a stable DM candidate $X^\prime$, along with dark radiation $R^\prime$. ``Dark radiation" refers to dark sector particles which are in thermal equilibrium and are relativistic at a given dark sector temperature $T^\prime$. Henceforth, visible (dark) sector quantities are denoted using unprimed (primed) variables. For simplicity and concreteness, we assume that no DM asymmetry is present, so DM particles and antiparticles need not be separately tracked in the Boltzmann equations. Relaxing this assumption is worth exploring in future studies, see for example~\cite{Blinov:2014nla}. Finally we make the assumption that $M_{X}, \, M_{X^\prime} \ll m_{\phi}$, which is naturally expected for the supersymmetric theories discussed in the introduction and in Section \ref{viable}. 

Before moving on to study the cosmological evolution of this system, it is worth mentioning that there are constraints on hidden sector relativistic degrees of freedom during BBN and during recombination, through their contribution to the expansion rate of the Universe. These constraints are typically presented in terms of the number of effective extra neutrino species $\Delta N_{\rm eff}$, which is related to the number of relativistic hidden sector degrees of freedom $g^\prime_*(T^\prime)$ by:\ba \label{Neff} \Delta\,N_{\rm eff} (T_{BBN}) = 0.57 \,\,g'_{\star}(T^\prime_{BBN})\,\xi^4(T_{BBN}), \hspace{3mm} \Delta\,N_{\rm eff} (T_{CMB}) = 2.2\,\,g'_{\star}(T^\prime_{CMB})\,\xi^4(T_{CMB})\ea where $T_{BBN} \sim 1$ MeV, $T_{CMB} \sim 1$ eV and $\xi(T) \equiv (T^\prime/T)^4$. The current 95\% CL bounds are $\Delta\,N_{\rm eff}(T_{BBN}) \leq 1.44$~\cite{Fields:2006ga} and $\Delta N_{\rm eff}(T_{CMB}) \leq 0.4$~\cite{Talk:2015}. We will discuss the implications of these constraints for the two sector models considered here in Section~\ref{Tmax}.

\subsection{Cosmological Evolution}\label{Cosmo}

The cosmology of the framework can be studied by writing down the Boltzmann equations for the time evolution of the relevant quantities which comprise the total energy density of the Universe. This includes the modulus energy density $\rho_{\phi}$, the energy density arising from  $X$ and $X'$ with number densities $n_{X}$ and $n_{X'}$ respectively, and the energy densities of radiation in the visible and dark sector, denoted by $\rho_{R}$ and $\rho_{R'}$ respectively.  The relevant parameters in the Boltzmann equations turn out to be: \ba \{T_{RH}, \Gamma_X, M_X, M_{X'}, \langle \sigma\,v\rangle, \langle \sigma\,v\rangle', B_X, B_{X'}, \eta, g_*(T), g^\prime_*(T^\prime)\}. \label{params}\ea  Here $\Gamma_X$ is the decay width of the unstable $X$ particle, $M_X$ and $M_{X'}$ denote the masses of $X$ and $X^\prime$ respectively, while $\langle \sigma\,v\rangle$ and $\langle \sigma\,v\rangle'$ denote the thermally averaged annihilation cross-section of $X$ and $X^\prime$ respectively. $g_*(T)$ and $g^\prime_*(T^\prime)$ are the relativistic degrees of freedom in the visible and dark sectors at a given temperature $T, T^\prime$. The quantities $B_X$ and $B_{X'}$ denote the branching fractions of the modulus to $X$ and $X'$ respectively\footnote{Note that $B_{X}$ also includes channels in which $\phi$ decays to $X$ through intermediate states; $B_{X^\prime}$ is similarly defined.}. Given the assumption $M_{X}, \, M_{X^\prime} \ll m_{\phi}$, $\eta$ approximately denotes the fraction of the energy density from the modulus going to dark radiation, with the remaining fraction $(1-\eta)$ going to visible radiation. Finally, following established convention we define $T_{RH}$ in terms of the decay width of the modulus $\Gamma_{\phi}$ as follows: \ba T_{RH} \equiv \sqrt{\Gamma_\phi M_{\rm pl}} \left(\frac{45}{4 \pi^3 g_*(T_{RH})}\right)^{1/4}, \label{TRH}\ea where $M_{\rm pl} = 1.22\times 10^{19}$ GeV is the Planck scale, and $g_{\star}(T_{RH})$ is the number of relativistic degrees of freedom in the visible sector at $T_{RH}$. We will discuss the physical intepretation of $T_{RH}$ in Section \ref{Tmax}.

{\it A priori}, the nine parameters in (\ref{params}) can vary over a wide range of values, and could affect the computation of the DM relic abundance in a variety of ways. However, we will show that for $\Gamma_X > {\cal O}(1)\Gamma_{\phi}$, the DM production mechanisms only depend on a subset of the parameters in (\ref{params}), in particular: \ba \{T_{RH},\,B_{tot},\,m_{\phi},\,\eta, M_{X^{\prime}},\,\langle\sigma v\rangle^{\prime},\,g_{\star}(T),\,g_{\star}^{\prime}(T^{\prime})\},\label{param-reduced}\ea where $B_{tot} \equiv B_X+B_{X^\prime}$ if $X$ decays to $X^\prime$ and  $B_{tot} \equiv B_{X^\prime}$ if $X$ does not decay to $X^\prime$. Note that there is no dependence on parameters measuring the attributes of the LOSP $X$-- $\{\Gamma_X, M_X, \langle \sigma v\rangle\}$! Furthermore, as will be discussed in Section \ref{viable}, the parameters $T_{RH}$ and $m_{\phi}$ are completely determined by the masses and couplings of the modulus $\phi$. Thus these parameters are insensitive to the details of the dark sector. In the forthcoming analysis, we find it useful to choose benchmark values for the following parameters: \ba  {\rm Benchmark}:\hspace{0.4cm}T_{RH} = 10\, {\rm MeV},\, B_{tot} = 0.1,\, m_{\phi} = 50 \, \mathrm{TeV}, \,\eta = 0.1,\;g_{\star}(T)= 10.75,\;g_{\star}^{\prime}(T^{\prime}) = 10.75 \label{benchmark}\ea The theoretical motivation for these benchmark values will be clear from the discussion in Section \ref{viable}. With these parameters fixed, the DM abundance will depend only on $M_{X^{\prime}}$ and $\langle \sigma v\rangle^{\prime}$, and we will see that these can take a wide range of values for viable DM production mechanisms.  As mentioned above, for most of the paper we take $\Gamma_X > \Gamma_{\phi}$ since this is naturally obtained if $\Gamma_{X}$ is not Planck suppressed. In Appendix~\ref{longlivedX}, however, we will briefly discuss the case $\Gamma_X \lesssim \Gamma_{\phi}$.

The Boltzmann equations  which describe this system are a natural generalization of those which are applicable to a single sector framework within a modulus dominated Universe, as studied in~\cite{Giudice:2000ex, Chung:1998rq}. As pointed out in these papers, it is more convenient to define dimensionless variables corresponding to the energy and number densities and also to convert derivatives with respect to time to those with respect to the (dimensionless) scale factor $A \equiv \frac{a}{a_I}$, with $a_I \equiv T^{-1}_{RH}$. Thus, following~\cite{Giudice:2000ex, Chung:1998rq} we define:
\ba
\Phi \equiv  \frac{\rho_{\phi} A^3}{T_{RH}^4}, \,\, R &\equiv& \rho_R \frac{A^4}{T_{RH}^4},\,\, X \equiv n_X \frac{A^3}{T_{RH}^3},\,\, R^\prime \equiv \rho_{R^\prime} \frac{A^4}{T_{RH}^4},\,\, X^\prime \equiv n_{X^\prime} \frac{A^3}{T_{RH}^3},\nonumber \\
\widetilde{H} &\equiv& \left(\Phi + \frac{R+R^\prime}{A} + \frac{E_{X^\prime}  X^\prime + E_X  X}{T_{RH}}\right)^{1/2}.
\ea $E_{X} \approx (M_{X}^2 + 3\, T^2)^{1/2}$ and $E_{X^\prime}\approx (M_{X^\prime}^2 + 3\, {T^\prime}^2)^{1/2}$ are the thermally averaged $X$, $X^\prime$ energies assuming that $X$ and $X^\prime$ are in kinetic equilibrium. The Boltzmann equations in terms of these comoving dimensionless variables are:
\ba \label{boltzmann}
\notag \widetilde{H} \frac{d\Phi}{dA} &=& -\,c_{\rho}^{1/2}\,A^{1/2} \Phi \\
\notag  \widetilde{H} \frac{d R}{dA} &=& c_{\rho}^{1/2}\,A^{3/2} \left(1 - \bar{B}\right) (1 - \eta) \Phi + c_1^{1/2}\,M_{\rm pl} \,\left[\frac{2 E_X \left<\sigma v \right>}{A^{3/2}}\left(X^2 - {X_{\rm eq}}^2\right)  +   A^{3/2} \left(\frac{E_X - E_{X^\prime}}{{T_{RH}}^3}\right) \left<\Gamma_X^{R} \right> X\right]\\
  \widetilde{H} \frac{d X}{d A } &=& \frac{c_\rho^{1/2} T_{RH} B_{X}}{m_{\phi}}A^{1/2} \Phi+ c_1^{1/2}\,M_{\rm pl} T_{RH} A^{-5/2} \,\left<\sigma v \right>\left({X_{\rm eq}}^2- X^2 \right)  - \frac{c_1^{1/2}M_{\rm pl}}{{T_{RH}}^2} A^{1/2} X \left< \Gamma_X\right>\\
\notag \widetilde{H} \frac{d X^\prime}{d A } &=& \frac{c_\rho^{1/2} T_{RH} B_{X^\prime}}{m_{\phi}}A^{1/2} \Phi +c_1^{1/2}\,M_{\rm pl} T_{RH} A^{-5/2} \, \left<\sigma v \right>'\left({X^\prime_{\rm eq}}^2- {X^\prime}^2 \right) + \frac{c_1^{1/2} M_{\rm pl}}{{T_{RH}}^2} A^{1/2}  X \left< \Gamma_X\right> \\
\notag  \widetilde{H} \frac{d R^\prime}{dA} & =& c_{\rho}^{1/2}\,A^{3/2} \left(1 - \bar{B}\right) \eta\,\Phi + c_1^{1/2}\,M_{\rm pl}\,\left[ \frac{2  E_{X^\prime} \left<\sigma v \right>'}{A^{3/2}}\left({X^\prime}^2 - {X^\prime_{\rm eq}}^2 \right) +   A^{3/2} \left(\frac{E_X - E_{X^\prime}}{{T_{RH}}^3}\right) \left<\Gamma_X^{R^\prime} \right> X\right]
\ea
with $c_{\rho} =\left(\frac{\pi^2 g^*(T_{RH})}{30}\right)$, $c_1 = (\frac{3}{8\pi})$ and
\begin{equation} \label{Bbar}
\bar{B} \equiv \frac{B_X E_X+  B_{X^\prime} E_{X^\prime}}{m_\phi} \, .
\end{equation} $X_{eq}$ and $X^\prime_{eq}$ are related to the $X$ and $X^\prime$ equilibrium number densities via:
\begin{align} X_{eq} & \equiv \left(\frac{A}{{T_{RH}}}\right)^3\frac{g T {M_{X}}^2}{2 \pi^2} K_{2}\left(\frac{M_{X}}{T}\right)  \hspace{3mm}\mathrm{if}\hspace{3mm} M_{X} \gg T, \hspace{3mm} \left(\frac{A}{{T_{RH}}}\right)^3\frac{c_{\xi} \,\zeta(3) T^3}{\pi^2}\hspace{3mm}\mathrm{if}\hspace{3mm} M_{X} \ll T \label{Xeq}, \end{align} where $g$ counts the degrees of freedom of $X$ and $c_{\xi} = g\, (3 g/4)$ for bosonic (fermionic) $X$. $X^\prime_{eq}$ is given by (\ref{Xeq}) with primed variables replacing unprimed variables. 

Note that we have assumed in (\ref{boltzmann}) that $X$ decays to $X^\prime$; we neglect $X^\prime + ... \rightarrow X$ inverse decays, as the dynamics which fix $\Omega_{X^\prime}$ occur when $T^\prime \lesssim M_{X}$ (see Section~\ref{parametrics}) at which point inverse decays are exponentially suppressed. The thermally averaged $X$ decay rate is given by:
\begin{equation}\label{Gamma}
\left<\Gamma_X\right> = \Gamma_X \frac{K_1(M_X/T)}{g_X K_2(M_X/T)}, \hspace{5mm} \left<\Gamma_X\right> \xrightarrow{M_{X} \gg \, T} \frac{\Gamma_X}{g_{X}}. 
\end{equation} 
where $\Gamma_X$ is the $X$ decay rate in the $X$ rest frame, and $K_1$ and $K_2$ are modified Bessel functions of the second kind. The quantities $\left<\Gamma_X^R\right>$ and $\left<\Gamma_X^{R^\prime}\right>$ are respectively the thermally averaged partial widths for $X \rightarrow X^\prime R$ and $X \rightarrow X^\prime R^\prime$. In the remainder of this work, we focus on the case where all $X$ decay channels yield $X^\prime$ such that (\ref{boltzmann}) is valid; this corresponds to $X$ and $X^\prime$ both being charged under the  DM stabilization symmetry. It is also possible for $X$ to instead decay directly to visible radiation, as is the case for R-parity violating SUSY models. In this case $X$ and $X^\prime$ are essentially decoupled in the Boltzmann equations, which significantly simplifies the analysis. In Section \ref{solutions} we focus on the more complicated case where $X$ decays to $X^\prime$, and discuss how relaxing this assumption affects our results. 

The above differential equations are solved subject to the following initial conditions: \begin{equation}
\label{initcond}
A = 1, \,\,\Phi = \Phi_I = \frac{3H_I^2\,M_{\rm pl}^2}{8\pi\,T_{RH}^4}, \,\, R = 0, \,\, R^\prime = 0,\,\, X = 0, \,\, X^\prime =  0\,\,
\end{equation} These initial conditions are somewhat unphysical as they imply $\rho_{R} = \rho_{R^\prime} = 0$ at $A=1$. However, at early times the visible and dark radiation energy densities are subdominant, so this approximation is justified. $H_I$ is the initial value of the Hubble parameter which fixes the initial energy density of the modulus field, parameterized by $\Phi_I$. As we will see, in most cases the DM relic abundance is largely insensitive to the initial condition $\Phi_I$.

\section{Solution of the Boltzmann Equations and the Dark Matter Abundance}\label{solutions}

Given the system of equations~(\ref{boltzmann}), it is possible to numerically solve it for various choices of the 
parameters in~(\ref{params}). However, in order to get a good physical intuition of the qualitatively different mechanisms at play, it is advisable to study various approximate (semi) analytic solutions which are applicable in different regions of the parameter space. We carry out such an exercise in this section. In Appendix \ref{accuracy}, we compare our approximations to the full numerical analysis and find very good agreement. 

\subsection{Useful Approximations}\label{approx}

We now derive useful approximations which allow us to obtain semi-analytic expressions for $\Omega_{DM} h^2$ in Section~\ref{parametrics}. To start with, it is worth noting that $\Phi$ remains constant until $H \sim \Gamma_\phi$ to a very good approximation. Thus in the following analysis we set $\widetilde{H} = {\Phi_I}^{1/2}$ throughout the period of modulus domination, considerably simplifying the Boltzmann equations. Our strategy will be to use physically well-motivated approximations to first solve for $\Phi$, $R$, $R'$ and $X$, and then use these solutions to study the equation for $X'$.

\subsubsection{Approximate solutions for $\Phi$, $R$ and $R^\prime$} 

Consider first the Boltzmann equation for $\Phi$. With $\widetilde{H} =\Phi_I^{1/2}$, it is straightforward to solve for $\Phi$: 
\ba
\Phi \approx \Phi_I\,\exp\left[-\frac{2}{3}\,\left(\frac{c_{\rho}}{\Phi_I}\right)^{1/2}\,(A^{3/2}-1)\right]\, .
\label{phiapprox}
\ea 
Thus, as expected, $\Phi$ remains approximately constant at $\Phi_I$, and only begins to decay vigorously when the dimensionless scale factor satisfies $A > A_{\star}$, with
\begin{equation} A_{\star} \equiv \left(\frac{3}{2}\left(\frac{\Phi_I}{c_{\rho}}\right)^{1/2}+1\right)^{2/3}\, .
\label{Astar}
\end{equation}

Now consider the equations for $R$ and $R^\prime$. As can be seen from~(\ref{boltzmann}), in addition to the modulus decay term these equations contain the $X$ and $X^\prime$ annihilation terms as well as the $X$ decay term. {\it However, it turns out that for $M_{X}, M_{X^\prime} \ll m_{\phi}$ all these terms are quite sub-dominant compared to the modulus decay term}. This is because if $M_{X}, \, M_{X^\prime} \ll m_{\phi}$, the energy densities of $X$ and $X^\prime$ are subdominant to $\rho_{\phi}$ during the modulus dominated era; a more detailed argument for this is presented in Appendix~\ref{justifyingapprox}.  Given this approximation, the solutions to (\ref{boltzmann}) do not depend on the branching fractions of $X$. Thus the approximate solutions for $R$ and $R^\prime$ can be found readily by integrating the modulus decay term:
\ba \label{Rapprox} 
R(A) &\approx& \left(\frac{c_{\rho}}{\Phi_I}\right)^{1/2} (1-\eta) \int_{1}^A (1-\bar{B}){A^\prime}^{3/2}\, \Phi(A^\prime)\, dA^\prime; \hspace{0.5cm}  R^\prime(A) \approx \frac{\eta}{1-\eta} R(A)\\ R_{\rm final} &\equiv& R(A \rightarrow \infty)\approx (1 - \eta)(1-B_{\rm eff})\,\Gamma\left(\frac{5}{3}\right)\,\left[\left(\frac{3}{2}\right)^{2/3}\left(\frac{\Phi_I}{c_{\rho}}\right)^{1/3}\,\Phi_I\right];\;\;\;R'_{\rm final} \approx \frac{\eta}{1-\eta}\,R_{\rm final}.\nonumber
\ea 
In the second line of~(\ref{Rapprox}), $R_{\rm final}$ represents the {\it late time} solution for $R$, i.e when the scale factor $A \gg A_*$. Note that $R \approx R_{\rm final}$ during the radiation dominated era. We have approximated $\bar{B}$ as 
\begin{equation} B_{\rm eff} \equiv \frac{B_{X} \left({M_{X}}^2 + 3 {T_{D}}^2\right)^{1/2} + B_{X^\prime} \left({M_{X^\prime}}^2 + 3 {T_{D}^\prime}^2\right)^{1/2}}{m_{\phi}}\, , \label{Beff} 
\end{equation}
where $T_{D}$ and $T_{D}^\prime$ approximately correspond to the temperatures at which the integrand~(\ref{Rapprox}) peaks. These temperatures characterize the transition between modulus and radiation domination, and are defined more precisely in Section \ref{Tmax}. To obtain the result above for $R_{\rm final}$, we have expanded the function obtained after the integration as a series expansion in $\frac{c_{\rho}}{\Phi_I}$ with $\frac{c_{\rho}}{\Phi_I} \ll 1$ and kept the leading term. This can be justified by taking $\Phi_I$ as given by~(\ref{initcond}), where $H_I$ is the Hubble parameter when the modulus $\phi$ starts dominating the energy density of the Universe. Thus, for $H_I = \gamma\,\Gamma_{\phi}$ with $\gamma \gg 1$,\footnote{We expect $\gamma \gg 1$ because the modulus dominates the energy density of the universe when $m_{\phi} \gtrsim H \gg \Gamma_{\phi}$.} one finds $\frac{c_{\rho}}{\Phi_I} = \frac{1}{\gamma^2} \ll 1$.

\subsubsection{Temperature-scale factor relation and the ``maximum'' temperature}\label{Tmax}

The temperature of a system is measured by the radiation energy density, and the relation between the two is given in general by: \ba\label{TvsR} T = \left(\frac{30}{\pi^2 g_*(T)}\right)^{1/4} \frac{{R}^{1/4}}{a} = \left(\frac{30}{\pi^2 g_*(T)}\right)^{1/4} \frac{{R}^{1/4}}{(A/T_{RH})}\, . \label{TR}\ea In a radiation dominated Universe, it is well known that $R^{1/4} = (\rho_R^{1/4}\,a)$ remains constant with time, giving $T \propto a^{-1}$. However, the situation is different within a modulus dominated Universe since $R^{1/4}$ does not remain constant with time.  It can be shown that at early times when $T \gg T_{RH}$, $\Phi \approx \Phi_I$ and the temperatures and scale factor are related approximately by \cite{Giudice:2000ex}:
\ba\label{tempa}
T\approx \left(\frac{8^8}{3^3 5^5}\right)^{1/20} \left(\frac{g_*(T_{\rm max})}{g_*(T)}\right)^{1/4} {T_{\rm max}} \left( A^{-3/2} - A^{-4}\right)^{1/4} ,
\ea  where $T_{\rm max}$, the maximum temperature attained during modulus domination, is given by:\begin{equation}\label{tmax}
T_{\rm max} \equiv (1-\eta)^{1/4} \left(\frac{3}{8}\right)^{2/5} \left(\frac{5}{\pi^3}\right)^{1/8} \left(\frac{g_*(T_{RH})^{1/2}}{g_*(T_{\rm max})}\right)^{1/4} (M_{\rm pl} H_I T_{RH}^2)^{1/4}\, .
\end{equation} Thus, we see that the temperature has a more complicated dependence on the scale factor compared to that in radiation domination. Using the fact that $H_I = \gamma \, \Gamma_{\phi}$ with $\gamma \gg 1$, one finds that $T_{\rm max} \sim \gamma^{1/4} T_{RH}$. From~(\ref{Rapprox}) it is straightforward to relate the visible and dark sector temperatures:
\ba \label{Tratio}
T^\prime \approx \left(\frac{\eta\, g_*(T)}{\left(1-\eta\right) g_*^\prime(T^\prime)}\right)^{1/4} T
\implies \xi \equiv \frac{T^\prime}{T} \approx \left(\frac{\eta\,g_{\star}(T)}{(1-\eta)g_{\star}^\prime(T^\prime)}\right)^{1/4}\, .
\ea  Combining (\ref{tmax}) and (\ref{Tratio}) gives $T^\prime_{\rm max}$ for the dark sector. As mentioned in Section~\ref{Cosmo}, bounds on $N_{\rm eff}$ at both $T_{BBN} \sim 1$ MeV and $T_{CMB} \sim 1$ eV constrain $T_{BBN}^\prime/T_{BBN}$ and $T_{CMB}^\prime/T_{CMB}$, which through (\ref{Tratio}) can be mapped into a constraint on $\eta$. Comparing~(\ref{Tratio}) with the $N_{\rm eff}$ bound~(\ref{Neff}), we see that the resulting constraint on $\eta$ is insensitive to $g^\prime_*(T^\prime)$ assuming $g^\prime_*(T^\prime)\neq 0$. Taking $g_*(T_{BBN}) = 10.75$ and $g_*(T_{CMB}) = 3$, the $\Delta N_{\rm eff}$ constraints~(\ref{Neff}) imply $\eta \lesssim 0.20$ (BBN) and $\eta \lesssim 0.06$ (CMB).

In the presence of dark radiation, $T_{RH}$ as defined in~(\ref{TRH}) no longer corresponds to the visible sector temperature when $H = \Gamma_{\phi}$, assuming the modulus has completely decayed ($\Phi = 0$). Instead, we define the temperatures $T_D$,\, $T^\prime_D$ as the visible and dark sector temperatures when $H\big|_{\Phi=0} = \Gamma_{\phi}$:
\begin{align}\label{TD}
\notag & H\big|_{\Phi=0} = \frac{(8 \pi/3)^{1/2}}{M_{\rm pl}}\left(\rho_R + \rho_{R^\prime}\right)^{1/2} = \frac{(8 \pi/3)^{1/2}}{M_{\rm pl}}\left(\frac{\rho_R}{1 - \eta}\right)^{1/2} = \Gamma_\phi \\
& \Rightarrow T_D \approx T_{RH} (1- \eta)^{1/4},\hspace{4mm} T^\prime_D \approx \left(\frac{g_*(T_D)}{g_*^\prime(T^\prime_D)}\right)^{1/4} \eta^{1/4}\,T_{RH}
\end{align} 
The bounds from $N_{\rm eff}$ discussed above imply $T_D \approx T_{RH}$. Hence, for simplicity we will set $g_*(T_{RH}) = g_*(T_D)$. It is also useful to compute the scale factor $A_{D}$ which corresponds to the temperature $T_{D},\, T_{D}^\prime$. We compute $A_D$ by substituting $T = T_D$ and $R=R_{\rm final}$ in (\ref{TR}):\begin{equation}\label{ad}
A_D = \left[\Gamma(5/3)\,(3/2)^{2/3}\,(1-B_{\rm eff})\right]^{1/4}\left(\frac{\Phi_I}{c_{\rho}}\right)^{1/3} \approx 1.5 (1- B_{\rm eff})^{1/4}\left(\frac{\Phi_I}{g_*(T_{RH})}\right)^{1/3}
\end{equation} From the definition of $A_{\star}$ in~(\ref{Astar}), we see that $A_{\star} \sim A_D$.

We caution the reader that the definitions of $T_{D}$, $T^\prime_{D}$ and $A_D$ established above are limited in the following sense. The above expressions for $T_{D}$, $T^\prime_{D}$ and $A_D$ were derived from $H = \Gamma_{\phi}$ assuming that the universe has reached radiation domination, i.e. $\Phi = 0$ and $R = R_{\rm final}, \,\,R^\prime = R^\prime_{\rm final}$. However, modulus decay is a continuous process which occurs when $H \sim \Gamma_{\phi}$, but does not have a well-defined start or end point. Upon solving the Boltzmann equations, one finds that when $H = \Gamma_{\phi}$, the modulus has not finished decaying and the radiation dominated phase has not yet been reached $(R \neq R_{\rm final})$. 
In the next subsection we will verify this fact graphically, utilizing the full numerical solutions for $\Phi$ and $R$ (see Figure~\ref{Fig-phiRX} below). Despite this ambiguity, we find $T_{D}$, $T_{D}^\prime$ and $A_D$ to be useful qualitative proxies for the temperature and scale factor at which the universe transitions from the modulus dominated to radiation dominated era.

\subsubsection{Approximate solution for $X$}\label{X-sol}

Now consider the Boltzmann equation for $X$. Motivated by earlier statements, we are interested in the case where $X$ is a LOSP with weak scale mass and annihilation cross section; thus $X_{eq}$ will be exponentially suppressed for temperatures of a few GeV. In our analysis, we will mostly consider the situation that the LOSP $X$ decays before the modulus (typically much before), i.e. $\Gamma_{X} > {\cal O}(1)\Gamma_{\phi}$. Such a condition can be naturally achieved since the modulus decays by Planck-suppressed operators. In Appendix~\ref{longlivedX}, we will briefly consider the case where $\Gamma_X \lesssim \Gamma_{\phi}$.

In the Boltzmann equation for $X^\prime$, the $X$ decay term grows like $A^{1/2}$; thus we are interested in the solution for $X$ in the low temperature regimes where $X_{\rm eq}$ can be neglected (this approximation is justified in Appendix~\ref{longlivedX}). With this approximation, the Boltzmann equation for $X$ can be written as: \ba 
\frac{d X}{d \log A} = - \left(\frac{{X}^2}{X_{\rm crit}}  + \frac{A^3}{X_{\rm crit} \left<\sigma v \right>}\frac{\Gamma_X}{g_X T_{RH}^3} X\right) + \left( \frac{A^3}{X_{\rm crit} \left<\sigma v \right>}\frac{c_{\rho}^{1/2} B_{X}}{m_{\phi} c_{1}^{1/2} M_{pl}}\Phi\right)\, ,
\label{Xeqn}
\ea 
where $X_{\rm crit}$ is the critical value required for annihilations to be efficient for a given value of the Hubble parameter. More precisely, $X_{\rm crit}$ is given by: \ba X_{\rm crit} \equiv (n_X)_{\rm crit}\,\frac{A^3}{T_{RH}^3} =  \frac{H A^3}{\left<\sigma v \right> T_{RH}^3 } = \frac{\widetilde{H}A^{3/2}}{c_1^{1/2}\,M_{\rm pl} T_{RH} \left<\sigma v \right>} \, .\label{Xcrit}\ea 
Now, if the processes for depletion of $X$ (the first and second terms on the right hand side of~(\ref{Xeqn})) and the production of $X$ (the third term in the right hand side of~(\ref{Xeqn})) are \emph{larger} than $X$ itself, then these are each faster than the Hubble rate and one rapidly reaches a situation where the two processes cancel each other, giving rise to what is known as \emph{quasi-static equilibrium} (QSE)~\cite{Cheung:2010gj}. The QSE solution is found by equating the right hand side of~(\ref{Xeqn}) to zero:\begin{equation} X_{\rm QSE} = \frac{\Gamma_X A^3}{2\, T_{RH}^3 g_X \left<\sigma v \right>}
\left[\left(1 + \frac{4 g_X^2 B_X c_{\rho}^{1/2}\, \Phi\, {T_{RH}}^6 \left<\sigma v \right>}{c_1^{1/2} A^3\, m_\phi M_{pl} \Gamma_X^2}\right)^{1/2} - 1 \right]\, .
\label{XQSE} \end{equation} 
Given the criteria described above~(\ref{XQSE}), QSE occurs when: 
\ba \left(X_{\rm QSE} +\frac{A^3}{\langle \sigma v\rangle} \frac{\Gamma_X}{g_X\,T_{RH}^3}\right) > X_{\rm crit}\,\hspace{0.7cm}\&\hspace{0.7cm}\frac{A^3}{\langle \sigma v\rangle}\left[\left(\frac{c_{\rho}^{1/2} B_X}{c_{1}^{1/2} m_{\phi}\,M_{pl}}\right)\,\left(\frac{\Phi}{X_{\rm QSE}}\right)\right] > X_{\rm crit} \, .\label{QSE-cond}\ea 
Upon inspection, one finds that the QSE condition~(\ref{QSE-cond}) is equivalent to the familiar condition $\Gamma_X/g_{X} > H$.
\emph{Thus, we see that as long as $\Gamma_X > g_X\,\Gamma_{\phi}$, the QSE  condition will be satisfied during the modulus dominated era such that $X \approx X_{\rm QSE}$ for $\Gamma_{X} > g_{X} H$.}

We can gain further insight into the QSE solution for $X$ by rewriting~(\ref{XQSE}) as:\ba X_{\rm QSE} &=& \frac{\Gamma_X A^3}{2 T_{RH}^3 g_X \left<\sigma v \right>}
\left[\left(1 + \frac{\langle \sigma v\rangle}{\langle \sigma v\rangle_*}\right)^{1/2} - 1 \right], \nonumber\\
\implies X_{\rm QSE} &\approx& \left( g_X\, b \, B_X \frac{c_{\rho}^{1/2} \,T_{RH}^3}{c_{1}^{1/2} \Gamma_X\,m_{\phi} M_{pl}}\right)\,\Phi\,;
\quad b \approx  \left\lbrace \begin{array}{lr}
  1 & ; \left<\sigma v \right> \ll \left<\sigma v \right>_* \\
  2\left(\frac{\left<\sigma v \right>^c}{\left<\sigma v \right>}\right)^{1/2} & ; \left<\sigma v \right> \gg \left<\sigma v \right>_*
 \end{array} \right. \label{Xapprox}.
\ea Physically, the QSE solution for $X$ occurs when moduli decay into $X$, and $X$ decay into $X^\prime$, balance one another; this explains the dependence of $X_{QSE}$ on $\Phi$. In the above expression, $\left<\sigma v \right>_*$ is defined as:\ba\label{sigmavxcrit}
 \left<\sigma v \right>_* &\equiv& \left(\frac{1}{4 g_{X}^2 B_{X}}\right) \left(\frac{{A}^3 }{\Phi_I }\right)\sqrt{\frac{c_{1}}{c_{\rho}}} \left(\frac{M_{pl} m_{\phi} {\Gamma_{X}}^2}{ T_{RH}^6}\right)\\
& \approx& 4.48\times10^{24}\,\,\mathrm{GeV}^{-2} \times\left(\frac{5}{g_X^2\, B_X}\right) \left(\frac{A}{A_D}\right)^3 \left(\frac{m_{\phi}}{50\,\mathrm{TeV}}\right)\left(\frac{10\,\mathrm{MeV}}{T_{RH}}\right)^6 \left(\frac{\Gamma_X}{10^{-5}\,{\rm GeV}}\right)^2\left(\frac{10.75}{g_*(T_{RH})}\right)^{3/2}\, . \nonumber 
\ea
Note that for the benchmark choice of parameters in~(\ref{benchmark}), and $\Gamma_{X}$ not extremely small, $\langle \sigma v\rangle_*$ is quite large (compared to a WIMP cross-section $\sim 10^{-7}-10^{-10}\,{\rm GeV}^{-2}$). We expect the same qualitative conclusion as long as the portal coupling is not extremely tiny. Thus for supersymmetric models where $X$ is the LOSP, we expect $\langle \sigma v\rangle \ll \langle \sigma v\rangle_*$, and hence $b \approx 1$ in the QSE solution for $X$ in the second line of~(\ref{Xapprox}). 

\begin{figure}[t!]
  \begin{center}
    \includegraphics[width=0.7\textwidth]{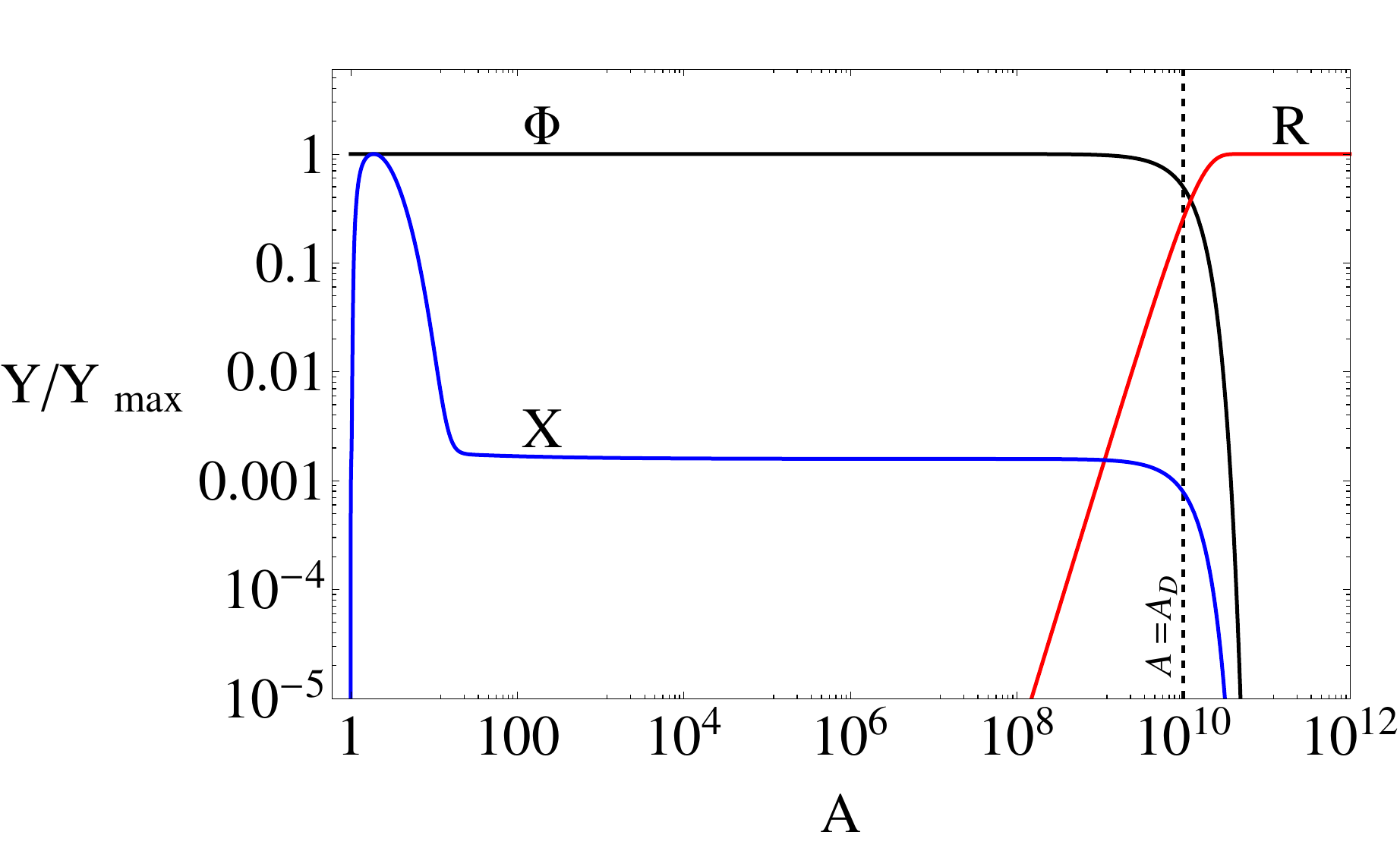}
  \end{center}
  \caption{\footnotesize{Plots of the exact solutions for $Y = \Phi,\,R$ and $X$ (normalized to their maximum values) as functions of the scale factor $A$. We have taken $H_I = 10^{15} \Gamma_{\phi}$, $B_{X} = 0.1$, $\left<\sigma v \right> = 10^{-7}$ GeV$^{-2}$ and $\Gamma_{X} = 10^{-5}$ GeV. All other parameters taken to their benchmark values~(\ref{benchmark}). The dashed vertical line represents the scale factor $A=A_D$ defined in~(\ref{ad}), which characterizes the transition between a modulus dominated and a radiation dominated universe.}}\label{Fig-phiRX}
\end{figure}

Figure~\ref{Fig-phiRX} shows a plot of the solutions for the values of $\Phi, R$ and $X$ (normalized to their maximum values) as functions of the scale factor $A$ for the choice of benchmark parameters as in~(\ref{benchmark}). 
As can be seen from~(\ref{phiapprox}), (\ref{Rapprox})~and~(\ref{Xapprox}), respectively, the solutions for  $\Phi, R$ and $X$ do not depend on $M_{X}$, $M_{X^{\prime}}$ or $ \left<\sigma v \right>^{\prime}$ to a good approximation. Moreover the solution for $X$ depends does not depend on $\left<\sigma v \right>$ for most models of interest in which $\left<\sigma v \right> \ll \left<\sigma v \right>_*$ as we have just discussed above.

\subsection{Classifying Production Mechanisms for Relic Dark Matter}\label{parametrics}

We now move on to studying the main quantity of interest -- the Boltzmann equation for $X^{\prime}$, whose solution will give us the expression for the relic abundance $\Omega_{DM} h^2$ of dark matter $X^{\prime}$
in terms of a subset of the parameters~(\ref{params}) appearing in the Boltzmann equations. More precisely, the $X^\prime$ relic abundance is given by:\begin{equation}\label{omega1}
\Omega_{DM} h^2 = \frac{\rho_{X^\prime}(T_f^\prime)}{\rho_{R}(T_f)} \frac{T_f}{T_{\rm now}} \Omega_{R} h^2 = M_{X^\prime} \frac{X^\prime(T_{f}^\prime)}{R(T_f)} \frac{A_f T_{f}}{T_{\rm now} T_{RH}} \Omega_{R} h^2 \, .
\end{equation} 
In the above expression, $T_f$ is the temperature at any very late time in which the universe has become radiation dominated ($T_f \ll T_D$) and the $X^\prime$ comoving abundance has become constant. The parameters $T_{\rm now} \approx 2.35 \times 10^{-13}$ GeV  and $\Omega_{R} h^2 \approx 4.17 \times 10^{-5}$ are the present day temperature and radiation relic density. Taking $R(T_f) \approx R_{\rm final}$ and using~(\ref{TvsR}) to relate $A_f$ and $T_f$, (\ref{omega1})~can be written as:\begin{equation}\label{omega2}
\Omega_{DM} h^2 \approx L^{-3/4} \frac{X^\prime(T_f^\prime)}{\Phi_I} \frac{M_{X^\prime}}{T_{\rm now}} \Omega_{R} h^2, \hspace{3mm} L \equiv (1-\eta)(1-B_{\rm eff}) \Gamma(5/3)\left(\frac{3}{2}\right)^{2/3}
\end{equation} In order to derive semi-analytic approximations for $X^\prime(T_f)$ and $\Omega_{DM} h^2$, we will solve the Boltzmann equation for $X^\prime$ given the approximations stated in the previous sections. In the following we will show that $X^\prime(T_f) \propto \Phi_I$, so $\Omega_{DM} h^2$ is insensitive to $\Phi_I$ as mentioned above.

Using the approximate solutions for $\Phi, R, R^{\prime}$ and $X$ in~(\ref{phiapprox}), (\ref{Rapprox})~and~(\ref{Xapprox}), respectively, we can reduce the system of Boltzmann equations in~(\ref{boltzmann}) to a single ordinary differential equation for the evolution of $X^\prime$:\begin{equation}
\frac{d X^\prime}{d A } \approx \frac{c_1^{1/2}\,M_{\rm pl} T_{RH} \left<\sigma v \right>^{\prime} A^{-5/2}}{\widetilde{H}}\left[{X^\prime_{\rm eq}}^2- {X^\prime}^2 \right] + \frac{c_1^{1/2}\,A^{1/2}}{\widetilde{H}}\left(  \frac{c_{\rho}^{1/2} T_{RH} B_{X^\prime} }{c_{1}^{1/2} m_{\phi} }\Phi + \frac{\Gamma_X M_{\rm pl}}{g_X T_{RH}^2} X_{\rm QSE}\right)\label{simplified}
\end{equation} where $X_{\rm QSE}$ is defined in~(\ref{Xapprox}). Note that if $X$ does not decay to $X^\prime$, the $X^\prime_{QSE}$ term in (\ref{simplified}) is absent. Using a similar definition for the critical annihilation for $X'$ as was used for $X$ in~(\ref{Xcrit}), one can rewrite (\ref{simplified}): 
\ba\label{loga}
\frac{d X^\prime}{d \log A} &\approx& -\left[\frac{ {X^\prime}^2}{X^\prime_{\rm crit}}\right]  + \left[\frac{{X_{\rm eq}^\prime}^2}{X^{\prime}_{\rm crit}}+ \frac{A^3}{X^\prime_{\rm crit} \left<\sigma v \right>^{\prime}}\left( \frac{c_{\rho}^{1/2} B_{tot}}{c_{1}^{1/2} m_{\phi} M_{pl}}\Phi \right)\right]\\
X^\prime_{\rm crit}(A) &\equiv& \frac{H A^3}{\left<\sigma v \right>^{\prime} T_{RH}^3 } = \frac{\widetilde{H}A^{3/2}}{c_1^{1/2}\,M_{\rm pl} T_{RH} \left<\sigma v \right>^{\prime}}, \nonumber
\ea 
where $B_{tot} \equiv B_{X} + B_{X^\prime}$ if $X$ decays to $X^\prime$\footnote{Note that as discussed below~(\ref{sigmavxcrit}), $\langle \sigma v\rangle \ll \langle \sigma v\rangle_*$ for most models where $X$ is a LOSP, for which $b \approx 1$ from~(\ref{Xapprox}). Therefore, we have used the expression for $X_{\rm QSE}$ with $b\approx 1$ in (\ref{loga}).}, and $B_{tot} \equiv B_{X^\prime}$ if $X$ does not decay to $X^\prime$. Just as for the case of $X$, if the processes of depletion of $X^{\prime}$ (first term on the right hand side of~(\ref{loga})) and production of $X^{\prime}$ (second, third and fourth terms on the right hand side of~(\ref{loga})) are each greater than $X^{\prime}$ itself, $X^\prime$ will rapidly reach a quasi-static equilibrium (QSE) attractor solution such that terms on the right hand side of~(\ref{loga}) cancel among themselves: \begin{equation}\label{xpqse}
X^\prime_{\rm QSE}(A) = \left[\frac{A^3}{\left<\sigma v \right>^{\prime}}\left( \frac{c_{\rho}^{1/2} B_{tot}}{c_{1}^{1/2} m_{\phi} M_{pl}}\Phi \right) + {X^\prime_{\rm eq}}^2 \right]^{1/2}\, .
\end{equation} Comparing~(\ref{loga}) and~(\ref{xpqse}), and using~(\ref{Xapprox}) for the QSE solution for $X$, we see that the QSE conditions hold when: \ba \label{xpqse-cond} X^\prime_{\rm QSE} > X^\prime_{\rm crit}\, .\ea 
{\it Note that in contrast to $X$ for $\left<\Gamma_{X}\right> > \Gamma_{\phi}$, $X^{\prime}$ does not necessarily enter QSE during the modulus dominated phase}. One reason for this is that in contrast to $X$, which is assumed to be a WIMP, we are exploring a much more general set of possibilities for the mass and interactions of the DM particle $X^{\prime}$. 

In order to understand better the broad possibilities that could arise for $X^{\prime}$, it is important to find the conditions necessary for QSE to hold at $A \approx A_D$. If the QSE conditions hold at $A \approx A_D$, then the positive contribution to $X^\prime$ from modulus decay is annihilated away such that $X$ maintains its QSE value. In this case, the final $X^\prime$ abundance is insensitive\footnote{Modulo logarithmic sensitivity, as will be discusssed in Section \ref{nr-qse}.} to modulus decay parameters such as $m_{\phi}$ and $B_{tot}$. Conversely if QSE does not hold at $A \approx A_{D}$, $\Omega_{X^\prime} h^2$ will be sensitive to contributions from modulus decay, along with other sources for $X^\prime$ production during the modulus dominated era. Comparing~(\ref{loga}) and~(\ref{xpqse}), we see that requiring $X^\prime_{\rm QSE}(A_D) > X^\prime_{\rm crit}(A_D)$ places a lower bound on $\left<\sigma v \right>^\prime$. Keeping the above statements in mind, it is useful to define a critical annihilation cross section such that $X^\prime_{\rm QSE} = X^\prime_{\rm crit}$ at $A = A_D$, to delineate the various possibilities:
\ba \label{critnr}
{\left<\sigma v \right>^{\prime}}_c &\equiv& \left(\frac{c_{\Gamma}^{1/2}}{c_1\,B_{\rm tot}}\right)\left(\frac{m_{\phi}}{T_{RH}^2 M_{\rm pl}}\right); \hspace{6.1cm}\,\, M_{X^\prime} \gg T_{D}^\prime\\ \label{critr} {\left<\sigma v \right>^{\prime}}_c &\equiv& \left(\frac{\pi^2\,c_{\Gamma}^{-1/2}}{\theta\,g^\prime\,\zeta(3)}\,\left(\frac{2}{3}\right)^{1/4}\frac{1}{{\Gamma(5/3)}^{3/8}}\right) \left(\frac{g^\prime_*(T_{D}^\prime)}{g_*(T_D)\,\eta}\right)^{3/4}\frac{1}{T_{RH}\,M_{\rm pl}};\hspace{5mm}\,\, M_{X^\prime} \ll T_{D}^\prime\nonumber\\ &\approx& 2.35\,\left(\frac{3.0}{\theta\,g}\right)\left(\frac{g_{\star}^{\prime}(T_D^{\prime})}{\eta}\right)^{3/4}\left(\frac{10.75}{g_{\star}(T_{RH})}\right)^{1/4}\frac{1}{T_{RH}\,M_{\rm pl}}
\ea where we have approximated $\widetilde{H} \approx \Phi_I$ at $A = A_D$. In the above expressions, $c_{\Gamma} \equiv \left(\frac{45}{4\pi^3\,g_{\star}(T_{RH})}\right)$, $g^\prime$ is the degrees of freedom of $X^\prime$, and $\theta = 1\, (3/4)$ for bosonic (fermionic) $X^\prime$. The above expressions were obtained by taking $X^\prime_{\rm eq} \rightarrow 0$ in the $M_{X^\prime} \gg T_{D}^\prime$ case and $X^\prime_{\rm QSE} = X^\prime_{\rm eq}$ in the $M_{X^\prime} \ll T_{D}^\prime$ case. 

In the following sections, we will classify production mechanisms for $X^\prime$ according to whether or not $\left<\sigma v \right>^{\prime}|_{T^\prime = T^\prime_{D}} > \left<\sigma v \right>^{\prime}_c$, or equivalently whether or not $X^\prime$ annihilations are efficient at $T_{D}^\prime$. \emph{To simplifiy the following analysis, we will assume that $\left<\sigma v \right>^\prime$ is temperature independent.} The generalization of our results to temperature dependent $\left<\sigma v \right>^\prime$ is presented in Appendix~\ref{tempsigmavd}.

\subsection{Efficient Annihilation at $T_{D}^\prime$: $\left<\sigma v \right>^{\prime} > \left<\sigma v \right>^{\prime}_c$} \label{efficient}

If $\left<\sigma v \right>^{\prime} > \left<\sigma v \right>^{\prime}_c$, $X^\prime$ tracks its QSE value until $X^\prime_{\rm QSE}$ drops below $X^\prime_{\rm crit}$ at $A \gtrsim A_{D}$. When $X^\prime_{\rm QSE}$ drops below $X^\prime_{\rm crit}$, annihilations are no longer efficient and the comoving $X^\prime$ abundance becomes constant. The dynamics of this process, along with the resulting parametrics for $\Omega_{X^\prime}$, depends on whether or not the  freeze-out temperature for $X^\prime$, $\hat{T}^\prime_{FO}$, is larger than $T_{D}^\prime$. Here $\hat{T}^\prime_{FO}$ is the $X^\prime$ freeze-out temperature, which is computed assuming a radiation dominated universe~(\ref{xfrad}). If $\hat{T}_{FO}^\prime > T_{D}^\prime$ we can neglect $X^\prime_{\rm eq}$ in $X^\prime_{\rm QSE}$ for $T \sim T_{D}$; in this case $X^\prime_{\rm QSE} \propto \Phi^{1/2}$, and $X^\prime_{\rm QSE}$ drops below $X^\prime_{\rm crit}$ when the modulus decays at $T^\prime \sim T_{D}^\prime$. If instead $T_{D}^\prime > \hat{T}_{FO}^\prime$, $X^\prime$ remains in thermal equilibrium during the onset of radiation domination ($X^\prime_{\rm QSE} \approx X^\prime_{\rm eq}$ for $T^\prime \lesssim T_{D}^\prime$). In this case the $X^\prime$ relic abundance is determined by the standard freeze-out mechanism.

\subsubsection{Non-relativistic quasi-static equilibrium\, ($QSE_{\rm nr}$)}\label{nr-qse} 

First consider the case where $\hat{T}^\prime_{FO}> T_{D}^\prime$ such that $X^\prime_{\rm eq}$ can be neglected for $T^\prime \gtrsim T_D^{\prime}$.  Assuming $\left<\sigma v \right>^{\prime} > \left<\sigma v \right>^{\prime}_c$, $X^\prime$ tracks $X^\prime_{\rm QSE} \propto \Phi^{1/2}$ until $A \gtrsim A_D$, after which the ratio $X^\prime_{\rm QSE}/X^\prime_{\rm crit}$ begins to drop exponentially due to the decay of $\Phi$ according to~(\ref{phiapprox}). The final $X^\prime$ value is given by $X^\prime_{\rm QSE}(A_c)$, where $A_c$ is determined by solving the transcendental equation:\ba \label{acrit}
\notag  X^\prime_{\rm QSE}(A_c) &=& \frac{1}{\kappa} X_{\rm crit}^\prime(A_c) \nonumber \Rightarrow \left(\frac{\Phi}{\widetilde{H}^2}\right) \Bigg|_{A_c} = \frac{\left<\sigma v \right>^{\prime}_c}{\kappa^2 \left<\sigma v \right>^{\prime}}\nonumber\\ \Rightarrow \log\left[\tilde{A}_c\right] &=& \frac{2}{3} c_{\rho}^{1/2}\left[\tilde{A}_c\right]^{3/2} +\log \left[{c_\rho}^{-1/3}\left(\frac{3}{2}\right)^{2/3} \Gamma(5/3)\right] - \log \left[\left(\frac{\kappa^2 \left<\sigma v \right>^{\prime}}{\left<\sigma v \right>^{\prime}_c} - 1\right) \right].\ea 
We have defined $\tilde{A}_c \equiv A_c\,{\Phi_I}^{1/3}$ and have used the approximation $R(A_c) \approx R_{\rm final}$. Taking $\kappa \approx 2$ gives close agreement with the full numerical result. We denote the above mechanism for DM production as $QSE_{\rm nr}$.

Upon solving~(\ref{acrit}) for $\tilde{A}_c$, it is straightforward to compute $\Omega_{DM} h^2$ using~(\ref{omega2}) with $X^\prime(T_f^\prime) = \kappa^{-1} X^\prime_{\rm crit}(A_c)$:
\ba
\Omega\,h^2 \;[QSE_{\rm nr}]  &\approx&  \frac{B_{\rm tot}^{1/2}}{L^{3/4}c_{\Gamma}^{1/4}} \frac{\tilde{A}_c^{3/2}\exp\left(-\frac{1}{3} c_\rho^{-1/2} {\tilde{A}_c}^{3/2}\right)}{(M_{\rm pl}m_{\phi} \left<\sigma v \right>^{\prime})^{1/2}} \left[\frac{M_{X^\prime}}{T_{\rm now}}\right] \left[\Omega_{R} h^2\right]\label{omegadm1}\\ 
&\approx&\left[\frac{(\Gamma(\frac{5}{3})\,(\frac{3}{2})^{2/3})^{1/2}}{\kappa\,c_{\rho}^{1/6}\,c_1^{1/2}L^{3/4}}\right] \left[\frac{\tilde{A}_c}{M_{X^\prime}M_{\rm pl}\left<\sigma v \right>^{\prime}}\frac{M_{X^\prime}}{T_{RH}}\right] \left[\frac{M_{X^\prime}}{T_{\rm now}}\right] \left[\Omega_{R} h^2\right].\nonumber
\ea In the above, we have made the approximation $e^{-\frac{2}{3}\left[\frac{c_\rho}{{\Phi_I}}\right]^{1/2}} \approx 1$, see discussion below (\ref{phiapprox}). Also, in the second line, we have used (\ref{acrit}) to get rid of the exponential factor in the first line. The factor $\tilde{A}_c$ in the numerator depends logarithmically on both $\left<\sigma v \right>^{\prime}$ and $\left<\sigma v \right>^\prime_c$. 

\begin{figure}[t!]
  \begin{center}
    \includegraphics[width=0.7\textwidth]{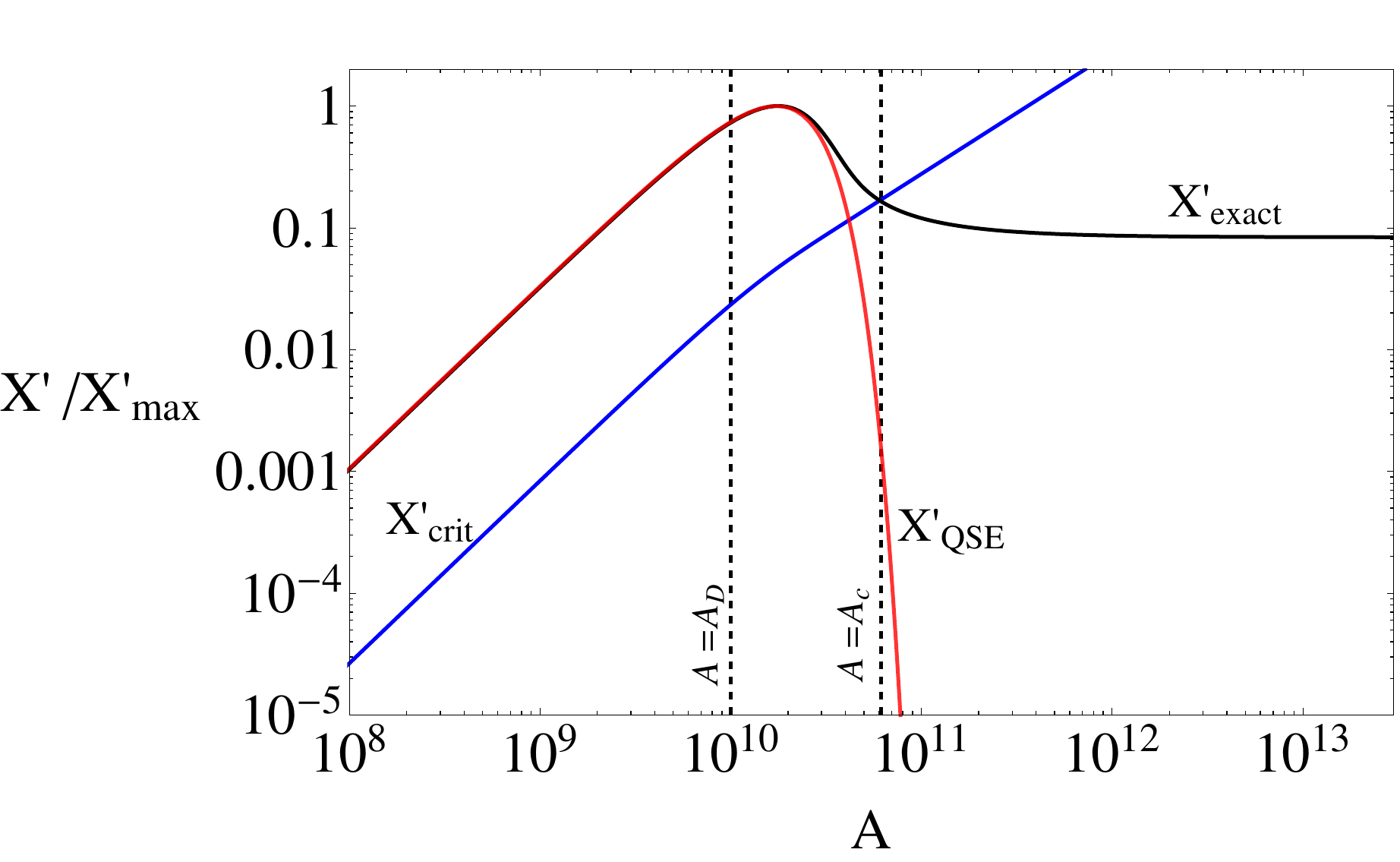}
  \end{center}
  \caption{\footnotesize{Plot of the exact solution of the Boltzmann equations for $X'$ (normalized to its maximum value) as a function of the scale factor $A$ corresponding to the $QSE_{\rm nr}$ mechanism. We have taken $H_I = 10^{15} \Gamma_\phi$, $\left<\sigma v\right>^{\prime} = 10^{-6}\,{\rm GeV^{-2}}$ and $\,M_{X^\prime} = 10$ GeV, with all other parameters set to the benchmark values (\ref{benchmark}). We have also plotted $X^\prime_{\rm crit}$ (\ref{loga}) and $X^\prime_{\rm QSE}$ (\ref{xpqse}). The horizontal dashed line corresponding to $A = A_{c}$ is determined by solving the transcendental equation (\ref{acrit}) for $A_{c}$.}}\label{Fig-QSEnr}
\end{figure}

\subsubsection{Standard freezeout during radiation domination\,($FO^{\rm rad}_{\rm r}$ $\&$ $FO^{\rm rad}_{\rm nr}$)}\label{std-fo}

Now consider the case where $\hat{T}^\prime_{FO} < T_{D}^\prime$ with $\left<\sigma v \right>^{\prime} > \left<\sigma v \right>^{\prime}_{c}$. Then, as discussed above, $X^\prime_{\rm QSE}(T_{D}^\prime) \approx X^\prime_{\rm eq}(T_{D}^\prime)$, which implies that $X^\prime$ is in thermal equilibrium at $T^\prime \approx T_{D}^\prime$ and freezes out at some $\hat{T}^\prime_{FO} < T_{D}^\prime$ when $X^\prime_{\rm eq}$ drops below $X^\prime_{\rm crit}$. {\it The universe is radiation dominated for $T^\prime \lesssim T^\prime_{D}$; thus the $X^\prime$ relic abundance is determined by the standard thermal freeze-out mechanism}. Furthermore, there are two possible sub cases - i) non-relativistic freezeout during radiation domination when $T_D^{\prime} > M_{X}^{\prime} > \hat{T}_{FO}^{\prime}$, which we denote as $FO^{\rm rad}_{\rm nr}$, and ii) relativistic freezeout when $T_D^{\prime}  > \hat{T}_{FO}^{\prime} > M_{X^{\prime}}$, which we denote as $FO^{\rm rad}_{\rm r}$. The relic abundance in the two cases are given by~(\ref{omega1}) with $T_{f} = \hat{T}_{FO}$, $T_{f}^\prime = \hat{T}_{FO}^\prime$ and $\rho_{X^\prime}(\hat{T}_{FO}^\prime)$ determined by the standard freeze-out calculation. Specifically, $\rho_{X^\prime}(\hat{T}_{FO}^\prime) = \rho_{{X}^\prime_{\rm eq}}(\hat{T}_{FO}^\prime)$, where $\hat{T}_{FO}^\prime$ is defined by $n^\prime_{\rm eq}(T_{FO}^\prime) \equiv H/\left<\sigma v\right>^\prime$. Assuming non-relativistic freeze-out, $\hat{T}^\prime_{FO}$ is given by solving the transcendental equation:\begin{equation}\label{xfrad}
\hat{x}_F^{\prime}\equiv \frac{M_{X^{\prime}}}{T^{\prime}_{FO}} = \log\left( \frac{3}{8 \pi^3} \sqrt{\frac{10\,\eta}{g^\prime_*(T^\prime_{FO})}} \left<\sigma v \right>^\prime g^\prime M_{X^\prime} M_{\rm pl} (\hat{x}_F^{\prime})^{1/2}\right)
\end{equation}and the resulting relic abundance is given by:\ba \label{radnr}
\Omega\,h^2\;[FO^{\rm rad}_{\rm nr}] &\approx& \left[\frac{4 \sqrt{5}}{\sqrt{\pi}}\right] \left[\frac{\eta^{1/4}} {\left(1-\eta\right)^{3/4}}\right] \left[\frac{1}{g_*(T_{FO}){g^\prime_{*}(T^\prime_{FO})}}\right]^{1/4} \left[\frac{\hat{x}_{F}^{\prime}}{M_{X^{\prime}}\,M_{pl} \left<\sigma v \right>^{\prime}}\right]\left[\frac{M_{X^{\prime}}}{T_{\rm now}}\right]\,[\Omega_{R}\,h^2]\, . \ea If instead $\hat{x}_F^{\prime} \lesssim 3$, $X^\prime$ freeze-out occurs relativistically, and:\ba \label{radr}
 \Omega\,h^2\;[FO^{\rm rad}_{\rm r}] &\approx& \left[\frac{30\,\zeta{(3)}}{\pi^4}\right]\left[\frac{\eta g_*(T_{FO})}{\left(1-\eta\right) g_*^\prime(T^\prime_{FO})}\right]^{3/4} \left[\frac{c_{\xi}}{g_{\star}(T_{FO})}\right]\left[\frac{M_{X^\prime}}{T_{\rm now}}\right]\;[\Omega_{R}\,h^2]
\ea where $c_\xi = g^\prime\, (3 g^\prime/4)$ for bosons (fermions).

Note that although the \emph{mechanism} for DM production discussed here is standard thermal freezeout, the relevant parametric region is very different compared to that of usual thermal WIMP freezeout. In particular, here $\hat{T}_{FO}^\prime$ is smaller than $T_D^{\prime} \approx T_{RH}\,(\eta)^{1/4}\,\left(\frac{g_{\star}(T_D)}{g_{\star}^{\prime}(T_D^{\prime})}\right)^{1/4} \lesssim (0.1-0.5)\,T_{RH}$ for reasonable choices of parameters. This implies that $M_{X^\prime} < \hat{x}_{F}^\prime T_{D}^\prime \lesssim 10 \times T_{RH}$. Furthermore, for the cosmological scenarios described in the introduction and in Section \ref{viable}, one expects $T_{RH}$ to be in the range: few~MeV $\lesssim T_{RH} \lesssim 100$ MeV. {\it Thus, the DM in this case is much lighter than a typical electroweak-scale WIMP, even if the underlying mechanism is non-relativistic freezeout during radiation domination ($FO^{\rm rad}_{\rm nr}$).} On the other hand, DM undergoing {\it relativistic} thermal freezeout in the dark sector ($FO^{\rm rad}_{\rm r}$) is qualitatively similar to the case of neutrino decoupling in the visible sector. We reiterate that in all other regions of $M_{X^\prime}$ and $\left<\sigma v \right>^\prime$ parameter space, the standard thermal freeze-out calculation will not be valid.

\subsection{Inefficient Annihilation at $T_{D}^\prime$: $\left<\sigma v \right>^{\prime} < \left<\sigma v \right>^{\prime}_c$}\label{inefficient}

We now consider the case where $\left<\sigma v \right>^{\prime} < \left<\sigma v \right>^{\prime}_c$ such that $X^\prime$ is not in QSE for $T \gtrsim T_{D}$. In contrast to the previous case, the $X^\prime$ relic abundance will be sensitive to both early-time $X^\prime$ production during the modulus dominated era and the modulus branching ratio $B_{tot}$. Because the annihilation rate $\Gamma(X^\prime) \sim n_{X^\prime}^2 \left<\sigma v \right>^\prime$ is much smaller than the Hubble parameter for $T^\prime \gtrsim T_{D}^\prime$, the ${X^\prime}^2$ term in~(\ref{simplified}) can be neglected for $T^\prime \gtrsim T^\prime_{D}$. The Boltzmann equation for $X^\prime$ becomes linear in this limit, and the contributions to $\Omega_{DM}$ can be separated into two sources:\begin{equation}
\Omega_{DM}\,h^2 = \Omega_{\rm ann}\,h^2 + \Omega_{\rm decay}\,h^2.
\end{equation} 
The first term, $\Omega_{\rm decay}\,h^2$, is the contribution from modulus and $X$ decays. This term can be computed by taking $\widetilde{H} = {\Phi_I}^{1/2}$ and integrating the second term in the RHS of (\ref{simplified}) to $A = A_{f} \gg A_{*}$. Taking $\exp(-2 c_{\rho}^{1/2}/3 {\Phi_I}^{1/2}) \approx 1$, equation~(\ref{omega2}) gives:\begin{equation}\label{moduliBR}
\Omega_{\rm decay}\,h^2 \approx L^{-3/4} \left[B_{\rm tot} \,\frac{T_{RH}}{m_{\phi}}\right] \,\frac{M_{X^\prime}}{T_{\rm now}}\,\left[\Omega_R\,h^2\right]
\end{equation} 
On the other hand, as the name suggests, $\Omega_{\rm ann}\,h^2$ parameterizes contributions to $X^\prime$ production which arise from the annihilation term in~({\ref{simplified}). This has been discussed in~\cite{Giudice:2000ex} in models with a single sector. There are two qualitatively different cases regarding the parameterics of $\Omega_{\rm ann}\,h^2$.

The first case arises when the DM particle $X^{\prime}$ attains equilibrium at high temperatures (but $\left<\sigma v\right>^{\prime}$ is still smaller than $\left<\sigma v\right>^{\prime}_c$) and freezes out during {\it modulus domination}; hence $T_{\rm max}^{\prime} > T_{FO}^{\prime} > T_{D}^{\prime}$. Here $T_{FO}^\prime$ is the $X^\prime$ freeze-out temperature computed assuming a \emph{modulus} dominated universe (\ref{xf}).  Now, one might naively think that both non-relativistic and relativistic thermal freezeout may be possible during modulus domination, just as they are during radiation domination (see section \ref{std-fo}). However, as noted in \cite{Giudice:2000ex}, relativistic freeze-out cannot occur during modulus domination if $\left<\sigma v \right>^\prime \propto \left(T^\prime\right)^n$ with $n < 6$. To see this, note that the term in~(\ref{loga}) corresponding to $R^\prime R^\prime \rightarrow X^\prime X^\prime$ inverse annihilations scales like ${X^\prime_{\rm eq}}^2 \left<\sigma v \right>^\prime /{X^\prime_{\rm crit}} \propto \left(T^\prime\right)^{(-6 + n)}$ when $X^\prime$ is relativistic. Thus if $X^\prime$ decouples from the thermal bath of dark radiation while relativistic at some temperature $T^\prime_{\rm dec}$, the $X^\prime$ comoving abundance will continue to grow for $T^\prime < T^\prime_{\rm dec}$ due to inverse annihilations, provided $n < 6$. In this work we will only consider $n < 6$; thus for the models considered here, freeze-out during modulus domination occurs only if $T_{\rm max}^{\prime} > T_{FO}^{\prime} > T_{D}^{\prime}$ \emph{and} $M_{X^\prime} > T^\prime_{FO}$.

The second case arises when $T^{\prime}_{FO} > T_{\rm max}^{\prime} >  T_D^{\prime}$ ($X^{\prime}$ never reaches equilibrium) or when $T^\prime_{\rm max} > T^\prime_{FO} > M_{X^\prime}$ ($X^\prime$ decouples while relativistic). In this case, it turns out that the contribution to DM abundance comes predominantly from {\it inverse annihilations} via $R^\prime R^\prime \rightarrow X^\prime X^\prime$, as will be seen shortly. 

\subsubsection{Non-relativistic freezeout during modulus domination\, ($FO^{\rm mod}_{\rm nr}$)}\label{nrFO}

Let us first consider the case where $X^\prime$ reaches chemical equilibrium and then undergoes freeze-out during modulus domination ($T_{\rm max}^{\prime} > T_{FO}^{\prime} > T_{D}^{\prime}$). From the arguments above, we note that freezeout can only occur when DM is non-relativistic, hence we denote this mechanism as $FO^{\rm mod}_{\rm nr}$. The $X^\prime$ freezeout temperature, defined as $T^{\prime}_{FO}$ such that $n_{X^\prime}^{\rm eq}(T^\prime_{FO}) \left<\sigma v \right> \equiv H(T^\prime_{FO})$, is given by solving the following transcendental equation for $x^{\prime}_F \equiv \frac{M_{X^{\prime}}}{T_{FO}^{\prime}}$:
\begin{equation}\label{xf}
x_{F}^{\prime} = \ln \left[\left(\frac{3}{2 \sqrt{10} \pi^3}\right) \left(\frac{g^\prime g_*(T_{RH})^{1/2}}{g^\prime_*(T^\prime_{FO})}\right)\left(\frac{M_{\rm pl}}{M_{X^{\prime}}}\right) [T^2_{RH} \left<\sigma v \right>^{\prime}] \,\eta\, {x_{F}^{\prime}}^{5/2}\right]
\end{equation} 
where $x_{F}^{\prime} \equiv \frac{M_{X^\prime}}{T^\prime_{FO}}$. 
Note that the above equation, and hence the parameters $T_{FO}^{\prime}$ and $x_F^{\prime}$, are valid only if $T_{\rm max}^{\prime} > T_{FO}^{\prime} > T_D^{\prime}$ and $M_{X^{\prime}} > T_{FO}^{\prime}$, i.e. $1 < x_F^{\prime} < \left(\frac{M_{X^{\prime}}}{T_D^{\prime}}\right)$. 

\begin{figure}[t!]
  \begin{center}
    \includegraphics[width=0.7\textwidth]{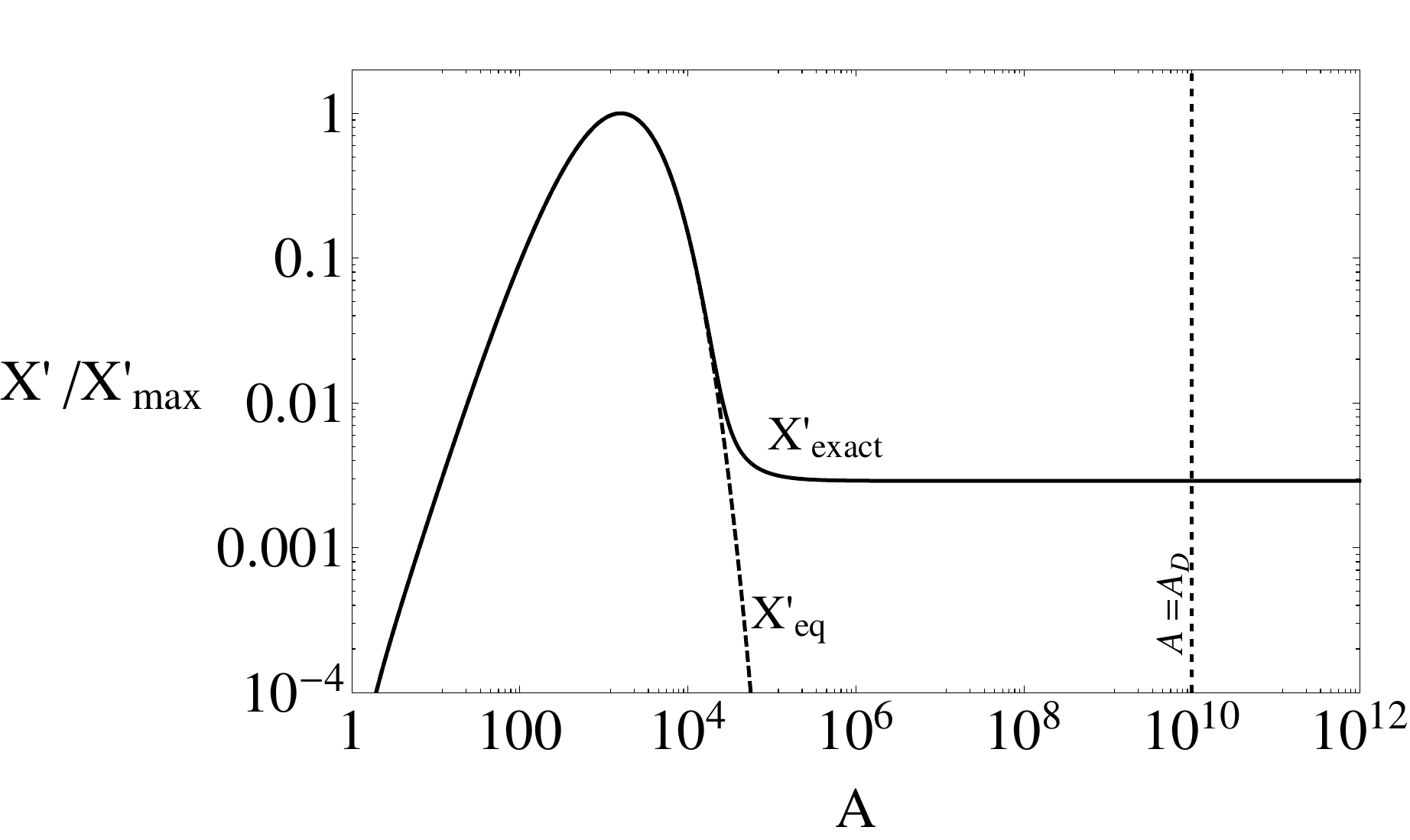}
  \end{center}
  \caption{\footnotesize{Plot of the exact solution for $X^{\prime}$ (normalized to its maximum value) as a function of the scale factor $A$ corresponding to the $FO^{\rm mod}_{\rm nr}$ mechanism. We have taken $H_I = 10^{15} \Gamma_\phi$, $\left<\sigma v\right>^{\prime} = 10^{-6}\,{\rm GeV^{-2}}, \,M_{X^\prime} = 10$ GeV as in Figure~\ref{Fig-QSEnr}, but have instead chosen $B_{tot}=0$ so that the QSE condition is not satisfied, see~(\ref{critnr}).  All other parameters set to benchmark values~(\ref{benchmark}). For comparison we have also plotted the comoving $X^\prime$ equilibrium number density $X^\prime_{eq}$.}}\label{Fig-FOmodnr}
\end{figure}

Then $\Omega_{\rm ann} h^2$ is given by~(\ref{omega2}) with $X^\prime(T_f) = X^\prime_{\rm eq} (T_{FO}^\prime)$:
\begin{equation}\label{nrfo}\Omega_{\rm ann}\,h^2\;[FO^{\rm mod}_{\rm nr}] \approx \left[\frac{8\, \eta}{\sqrt{5\pi}\,L^{3/4}}\right] 
 \left[\frac{g_*(T_{RH})^{1/2}}{g^\prime_{*}(T^\prime_{FO})}\right] \left[\frac{T_{RH}}{M_{X^\prime}}\right]^3
\left[\frac{{x_F^{\prime}}^4}{M_{X^{\prime}}\,M_{\rm pl}\left<\sigma v\right>^{\prime}}\right]\left[\frac{M_{X^{\prime}}}{T_{\rm now}}\right]\;[\Omega_{R}\,h^2]\end{equation} where $x_F^{\prime}$ is the solution of~(\ref{xf}). From~(\ref{xf}), it can be seen that the condition $x_{F}^{\prime} > 1$ is equivalent to $\left<\sigma v \right>^{\prime} > \left<\sigma v \right>^{\prime}_0$ or $M_{X^{\prime}} < M_{0}$ where:
\ba 
\left<\sigma v \right>^{\prime}_0 (M_{X^{\prime}})&\equiv& \left[\frac{2 e \sqrt{10} \pi^3}{3}\right] \left[\frac{g^\prime_*(T^\prime_{FO})}{g^\prime g_*(T_{RH})^{1/2}}\right] \left[\frac{M_{X^\prime}}{M_{\rm pl}\,T_{RH}^2\,\eta}\right]\nonumber\\
M_0\,(\left<\sigma v\right>^{\prime})&\equiv&  \left[\frac{3}{2 e \sqrt{10} \pi^3}\right] \left[\frac{g^\prime g_*(T_{RH})^{1/2}}{g^\prime_*(T^\prime_{FO})}\right] \left[M_{\rm pl} T_{RH}^2 \left<\sigma v \right>^{\prime}\, \eta\right] \label{cond}
\ea In addition, $x_{F}^{\prime}$ must be smaller than $M_{X^{\prime}}/T_D^{\prime}$, which puts an additional constraint on the parameters. Thus the parameter space for viable $FO^{\rm mod}_{\rm nr}$ is rather limited, as we will show in Section~\ref{viable}.

\subsubsection{Non-relativistic and relativistic inverse annihilation\, ($IA_{\rm nr}$\,$\&$ $IA_{\rm r}$)}\label{IA}

Finally, let us consider the situation when one of the conditions in the previous subsection, i.e. $T_{\rm max}^{\prime} > T_{FO}^{\prime} > T_D^{\prime}$ or $M_{X^{\prime}} > T_{FO}^{\prime}$, is \emph{not} satisfied. In this case, $X^\prime$ is instead populated by $R^\prime R^\prime \rightarrow X^\prime X^\prime$ inverse annihilations. This occurs if $X^\prime$ never reaches equilibrium for $T^\prime < T^\prime_{max}$, or if $X^\prime$ decouples from the thermal bath while relativistic. In either case ${X^\prime}^2 \ll {X^\prime_{eq}}^2$ for $T^\prime \lesssim M_{X^\prime}$, allowing us to neglect the ${X^\prime}^2$ term in~(\ref{simplified}). Integrating the first term on the right hand side of~(\ref{simplified}) from $A = A_0 \equiv (8/3)^{2/5}$ to some scale factor $A=A_f$, one gets\footnote{$A_0$ corresponds to the scale factor at which $T = T_{\rm max}$, see~(\ref{tmax}).}: 
\begin{equation} \label{invann1} X^{\prime}(A_f)  \approx c_1^{1/2}\,M_{\rm pl}\left<\sigma v\right>^{\prime} {T_{RH}}^{-5}\,\int_{A_0}^{A_f}\,dA\,\frac{A^{7/2}\,{n^{\prime}_{\rm eq}}^2}{\widetilde{H}}\, . \end{equation}
While $X^\prime$ is relativistic, the integrand of~(\ref{invann1}) grows like $A^{5/4}$ in the modulus dominated phase ($\widetilde{H} \approx {\Phi_{I}}^{1/2}$) and falls like $A^{-3}$ in the radiation dominated phase $(\widetilde{H} \approx \sqrt{R/A})$. Thus if $M_{X^\prime} > T_{D}^\prime$, $X^\prime$ production occurs predominantly when $X^\prime$ first becomes non-relativistic, while if $M_{X^\prime} < T_{D^\prime}$ $X^\prime$ production occurs predominantly at the transition between modulus domination and radiation domination.

In either case the important dynamics for $X^\prime$ production approximately occurs during modulus domination; thus taking $\widetilde{H} \approx {\Phi_I}^{1/2}$ we can use~(\ref{tempa}) to rewrite~(\ref{invann1}) as:\begin{equation}
X^\prime(T_f^\prime) \approx \,\eta^3\left[\frac{192}{(125\pi^7)^{1/2}}\right]\left[\frac{{g_{\star}}^{3/2}(T_{RH})}{{g^{\prime}_{\star}}^3(T_{\star}^{\prime})}\right]\,\left[\frac{{T_{RH}}^7\,M_{\rm pl}\left<\sigma v\right>^{\prime}\,\Phi_I}{M_{X^{\prime}}^{12}}\right]\int_{\frac{M_{X^\prime}}{T^\prime_{\rm max}}}^{\frac{M_{X^\prime}}{T^\prime_{f}}} \,dx^{\prime}\,{x^{\prime}}^{11}\,{n^{\prime}_{\rm eq}}^2\, ,
\label{IA-Xprime}\end{equation} where we have defined $x^\prime \equiv M_{X^\prime}/T^\prime$. Here $T^\prime_*$ is defined as the temperature at which the integrand of $\int dx^\prime {x^\prime}^{11} {n_{eq}^\prime}^2$ is peaked, and $T^\prime_{f}$ is a temperature chosen such that $X^\prime(T^\prime)$ is essentially constant for $T^\prime < T^\prime_{f}$. For relativistic $X^\prime$, the integrand of~(\ref{invann1}) peaks at $T^\prime \approx T^\prime_{D}/1.75$. Thus we will henceforth take $T^\prime_{f} \approx T^\prime_{D}/1.75$, though if $M_{X^\prime} \gg T_{D}^\prime$ the integrand of~(\ref{invann1}),~(\ref{IA-Xprime}) falls rapidly well before $T_{f}^\prime$.

\begin{figure} 
\begin{center}
    \includegraphics[width=0.7\textwidth]{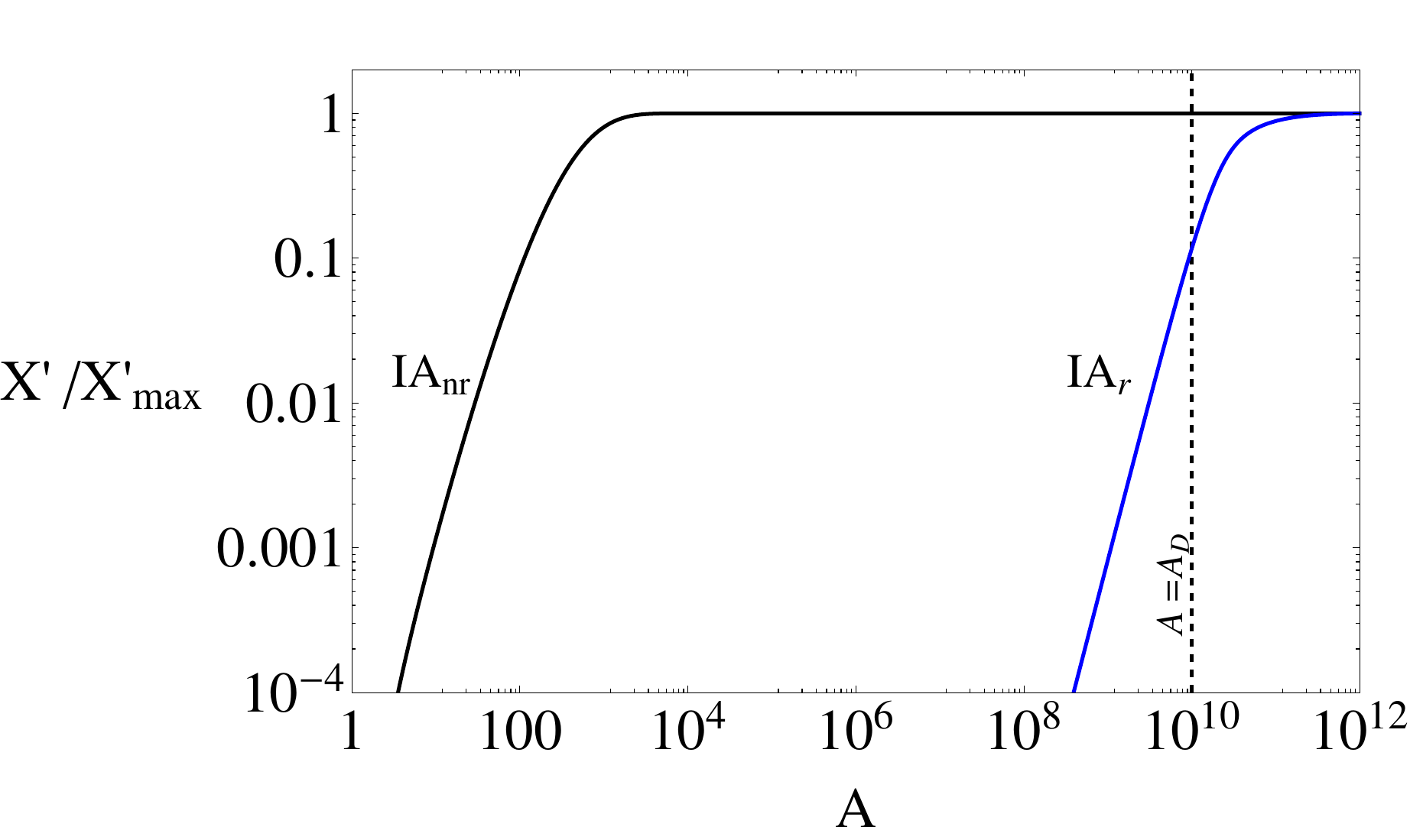}
  \end{center}
\caption{\footnotesize{Plot of the exact solution of the Boltzmann equations for $X'$ (normalized to its maximum value) as a function of the scale factor $A$ corresponding to the $IA_{\rm nr}$ and $IA_{\rm r}$ mechanisms. We have chosen $\left<\sigma v\right>^{\prime} = 10^{-16}\,{\rm GeV^{-2}}, \,M_{X^\prime} = 10$ GeV for $IA_{\rm nr}$, and $\left<\sigma v\right>^{\prime} = 10^{-16}\,{\rm GeV^{-2}}, \,M_{X^\prime} = 10^{-4}$ GeV for $IA_{\rm r}$, with $B_{tot} = 0$ and $H_I = 10^{15} \Gamma_\phi$; all other parameters taken to their benchmark values~(\ref{benchmark}).}}
\label{Fig-IArnr}
\end{figure}

The evaluation of the integral in~(\ref{IA-Xprime}) is different in different regimes. If $M_{X^{\prime}} > T_D^{\prime}$, we can evaluate~(\ref{IA-Xprime}) assuming $X^\prime$ satisfies Maxwell-Boltzmann statistics. The integral in (\ref{IA-Xprime}) can then be expressed as:
\begin{equation}\label{integral}\int_{\frac{M_{X^\prime}}{T^\prime_{\rm max}}}^{\frac{M_{X^\prime}}{T^\prime_{f}}} dx^\prime {x^\prime}^{11} {n^\prime_{\rm eq}}^2 = \frac{{g^\prime}^2M_{X^\prime}^6}{4 \pi^4} \int_{\frac{M_{X^\prime}}{T^\prime_{\rm max}}}^{\frac{M_{X^\prime}}{T^\prime_{f}}} dx^\prime {x^\prime}^9 K_2(x^\prime)^2\, .
\end{equation} 
The function ${x^\prime}^9 K_{2}(x^\prime)^2$ peaks at $x^\prime_{\star} \approx 3.6$, corresponding to $T_*^\prime \approx 0.28\,M_{X^\prime}$. Thus the maximum $X^\prime$ production takes place when $X^\prime$ is non-relativistic, justifying our assumption of Maxwell-Boltzmann statistics. Finally, if $\frac{M_{X^\prime}}{T^\prime_{\rm max}} > x^{\prime}_{\star}$, then~(\ref{integral}) will be exponentially suppressed, in particular by $\exp(- 2 M_{X^\prime}/T^\prime_{\rm max})$. We denote the above mechanism of DM production via non-relativistic inverse annihilations as $IA_{\rm nr}$. We remind the reader that for $M_{X^\prime} > T_{D^\prime}$,~(\ref{IA-Xprime}) is valid if $T_{FO}^\prime > T^\prime_{\rm max}$ or if $T^\prime_{FO} > M_{X^\prime}$ where $T_{FO}^\prime$ is given by~(\ref{xf}); otherwise $\Omega_{\rm ann} h^2$ is determined by non-relativistic freeze-out during modulus domination as described in Section~\ref{nrFO}. 

What happens when $M_{X^\prime} < T_{D}^\prime$? In this case, $X^\prime$ production peaks when $X^\prime$ is relativistic at $T^\prime_* = T_{D}^\prime/1.75$, and Fermi-Dirac or Bose-Einstein statistics must be taken into account. The integral in~(\ref{integral}) can then be expressed as:\begin{equation}
\int_{\frac{M_{X^\prime}}{T^\prime_{\rm max}}}^{\frac{M_{X^\prime}}{T^\prime_{f}}} dx^\prime {x^\prime}^{11} {n^\prime_{\rm eq}}^2 = \frac{\zeta(3)^2 {c_\xi}^2  {M_{X^\prime}}^6}{\pi^4} \int_{\frac{M_{X^\prime}}{T^\prime_{\rm max}}}^{\frac{M_{X^\prime}}{T^\prime_{f}}} dx^\prime {x^\prime}^{5} \approx \frac{1.75^6\,\zeta(3)^2 {c_\xi}^2 {M_{X^\prime}}^{12}}{6 \pi^4 {T_{D}^\prime}^6}\label{integral2}
\end{equation} where again $c_\xi = g^\prime\, (3 g^\prime/4)$ for bosons (fermions). We denote the above mechanism of DM production via relativistic inverse annihilations as $IA_{\rm r}$.  

Given~(\ref{omega2}) and (\ref{IA-Xprime})-(\ref{integral2}), the relic abundance from inverse annihilations can be readily computed:\ba
\label{invannNR} \Omega_{\rm ann}\, h^2\;[IA_{\rm nr}] &\approx&  \left[\frac{48\,{g^{\prime}}^2\,\chi\, \eta^3}{125^{1/2}\pi^{15/2}\,L^{3/4}}\right]\left[\frac{{g_{\star}}^{3/2}(T_{RH})}{{g^{\prime}_{\star}}^3(T_{\star}^{\prime})}\right]\,\left[\left(\frac{T_{RH}}{M_{X^{\prime}}}\right)^7\,M_{\rm pl}M_{X^{\prime}}\left<\sigma v\right>^{\prime}\right]\left[\frac{M_{X^\prime}}{T_{\rm now}}\right]\,[\Omega_{R}\,h^2]  \\
\label{invannR} \Omega_{\rm ann}\,h^2\;[IA_{\rm r}]  &\approx& \left[\frac{32\,{c_{\xi}}^2\, \zeta(3)^2 \,(1.75)^6}{125^{1/2}\pi^{15/2}\,L^{3/4}}\right]\left[\frac{\eta^{3/2}}{{g^{\prime}_{\star}}^{3/2}(T_{D}^{\prime})}\right]\,\left[\left(\frac{T_{RH}}{M_{X^\prime}}\right)M_{\rm pl}M_{X^{\prime}}\left<\sigma v\right>^{\prime}\right]\left[\frac{M_X^\prime}{T_{\rm now}}\right]\,[\Omega_{R}\,h^2]
\ea where $\chi \equiv \int_{\frac{M_{X^\prime}}{T^\prime_{\rm max}}}^{\frac{M_{X^\prime}}{T^\prime_{D}}} dx^\prime {x^\prime}^{9} {K_2(x^{\prime})}^2$ and we have taken $g_*(T_{D}) = g_*(T_{RH})$ and $g^\prime_*(T_{D}^\prime) = g^\prime_*(T_{f}^\prime)$ in~(\ref{invannR}). Of all the production mechanisms we have studied, the only scenario where the $X^\prime$ relic abundance depends on $T^\prime_{max}$ is $IA_{\rm nr}$ in the case where $M_{X^\prime} > T^\prime_{max}$.

Note that we have assumed above that $\left<\sigma v\right>^\prime$ is independent of temperature. For the $QSE_{\rm nr}$, $FO^{\rm rad}_{\rm nr}$, $FO^{\rm mod}_{\rm nr}$ and $IA_{\rm nr}$, the processes which determine the DM relic abundance occur when $X^\prime$ is non-relativistic. Thus for these mechanisms, a temperature-independent $\left<\sigma v\right>^\prime$ is typically a good assumption for s-wave annihilation (p-wave annihilations are considered in Appendix~\ref{tempsigmavd}). However for $IA_{\rm r}$, the relevant process responsible for the DM abundance (inverse annihilation) takes place when $X^\prime$ is relativistic\footnote{The process responsible for DM abundance for $FO^{\rm rad}_{\rm r}$ does take place when $X^{\prime}$ is relativistic, but in this case the DM abundance is independent of $\left<\sigma v\right>^\prime$, see~(\ref{radr}).}. Since $IA_{\rm r}$ requires $\left<\sigma v\right>^\prime < \left<\sigma v\right>^\prime_c$, it is expected that $\left<\sigma v\right>^\prime$ in this case is schematically given by $\left<\sigma v\right>^\prime = \frac{T^n}{\Lambda^{n+2}}$ for some heavy mediator scale $\Lambda$ and positive integer $n$. The temperature-independent $\left<\sigma v\right>^\prime$ case studied here corresponds to $n=0$. Another well motivated case is $n=2$, corresponding to fermionic $X^\prime$ annihilating via a heavy bosonic mediator. We consider this possibility in Appendix~\ref{tempsigmavd}, and show that the $n=2$ case can be recovered from (\ref{invannR}) by making the replacement $\left<\sigma v \right>^\prime \rightarrow 0.17\times {T_{D}^\prime}^2/\Lambda^4 $.

\subsection{Summary of Results}\label{summary}

\begin{table}[h!]
\centering
\begin{tabular}{| c | l | c |}
\hline
& \hspace{0.3cm}DM Production Mechanism & Parametric Region \\
\hline
\multirow{2}{*}{I. Efficient Annihilation at $T_{D}^{\prime}$} & \multirow{2}{*}{A. Non-Relativistic QSE } & \multirow{2}{*}{$M_{X^\prime} > \hat{x}_{F}^\prime T_{D}^\prime$} \\
& & \\
 & \hspace{1cm}[$QSE_{\rm nr}$]  & \\
\cline{2-3}
\multirow{2}{*}{\vspace{10mm}$\left<\sigma v \right>^{\prime} > \left<\sigma v \right>^{\prime}_{c}$} & \multirow{2}{*}{B. FO During Radiation Domination } & \multirow{2}{*}{$M_{X^\prime} < \hat{x}_{F}^\prime T_{D}^\prime$}\\
 &  & \\
  &\hspace{0.7cm} [$FO^{\rm rad}_{\rm nr}\,\&\,FO^{\rm rad}_{\rm r}$] &\\ & & \\
\hline
\multirow{2}{*}{II. Inefficient Annihilation at $T_{D}^{\prime}$} & \multirow{2}{*}{A.} FO During Matter Domination + & $\{T^\prime_{\rm max},M_{X^{\prime}}\} > T_{FO}^\prime > T_{D}^\prime$ \\
 & \hspace{4mm} Production from Modulus Decay &  $\left[\left<\sigma v\right>^{\prime} > \left<\sigma v\right>^{\prime}_0\right]$\\
 & \hspace{0.8cm} [$FO^{\rm mod}_{\rm nr}$] & \\ & & \\
\cline{2-3}
\multirow{2}{*}{\vspace{10mm}$\left<\sigma v \right>^{\prime} < \left<\sigma v \right>^{\prime}_{c}$} & \multirow{2}{*}{B.} Inverse Annihilations($R^\prime R^\prime \rightarrow X^\prime X^\prime$) +
 &  \\
 & \hspace{4mm} Production from Modulus Decay  &  IIA condition not satisfied \\
 & \hspace{0.7cm} [$IA_{\rm nr}\,\&\,IA_{\rm r}$] &$\left[\left<\sigma v\right>^{\prime} < \left<\sigma v\right>^{\prime}_0\right]$ \\ & & \\
\hline
\end{tabular}
\caption{Summary of the different parametric regimes for $\Omega_{DM} h^2$ as discussed in Section~\ref{parametrics}. The quantity $\left<\sigma v \right>^{\prime}_{c}$ is defined in~(\ref{critnr}) and $\left<\sigma v \right>^{\prime}_{0}$ in~(\ref{cond}). The temperatures $T^{\prime}_{\rm max}$ is defined in~(\ref{tmax}), $T^{\prime}_D$ in~(\ref{TD}), and $T^{\prime}_{FO}$ above~(\ref{xf}). \label{tab1}}
\end{table}

In this section, we summarize the results of this section for the benefit of the reader.  There are four qualitatively distinct parametric regimes for $\Omega_{DM} h^2$ in the framework considered. These different regimes are summarized in Table~\ref{tab1}; $\left<\sigma v \right>^\prime$ is defined in (\ref{critnr}), $T_{D}^\prime$ is defined in (\ref{TD}), and $T_{max}$ is defined in (\ref{tmax}). The quanitites $\hat{T}^\prime_{FO}$ and $T^\prime_{FO}$ are respectively the $X^\prime$ freezeout temperatures during radiation domination (\ref{xfrad}) and modulus domination (\ref{xf}). Here we briefly review the parametrics for $\Omega_{DM} h^2$ in these different regimes, and collect the semi-analytic expressions for $\Omega_{DM} h^2$ derived earlier. In the following expressions we will set $g_*(T_{RH}) = g_*(T_D) = 10.75$, which is the SM value for $g_*(T)$ at $T \sim 10$ MeV. We also assume a fermionic DM candidate and set $g^\prime = 2$. Note that the various mechanisms are valid in different parameteric regions; this is reflected in the different fiducial values for $M_{X^\prime}$ and $\left<\sigma v \right>^\prime$ chosen in the expressions below. In Appendix~\ref{accuracy} we compare our approximate expressions with numerical solutions to the Boltzmann equations (\ref{boltzmann}) and find close agreement.

\begin{itemize}

\item {\bf I.A: Non-Relativistic QSE ($QSE_{\rm nr}$)}:\newline
 DM annihilations are large enough to drive $X^\prime$ to its \emph{quasi-static equilibrium} (QSE) value until $T^\prime$ is close to $T_{D}^\prime$, soon after which QSE is lost and the comoving DM abundance becomes constant. The relic abundance in this regime is given by (\ref{omegadm1}):\ba\label{approxnrqse}
\Omega\, h^2\,[QSE_{\rm nr}] &\approx& 5.2 \times \left(1-\eta\right)^{-3/4}\left(\frac{\tilde{A}_c}{3}\right)\left(\frac{M_{X^\prime}}{10\, \mathrm{GeV}}\right) \left(\frac{10 \, \mathrm{MeV}}{T_{RH}}\right)\left(\frac{10^{-8} \, \mathrm{GeV}^{-2}}{\left<\sigma v \right>^\prime} \right) \ea $\tilde{A}_c$ is defined below (\ref{acrit}), and lies in the range : $1 \lesssim \tilde{A}_c \lesssim 5$ for $\left<\sigma v \right>^\prime_{c}  \lesssim \left<\sigma v \right>^\prime \lesssim 10^{5} \left<\sigma v \right>^\prime_{c}  $. {\it $QSE_{\rm nr}$ is the precise generalization of the ``non-thermal WIMP miracle" studied in~\cite{Moroi:1999zb, Acharya:2009zt}, and also captures the sub-dominant logarithmic dependence on $\left<\sigma v \right>^{\prime}$ and $\left<\sigma v\right>^\prime_{c}$ via $\tilde{A}_c$ which was not considered in~\cite{Moroi:1999zb, Acharya:2009zt}}.

\item {\bf I.B: Freeze-out during radiation domination ($FO^{\rm rad}_{\rm nr}\,\&\,FO^{\rm rad}_{\rm r}$)}:\newline
$X^\prime$ tracks its equilibrium value until after $T^\prime \approx T_{D}^\prime$, and freezes-out \emph{after} the modulus decays and the Universe becomes radiation dominated. Both non-relativistic ($FO^{\rm rad}_{\rm nr}$) and relativistic ($FO^{\rm rad}_{\rm r}$) thermal freezeout are possible. $FO^{\rm rad}_{\rm nr}$ is the dark sector analogue of standard WIMP freeze-out during radiation domination, while $FO^{\rm rad}_{\rm r}$ is the dark analogue of neutrino decoupling in the visible sector. This mechanism occurs only for $M_{X^\prime} \lesssim T_{D}^\prime$; see Table \ref{tab1}. The relic abundances are given by~(\ref{radnr}) and~(\ref{radr}): \ba 
\Omega\,h^2\;[FO^{\rm rad}_{\rm nr}] &\approx& 0.13 \times \left(\frac{\eta}{\left(1-\eta\right)^{3} g_*(\hat{T}_{F}) g^\prime_*(\hat{T}_{F}^\prime)}\right)^{1/4}\left(\frac{\hat{x}^\prime_{F}}{17.5}\right)\left(\frac{10^{-8} \, \mathrm{GeV}^{-2}}{\left<\sigma v \right>^\prime}\right) \label{OFOradnr}\\ 
 \Omega\,h^2\;[FO^{\rm rad}_{\rm r}] &\approx& 100 \times \left(\frac{\eta^3}{\left(1-\eta\right)^{3} g_*(\hat{T}_{F}) g^\prime_*(\hat{T}_{F}^\prime)^3}\right)^{1/4}\left(\frac{M_{X^\prime}}{1 \, \mathrm{KeV}}\right)
\ea $\hat{x}_F^\prime$ is defined in~(\ref{xfrad}) and captures the standard logarithmic sensitivity to $\left<\sigma v\right>^{\prime}$ for thermal freezeout.

\item {\bf II.A: Freeze-out during modulus domination and production from modulus decay ($FO^{\rm mod}_{\rm nr}$)}:\newline $X^\prime$ reaches its equilibrium value and then freezes out during the modulus dominated phase. After freeze-out, modulus decay continues to populate $X^\prime$ until  $T \lesssim T_{D}$. As discussed in Section~\ref{inefficient}, non-relativistic freeze-out during modulus domination occurs only if $T_{\rm max}^{\prime} > T_{FO}^{\prime} > T_{D}^\prime$ and $M_{X^{\prime}} > T^{\prime}_{FO}$. This implies $1 < x_F^{\prime} < \frac{M_{X^{\prime}}}{T_D^{\prime}}$, and $\left<\sigma v \right>^{\prime}_{0} < \left<\sigma v \right>^{\prime} < \left<\sigma v \right>^{\prime}_c$ where $\left<\sigma v \right>^\prime_0$ is given in~(\ref{cond}). The relic abundance is given by $\Omega_{DM}\,h^2 = \Omega_{\rm decay}\, h^2 + \Omega_{\rm ann}\,h^2$ where $\Omega_{\rm decay}\, h^2 $ and $\Omega_{\rm ann}\,h^2$ are given respectively by~(\ref{moduliBR}) and~(\ref{nrfo}) :\ba
\Omega_{\rm decay}\,h^2 &\approx& 0.31 \times \left(\frac{B_{tot}}{\left(1-\eta\right)^{3/4}}\right)\left(\frac{M_{X^\prime}}{10 \, \mathrm{MeV}}\right)\left(\frac{T_{RH}}{10 \, \mathrm{MeV}}\right)\left(\frac{50 \, \mathrm{TeV}}{m_{\phi}}\right)\label{omegadecay}\\
\Omega_{\rm ann}\,h^2\;[FO^{\rm mod}_{\rm nr}] &\approx& \left(\frac{1.1 \times 10^{-6}\,\eta}{g^\prime_*(T^\prime_{FO})\left(1-\eta\right)^{3/4}}\right)\left(\frac{x^\prime_{F}}{19}\right)^4 \left(\frac{T_{RH}}{10 \, \mathrm{MeV}}\right)^3\left(\frac{10 \, \mathrm{GeV}}{M_{X^\prime}}\right)^3 \left(\frac{10^{-8} \, \mathrm{GeV}^{-2}}{\left<\sigma v \right>^\prime}\right)
\label{omegafomodnr}
\ea  ${x}_F^\prime$ in~(\ref{omegafomodnr}) is defined in~(\ref{xf}) and is logarithmically sensitive to $\left<\sigma v\right>^{\prime}$.

\item {\bf II.B: Inverse annihilation and production from modulus decay ($IA_{\rm nr}\,\&\,IA_{\rm r}$)}: \newline
$X^\prime$ does not undergo freezeout during modulus domination. DM production takes place predominantly by inverse annihilations as well as production from modulus decay. Specifically, $\Omega_{DM}\,h^2 = \Omega_{\rm decay}\, h^2 + \Omega_{\rm ann}\,h^2$ where $\Omega_{\rm decay}\, h^2$ is given by~(\ref{omegadecay}), while $ \Omega_{\rm ann}\,h^2$ gets contributions from inverse annihilations. There are two different parametrics for $\Omega_{\rm ann}\,h^2$ depending on whether $M_{X^{\prime}} > T^\prime_D$ or vice versa.
\vspace{2mm}
\newline
(i) $M_{X^\prime} > T_{D}^\prime$: The inverse annihilation contribution peaks at $T^\prime_* \approx 0.28 \,M_{X^\prime}$ and:
\ba
\Omega_{\rm ann}\, h^2[IA_{\rm nr}]\approx \left(\frac{ 6.2 \times 10^{-7} \,\eta^3}{(1-\eta)^{3/4} {g^\prime_*(T^\prime_*)}^3} \right) \left(\frac{\chi}{292}\right)\left(\frac{T_{RH}}{10 \, \mathrm{MeV}}\right)^7 \left(\frac{10 \, \mathrm{GeV}}{M_{X^\prime}}\right)^5 \left(\frac{\left<\sigma v \right>^\prime}{10^{-16} \, \mathrm{GeV}^{-2}}\right)
\ea
where $\chi$ is defined below (\ref{invannR}). To a good approximation, $\chi \approx 292$ if $T^\prime_{\rm max} > T^\prime_* > T^\prime_{D}$. On the other hand, if $T^\prime_{*} > T^\prime_{\rm max}$ ($M_{X^\prime}$ is very large), $\chi$ will become suppressed by a factor of $\exp(-2 M_{X^\prime}/T^\prime_{\rm max})$. 
\\ \\
ii) $M_{X^\prime} < T_{D^\prime}$: The inverse annihilation contribution peaks at $T^\prime_* \approx T_{D}^\prime/1.75$, and:\begin{equation} \Omega_{\rm ann}\,h^2\;[IA_{\rm r}]  \approx 95 \times \left(\frac{\eta^{3/2}}{(1-\eta)^{3/4}g^\prime_*(T^\prime_D)^{3/2}}\right) \left(\frac{T_{RH}}{10 \, \mathrm{MeV}}\right) \left(\frac{M_{X^\prime}}{1 \, \mathrm{KeV}}\right) \left(\frac{\left<\sigma v \right>^\prime}{10^{-16} \, \mathrm{GeV}^{-2}}\right) \end{equation}
\end{itemize}

\subsubsection*{Reducing to a Single Sector}

Though the results derived in this Section assume a two-sector cosmology as described in Section \ref{overview}, it is straightforward to reduce these expressions to the single sector case. To see this, we define a temperature $T^{0} \equiv T^0_{\rm \, max}\left(A^{-3/2} - A^{-4}\right)^{-1/4}$ where:\begin{equation}\label{T0}
T^0_{\rm max} \equiv \left(\frac{3}{8}\right)^{2/5} \left(\frac{5}{\pi^3}\right)^{1/8} \left(\frac{g_*(T_{RH})^{1/2}}{g_*(T_{\rm max})}\right)^{1/4} (M_{\rm pl} H_I T_{RH}^2)^{1/4}\, .\end{equation} $T^0$ corresponds to the temperature for a given value of $A$ in single sector cosmologies (see eq. (15) in~\cite{Giudice:2000ex}). Reducing our expressions to the single sector case amounts to replacing both $T$ and $T^\prime$ with $T^0$ in the above expressions for $\Omega_{DM} h^2$. Comparing (\ref{T0}) to (\ref{tempa})-(\ref{TD}), this amounts to making the replacements $(1-\eta) \rightarrow 1$ and $\eta/g^\prime_*(T^\prime) \rightarrow 1/g_*(T)$ in the above expressions. 

\section{Implications for UV-motivated Supersymmetric Theories}\label{viable}

In this section, we examine the implications of the results obtained in Section \ref{solutions} for UV-motivated supersymmetric theories that contain moduli fields, and identify regions of parameter space which yield suitable DM candidates.  As discussed in Section \ref{overview} the DM relic abundance in these models is fixed by the following parameters:\begin{equation}
{T_{RH}, \, m_{\phi} , B_{tot}, \, \eta, \, g_*(T), \, g^\prime_*(T^\prime), M_{X^\prime}, \left<\sigma v \right>^\prime}
\end{equation} To simplify our analysis, we will henceforth assume that $g_*(T)$ and $g^\prime_*(T^\prime)$ are constant, and take $g_*(T) = g^\prime_*(T^\prime) = 10.75$. We also fix $\eta = 0.1$, which is a reasonable value assuming the modulus couplings are not sequestered from the dark sector\footnote{For this value of $\eta$, the latest CMB bound on $N_{eff}$ requires that all dark radiation particles have masses greater than $\sim 1$~eV. Otherwise, $\eta$ must be smaller. The qualitative features of our results will be the same for smaller $\eta$ as well.}. Relaxing these assumptions will change the computed relic abundance as per the formulae in Section \ref{parametrics}, but will not qualitatively effect the results presented here. 

As mentioned in Section \ref{overview}, the parameters $T_{RH}$, $m_{\phi}$ and $B_{tot}$ can be viewed as inputs from the UV theory, and are fixed by the couplings and masses of the moduli fields. For a particular UV framework, these quantities are constrained to lie within a particular range of values. We will focus here on UV completions which contain gravitationally coupled moduli fields while also yielding TeV scale supersymmetry. If the modulus interacts gravitationally, dimensional analysis suggests that $\Gamma_{\phi} = c_{1} {m_{\phi}}^3/M_{pl}^2$, and $T_{RH}$ as defined in (\ref{TRH}) is related to the modulus mass via:\begin{equation}
T_{RH} \approx 14 \, \mathrm{MeV} \times \left(\frac{m_{\phi}}{50 \, \mathrm{TeV}} \right)^{3/2} {c_{1}}^{1/2}\label{TRHapprox}
\end{equation} Thus the BBN bound $T_{RH} \gtrsim $ MeV places a lower bound on $m_{\phi}$ in the tens of TeV range. 

The range of values for $m_{\phi}$ is further restricted by imposing the requirement of TeV scale supersymmetry. We focus here on models in which SUSY breaking is mediated to the visible sector via gravitational interactions; this arises naturally in theories containing moduli. In the minimal case (i.e. no sequestering or large volume suppression of SUSY breaking), the lightest modulus mass is order the gravitino mass $m_{3/2}$, which sets the scale of the SUSY breaking parameters~\cite{Acharya:2012tw, Acharya:2010af, Denef:2004cf, GomezReino:2006dk}. For many such models, the scalar superpartner masses will be comparable to $m_{3/2}$, while the gauginos may be parametrically lighter by roughly a loop factor.  The lightest superpartners in the visible sector will then be gauginos whose masses are suppressed with respect to $m_{3/2}$. This is true for Type II and heterotic models with KKLT-type moduli stabilization, M-theory compactifications with stabilized moduli, and also for spectra with pure anomaly mediation. Thus for these SUSY models, the requirement of TeV scale supersymmetry along with constraints from BBN imply:\begin{equation}\label{approxtrh}
30 \, \mathrm{TeV} \lesssim m_{\phi} \lesssim \mathcal{O}(100) \, \mathrm{TeV}, \hspace{4mm} 5\, \mathrm{MeV} \lesssim T_{RH} \lesssim \mathcal{O}(100) \, \mathrm{MeV},
\end{equation} assuming $c_{1} \sim \mathcal{O}(1)$. This justifies our choice of benchmark parameters in (\ref{benchmark}). If the DM is an MSSM particle there is a tension between (\ref{approxtrh}) and indirect detection constraints, which require $T_{RH} \gtrsim 1$ GeV \cite{Fan:2013faa,Blinov:2014nla}.

The quantity $B_{tot}$ is more difficult to constrain from a theoretical point of view, as it depends on the precise interactions between the modulus and visible/dark sector particles. Nevertheless, if the canonically normalized lightest modulus contains a non-trivial fraction of the modulus that determines the gauge coupling of the visible and/or dark sector, then one expects a contribution to $B_{tot}$ by operators of the form $\int d^2\theta \,\Phi\,W_{\alpha}\,W^{\alpha}$ where $W_{\alpha}$ is the chiral gauge superfield of either the visible or dark sector\footnote{This allows the lightest modulus to decay to visible or dark sector gauginos, which would then cascade decay to the DM $X^\prime$.}. Therefore, in M-theory compactifications \cite{Acharya:2008bk} and also roughly isotropic Type II compactifications, $B_{tot}$ is expected to be $\mathcal{O}(0.1)$. However, in anisotropic compactifications in which the visible and dark sectors are localized at different regions of the internal manifold, it is possible that $B_{tot}$ is suppressed, see \cite{Cicoli:2012aq, Higaki:2012ar} for example. We will consider below a wide range of values for $B_{tot}$ to perform as general an analysis as possible. 

In the following, we fix $T_{RH}$, $m_{\phi}$ and $B_{tot}$ to particular values, and scan over $\left<\sigma v \right>^\prime$ and $M_{X^\prime}$ to give a fairly model-independent characterization of the viable regions of DM parameter space. All other parameters are taken to their benchmark values (\ref{benchmark}). In Figure \ref{Fig5}, we have scanned over the $\left<\sigma v \right>^\prime$,\, $M_{X^\prime}$ parameter space for various values of $B_{tot}$, with $T_{RH} = 10$ MeV, $m_{\phi} = 50$ TeV for the left column and $T_{RH} = 100$ MeV, $m_{\phi} = 150$ TeV for the right column (consistent with the $T_{RH} \propto {m_{\phi}}^{3/2}$ scaling manifest in (\ref{TRHapprox})).

$B_{tot}$ determines both the cross section required for $QSE_{nr}$ (see (\ref{critnr})), and the size of the modulus decay contribution (\ref{moduliBR}) in the inefficient annihilation region. Thus the available parameter regions are quite sensitive to orders of magnitude changes in $B_{tot}$. For $B_{tot} =0.1$, the viable parameter space effectively splits into two regions. In the upper region $\left<\sigma v \right>^\prime \gtrsim 10^{-9}\, \mathrm{GeV}^{-2}$, the relic DM abundance is produced via either $QSE_{nr}$ or $FO^{rad}_{nr}$, while in the lower region $\left<\sigma v \right>^\prime \lesssim 10^{-17}$ and the relic DM abundance is populated via inverse annihilations and/or modulus decay. In the inefficient annihilation regime, most of the parameter space with $M_{X^\prime} > T_{RH}$ results in an overabundance of DM due to the modulus decay contribution for $B_{tot} = 0.1$ (see (\ref{moduliBR})). The value of $M_{X^\prime}$ where the modulus decay contribution (\ref{moduliBR}) saturates $\Omega_{DM} h^2 = 0.12$ scales like ${B_{tot}}^{-1}$; thus for smaller values of $B_{tot}$, much more of the $ M_{X^\prime} > T_{RH}$ parameter space becomes available. Particularly, for $B_{tot} \lesssim 10^{-3}$ both the $FO^{mod}_{nr}$ and $IA_{nr}$ mechanisms can give the correct relic abundance for a significant portion of the parameter space. These mechanisms are absent for $B_{tot} = 0.1$, as the DM masses required would result in too large a contribution from modulus decay.

\begin{figure}
\centering
\includegraphics[scale=0.25]{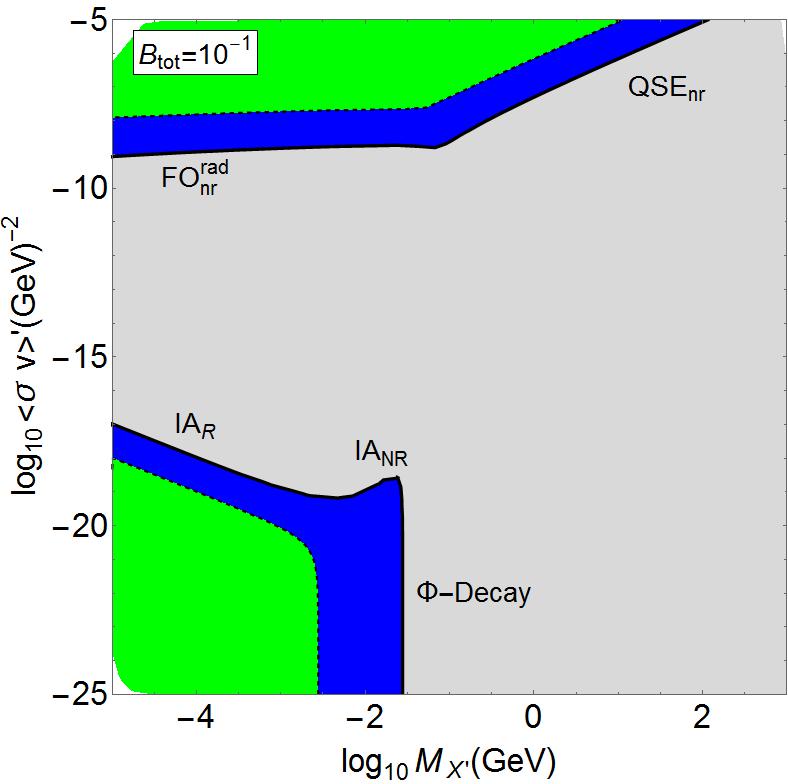}
\includegraphics[scale=0.25]{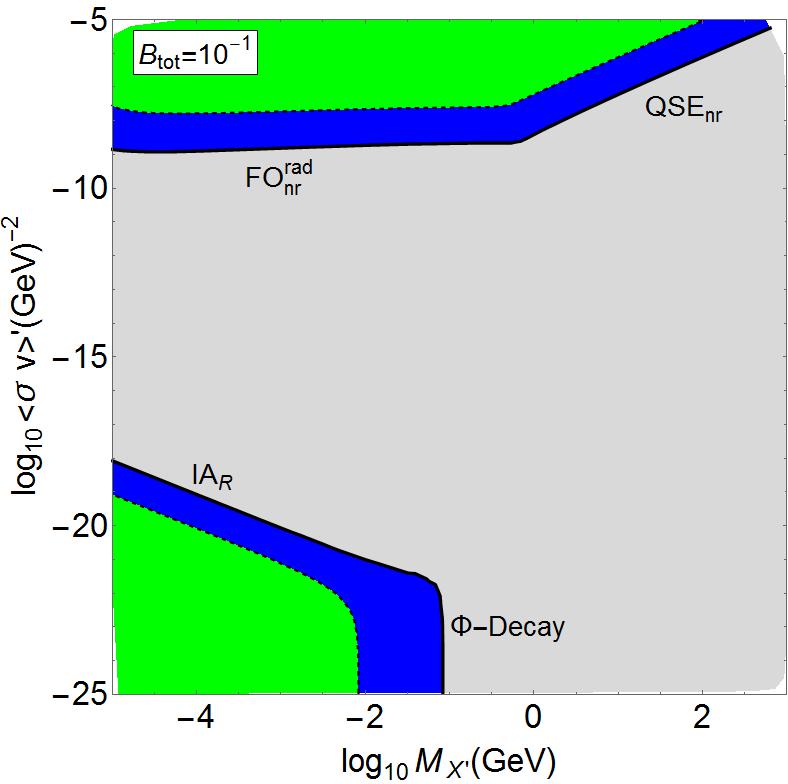}
\includegraphics[scale=0.25]{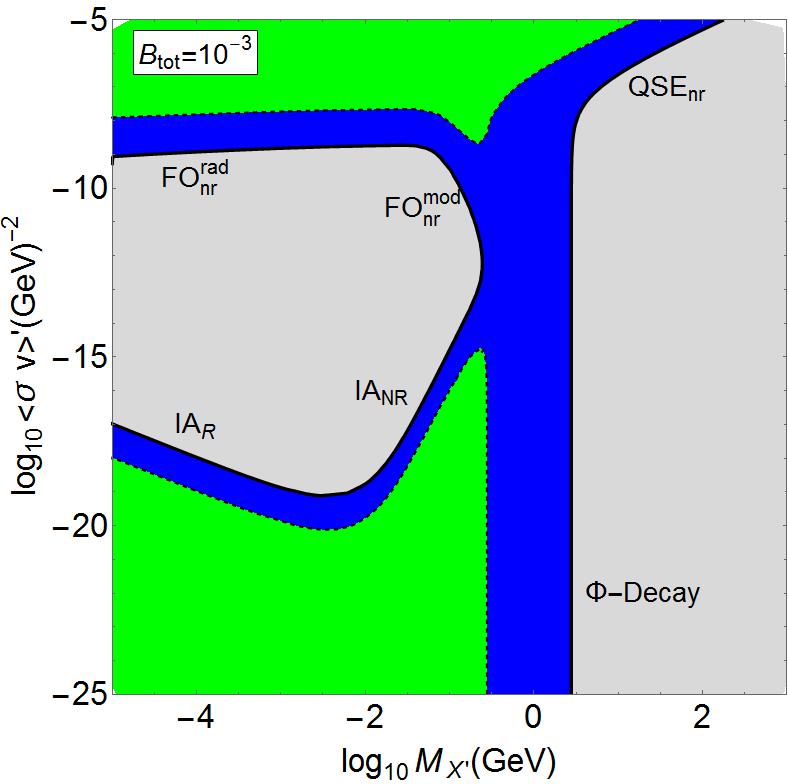}
\includegraphics[scale=0.25]{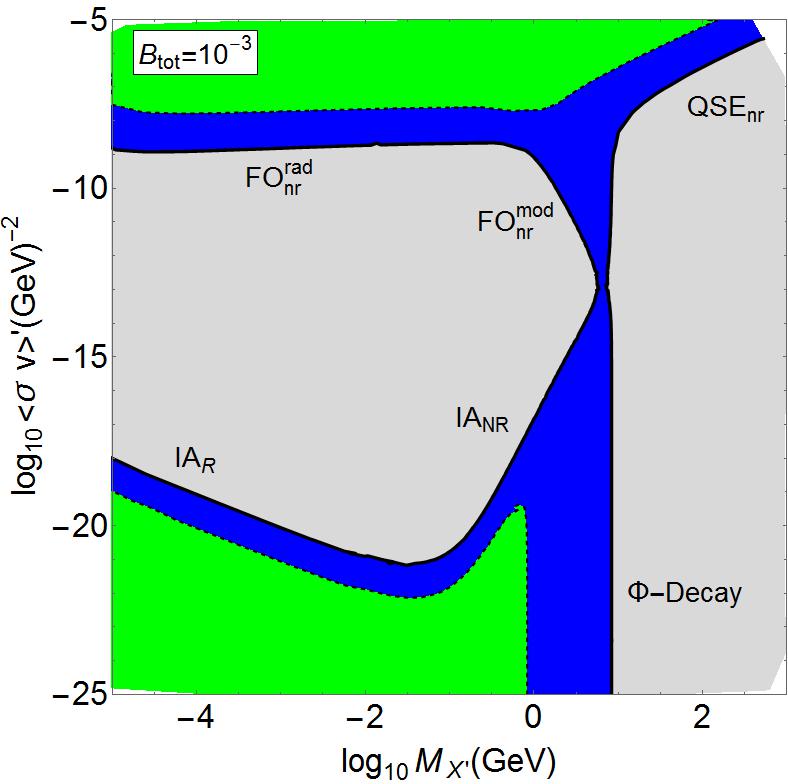}
\includegraphics[scale=0.25]{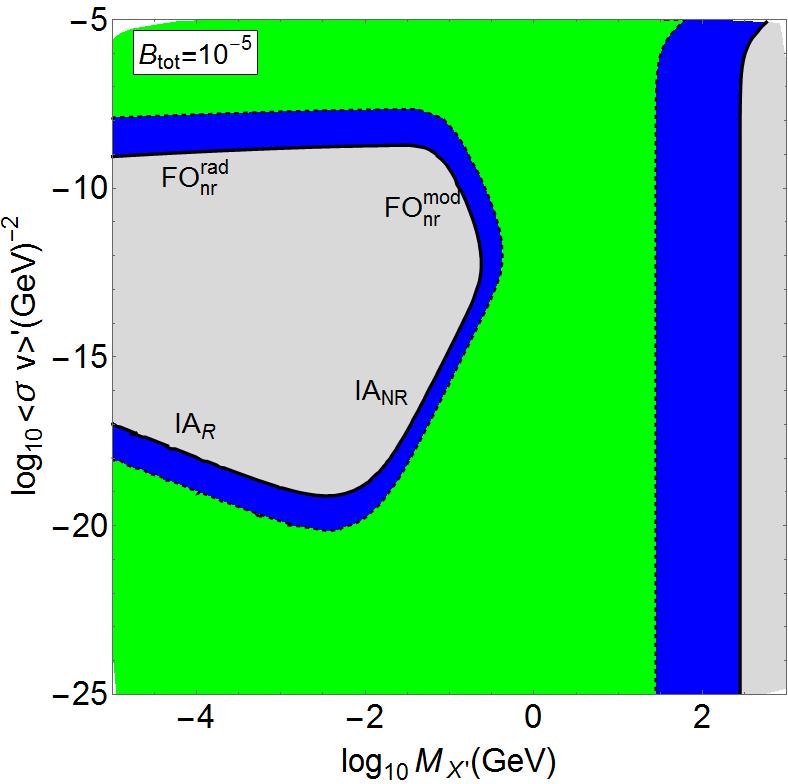}
\includegraphics[scale=0.25]{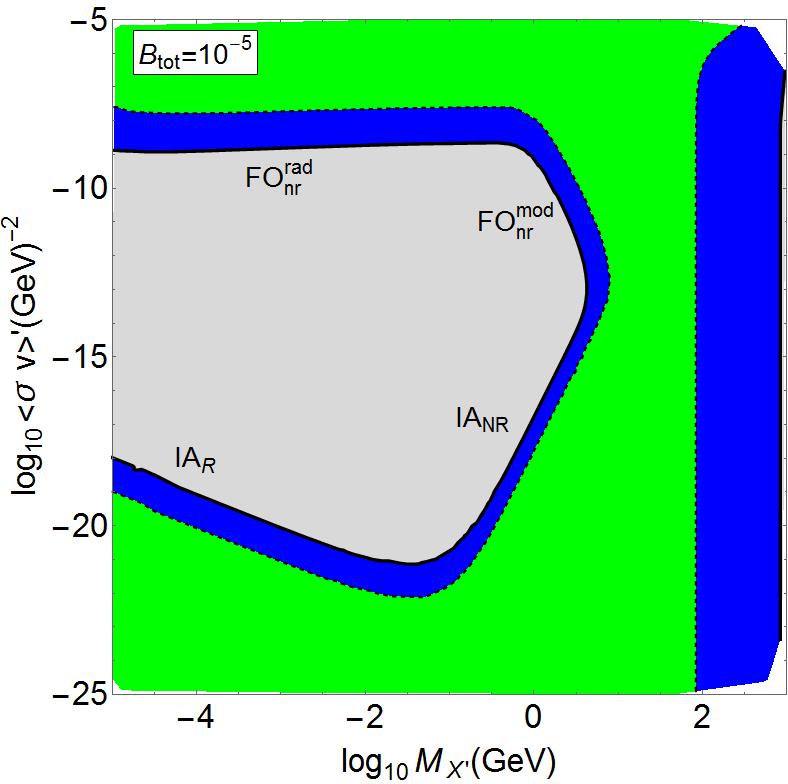}
\caption{\footnotesize{Left column: scan of the $\left<\sigma v \right>^\prime$, $M_{X^\prime}$ parameter space with $T_{RH} = 10$ MeV, $m_{\phi} = 50$ TeV, and various values of $B_{tot}$. Right column: similar plots with $T_{RH} = 100$ MeV, $m_{\phi} = 150$ TeV. All other parameters are fixed to the benchmark values (\ref{benchmark}). Solid (dashed) contours correspond to $\Omega_{DM} h^2 = 0.12 \, (0.012)$. Green, blue and gray regions represent  $\Omega_{DM} h^2 < 0.012$, $0.012 < \Omega_{DM} h^2 < 0.12$ and $ 0.12 < \Omega_{DM} h^2$. For these plots we have taken $H_{I} = 10^{20} \Gamma_{\phi}$, corresponding to $T^\prime_{max} \sim 3$ TeV (\ref{tmax}).} \label{Fig5}}
\end{figure}

\section{Experimental/Observational Consequences}\label{conseq}

In this section, we discuss potential experimental probes of the framework analyzed above. As one can imagine, since our analysis covers a large range of values for the mass and couplings of DM in $M_{X^{\prime}}$ and $\left<\sigma v \right>^\prime$, there are variety of interesting possibilities for observations. A detailed analysis of the various experimental signatures which can arise in this framework is beyond the scope of this paper. Instead, we will limit ourselves here to making some general and preliminary remarks which will be relevant for future studies. 

A nice schematic illustration of the framework studied here is provided in Figure \ref{twosector}. From there we see that there are three different kinds of couplings, denoted as: $\{\lambda_{V-V},\, \lambda_{V-D}$ and $\lambda_{D-D}\}$. Now, the very assumption that the visible and dark sectors are `separate' sectors implies that the ``portal" couplings of type $\lambda_{V-D}$  are parametrically smaller than the $\{\lambda_{V-V}, \lambda_{D-D}\}$ couplings. When this is true, $\left<\sigma v \right>$ dominantly depends on $\lambda_{V-V}$, while $\left<\sigma v \right>^\prime$ depends mostly on $\lambda_{D-D}$. {\it However, within this framework, it is the portal couplings of type $\lambda_{V-D}$ that determine the signals for all ``standard" searches for dark matter, such as direct-detection, indirect-detection, and collider experiments}.  The portal couplings $\lambda_{V-D}$ can cover a huge range. At one extreme, it is possible to have $\lambda_{V-D} \simeq \lambda_{grav}$, the latter corresponding to gravitational strength couplings suppressed by the Planck scale. In this case, the decay width of the LOSP $X$, $\Gamma_X$,  is comparable to that of the modulus $\Gamma_{\phi}$ \footnote{This is because the modulus also couples with gravitational strength to both the visible and dark sectors}. In our work, we have \emph{not} focused on this case for both theoretical and experimental reasons, see Appendix \ref{longlivedX}. At the other extreme, it is possible that $\lambda_{V-D}$ is large enough so that the two sectors are in thermal equilibrium with each other and thus combine to form one sector. As mentioned above, we have also not focused on such a regime. 

Nevertheless, the models considered here can still accommodate a huge range of values $1 \gg \lambda_{V-D} > \lambda_{grav}$ for the portal coupling $\lambda_{V-D}$, which in turn allows for a wide variety of DM signals (or lack thereof) in direct detection, indirect detection and collider searches. For this range of portal couplings, our results from Section \ref{parametrics} show that the relic abundance for the dark matter $X^\prime$ does not depend on the properties of the LOSP $X$ -- $\{M_{X}, \Gamma_X, \left<\sigma v \right>\}$, or equivalently the portal couplings $\lambda_{V-D}$. Thus, in order to characterize the ``standard" DM signals which arise in this framework, one must consider explicit dark sector models in which the size of the portal couplings $\lambda_{V-D}$ are calculable. We save this exercise for future work, except for making some comments about the consequences of a decaying LOSP.  

The LOSP $X$, being a visible sector particle, can be produced at colliders. Since it is unstable, it is possible that the LOSP is charged and/or colored. Prospects for detecting a charged/colored LOSP at the LHC are much better than that for a neutral LOSP, as a charged/colored LOSP will interact with detector materials and slow down considerably relative to a neutral LOSP. Charged/colored LOSP decay widths in the range:  $10^{-13}\,{\rm GeV} \gtrsim \Gamma_X \gtrsim 10^{-31}\,{\rm GeV}$ can be measured in principle. However, subject to model-dependent details, large windows in the above range are now disfavored \cite{Khachatryan:2015jha}. On the other hand, only decay widths larger than around $10^{-17}$ GeV ($\tau_X \lesssim 10^{-9}$ s) can be measured for a neutral LOSP because then a sizable fraction of LOSP particles decay inside the detector.  The LOSP decay width $\Gamma_X$ can be parameterized as: \ba \Gamma_X \sim \frac{\lambda_{V-D}^2}{16\pi}\,M_X.\ea Thus, one requires $\lambda_{V-D} \gtrsim 10^{-9}$ in order for a weak scale neutral LOSP to significantly decay inside the detector so that its decay products could be measured in principle. Otherwise the neutral LOSP is stable for collider purposes, and manifests itself as missing energy. 

We now describe some possible signatures of the framework that do not depend on portal couplings between the visible and dark sectors.

\subsection{Cosmological/Astrophysical Effects}\label{astro}

Here, we comment on astrophysical and cosmological effects arising from two sources -- i) that from DM couplings of type $\lambda_{D-D}$, i.e. from interactions within the dark sector, and ii) from the presence of a modulus-dominated phase in the early Universe. {\it Since these effects are independent of $\lambda_{V-D}$ couplings, the observables which arise are independent of the pattern of ``standard" signals for DM}. As such, they provide additional observables to probe DM and its properties. Some interesting examples of such effects include:
\begin{itemize}
\item Observables sensitive to power spectrum of density fluctuations of dark matter.  
\item Observables sensitive to the morphology of galactic DM halos. 
\end{itemize} 
Understanding these and other observables is becoming increasingly important, both because of the realization that interactions in dark sector can affect these observables, as well as from the fact that the quantity and quality of cosmological and astrophysical data has been getting steadily better. Here, we briefly discuss the following issues:
\begin{itemize}
\item {\it Sensitivity to Modulus-Dominated Era}:\newline
The presence of a modulus-dominated era in the early Universe can have important implications. As pointed out in \cite{Erickcek:2011us, Fan:2014zua}, this can lead to substantial linear growth of sub-horizon DM perturbations during the modulus-dominated era. More precisely, the presence of a (low) reheat temperature sets a new cosmological length scale, $L_{RH} \equiv (a_{RH}\,H_{RH})^{-1}$, the comoving horizon at the time of reheating. Therefore, {\it in the absence of other effects}, DM perturbations on length scales $l < L_{RH}$ grow linearly during the modulus dominated phase, and could have interesting observable effects. However, presence of other relevant scales can affect whether such sub-horizon growth of DM perturbations are observable or not. These scales are described below.

\item {\it Damping of DM Perturbations due to Acoustic Oscillations $\&$ Free-Streaming}: \newline
It is well known that \emph{chemical} equilibrium is in general different from \emph{kinetic} equilibrium. In the context of DM interactions, the former is set by \emph{number-changing} interactions in which DM number is not preserved, while the latter is set by \emph{number-preserving} interactions in which DM number is conserved. For example, within the standard WIMP paradigm, chemical decoupling leaving to thermal freezeout happens much earlier than kinetic decoupling since the interaction rate for the latter is enhanced by the relativistic abundance of light SM species in interactions of the type: $DM + SM \rightarrow DM + SM$. 

There are two important scales related to kinetic decoupling that determine the length scale at which DM perturbations get damped or suppressed:

{\bf i)} Scale arising due to the coupling of DM to the dark radiation fluid (and also to the visible radiation and baryons in general). The effect of coupling of DM to \emph{visible} baryons and radiation is also present for standard WIMPs in general \cite{Loeb:2005pm}, but qualitatively different effects may arise here due to the presence of dark radiation (DR) in addition \cite{Feng:2009mn,Cyr-Racine:2013fsa}. It is expected that the DM-DR interactions will give rise to damped oscillatory features in the DM power spectrum with a characteristic length scale denoted as $L_{d}$, given by. \ba \label{dao} L_{d} = \frac{\eta_{kd}}{x_d}, \ea where $\eta_{kd}$ is the conformal time at kinetic decoupling, and $x_d$ is a numerical factor of ${\cal O}(1)$ 
(we take $x_{d} \approx 7$, see \cite{Loeb:2005pm, Feng:2009mn} for example).\newline
{\bf ii)} Scale arising due to the free-streaming of particles after kinetic decoupling. This length scale is defined as $L_{fs} \equiv \int_{t_*}^{t_0} v/a\,dt $, where $v$ is the average DM velocity, $a$ is the scale factor, $t_0$ is the current age of the Universe, and $t_*$ is a characteristic time which is different for different mechanisms and will be discussed shortly. If the universe is radiation dominated at $t_*$, then $L_{fs}$ is given by (see e.g. \cite{Fan:2014zua}): \ba \label{fsrad} L_{fs}^{rad} \approx \frac{1}{H_0 \sqrt{\Omega_R}}\int_{a_*}^{1} \left[\left(1 + \left(\frac{M_{X^\prime} a}{p_* a_*}\right)^2\right)\left(1 + \frac{a}{a_{eq} }\right)\right]^{-1/2} da \ea where $a_{eq} \approx 2.9 \times 10^{-4}$ and $H_0\approx 1.5 \times 10^{-42}$ GeV. If the universe is modulus dominated at $t_*$, $L_{fs}$ is instead given by:\begin{equation} \label{fsmod} L_{fs}^{mod} \approx \frac{a_{RH}^{1/2}}{H_0\sqrt{\Omega_R}} \int_{a_*}^{a_{RH}} a^{-1/2}\left(1 + \left(\frac{M_{X^\prime} a}{p_* a_*}\right)^2\right)^{-1/2} da + L_{fs}^{rad} \Big(a_* \rightarrow a_{RH}, \, p_* \rightarrow p_{rh}\Big)\end{equation} where we have taken $H = H_{RH} \left(a_{RH}/a\right)^{3/2}$ during modulus domination. Here $a_{RH}$ corresponds to the scale factor at which $H = \Gamma_{\phi}$, normalized such that $a = 1$ today.
\end{itemize}
Both scales above are present in general, and the damping scale is determined by $L_{cut}={\bf max}(L_{d},L_{fs})$. The scale $L_{cut}$ is relevant in determining the mass of the smallest DM proto-halos: $M_{proto} \propto L_{cut}^3$. 

As discussed above, DM perturbations on length scales $l$ such that $L_{cut}< l < L_{RH}$ grow linearly during modulus domination \emph{and} the growth during this era is \emph{not} washed out by free-streaming and/or acoustic damping effects. Thus, these perturbations could have interesting and novel effects. For example, as pointed out in \cite{Erickcek:2011us}, a low reheat temperature of order 10 MeV or so can give rise to an abundance of earth-mass dark matter microhalos in the early Universe containing a significant fraction of dark matter. A possible way to observe these microhalos is via their strong gravitational lensing effects on quasars \cite{Schmidt:1998ah,Chen:2010ae}, or via their impact on pulse arrival times from millisecond pulsars \cite{Baghram:2011is}. Furthermore, if the portal couplings $\lambda_{V-D}$ are large enough, these DM microhalos can annihilate to $\gamma$-rays, thereby acting as $\gamma$-ray point sources and contributing to the $\gamma$-ray background \cite{Oda:2005nv,Scott:2009tu,Afshordi:2009hn}. It is worthwhile to explore these possibilities in more detail.  

In the case where $L_{cut} > L_{RH}$ such that the growth of DM perturbations during modulus domination is washed out, the damping of DM perturbations below the scale $L_{cut}$ can still give rise to observable effects. Notably, ``warm" dark matter with $L_{cut} = L_{fs} \sim 1 - 100$ Kpc can reconcile many of the discrepancies between $\Lambda$CDM cosmology and observations on galactic/sub-galactic scales \cite{deVega:2010yk,Lovell:2011rd,deVega:2013ysa}. If the damping scale becomes too large i.e. $L_{cut} \gtrsim 1$ Mpc, bounds from Lyman-$\alpha$ will start to apply \cite{Viel:2005qj}. 

\subsection{Prospects for the Framework}\label{prospects}

What can be said about the effects mentioned above \emph{vis-a-vis} the framework considered? Qualitatively, there are two different scenarios which are determined by whether or not $T^\prime_{kd}$ is larger than $T_{D}^\prime$. If $T^\prime_{kd} < T_{D}^\prime$, then $X^\prime$ kinetically decouples during radiation domination after the modulus has decayed. Depending on the mass and kinetic decoupling temperature, either $L_{fs}$ or $L_{d}$ will determine the damping scale $L_{cut}$. Alternatively, if $T^\prime_{kd} > T_{D}^\prime$ then the DM kinetically decouples during the modulus dominated phase; $L_{fs}$ will then determine $L_{cut}$ for most of the relevant parameter space. 

In order to discuss observational signatures for the framework considered here, it is pertinent to consider what range of values for $T^\prime_{kd}$ is expected, given the DM production mechanisms discussed in Section \ref{parametrics}. Generically, one expects that crossing symmetry relates the $X^\prime X^\prime \rightarrow R^\prime R^\prime$ annihilation cross section ($\left<\sigma v\right>^\prime$) to the ($X^\prime R^\prime \rightarrow X^\prime R^\prime$) elastic scattering cross section ($\sigma^\prime_{el}$). In the case of fermionic DM annihilating into fermionic $R^\prime$ through a massive bosonic mediator, $\left<\sigma v \right>^\prime \sim ({M_{X^\prime}}^2+ {T^\prime}^2)/\Lambda^4$ and $\sigma^\prime_{el}\sim {T^\prime}^2/\Lambda^4$ where $\Lambda$ is the mediator mass scale (see e.g.\cite{Chen:2001jz,Hofmann:2001bi}). If this is the only $X^\prime -R^\prime$ scattering process, $X^\prime$ kinetically decouples when the scattering rate $\Gamma_{el} \sim \sigma^\prime_{el} n^\prime_{eq} \,\omega$ drops below the Hubble rate, where $\omega = 1 \,\, (T^\prime/M_{X^\prime})$ for relativistic (non-relativistic) $X^\prime$. Taking the benchmark parameters (\ref{benchmark}) for this example, $T^{\prime}_{kd} < T_D^\prime$ implies $\Lambda \lesssim 800$ GeV \,($250 {M_{X^\prime}}^{-1/4}$ GeV) if $X^\prime$ kinetically decouples while relativistic (non-relativistic).

However, in a more realistic model there may be other (e.g. inelastic) processes which also keep $X^\prime$ in kinetic equilibrium; thus the precise relationship between $\left<\sigma v \right>^\prime$ and $T^\prime_{kd}$ is fairly model-dependent. In the following, we will treat $T^\prime_{kd}$ as a free parameter, though it will be useful to keep the above toy example in mind as a benchmark scenario.

{\it \underline{Kinetic Decoupling During Radiation Domination}}: In this case, the DM particle $X^\prime$ is in kinetic equilibrium until $T^{\prime} < T_D^\prime$. The kinetic decoupling temperature will then determine the length scales $L_d$ in (\ref{dao}) and $L_{fs}$ in (\ref{fsrad}). Specifically, $L_{fs}$ is computed using (\ref{fsrad}) with $T^\prime(t_*) = T^\prime_{kd}$ and $p_* = \sqrt{3} T^\prime_{kd} \,\omega^{1/2}$ where $\omega = 1 \, (M_{X^\prime}/T^\prime_{kd})$ for $M_{X^\prime} < T^\prime_{kd}$ ($M_{X^\prime} > T^\prime_{kd}$). From Figure \ref{FS-Eff}, we see that for larger DM masses $10^{-2}\,{\rm GeV} \lesssim M_{X^\prime} \lesssim 10^2$ GeV and smaller kinetic decoupling temperatures $T_{KD}^\prime < 0.1\,T_{D}^\prime$, $L_D$ is larger than $L_{fs}$ and determines $L_{cut}$ and the mass of the smallest proto-halos. All of this parameter space is consistent with the upper bounds arising from the observables studied in \cite{Cyr-Racine:2013fsa}. In the complementary parameter space, $L_{cut}$ is determined by $L_{fs}$. A large region of this parameter space is consistent with the Lyman-$\alpha$ forest upper bound on $L_{fs}$ of about 1 Mpc. Finally, for most of the parameter space $L_{cut}={\bf max}(L_d,L_{fs})$ is greater than $L_{rh}$, implying that growth of DM perturbations in the modulus-dominated era is washed out. Only in a very small region of parameter space with $1 \lesssim M_{X^{\prime}} \lesssim 100$ GeV and $0.1\,T_{D}^\prime \lesssim T_{KD}^\prime \lesssim T_D^\prime$, one has $L_{cut} < L_{RH}$ so that the memory of growth of DM perturbations on length scales $l$ with $L_{cut} < l < L_{RH}$, is retained. This can have interesting implications as mentioned previously.

\begin{figure}[t!]
\centering
\includegraphics[scale=0.4]{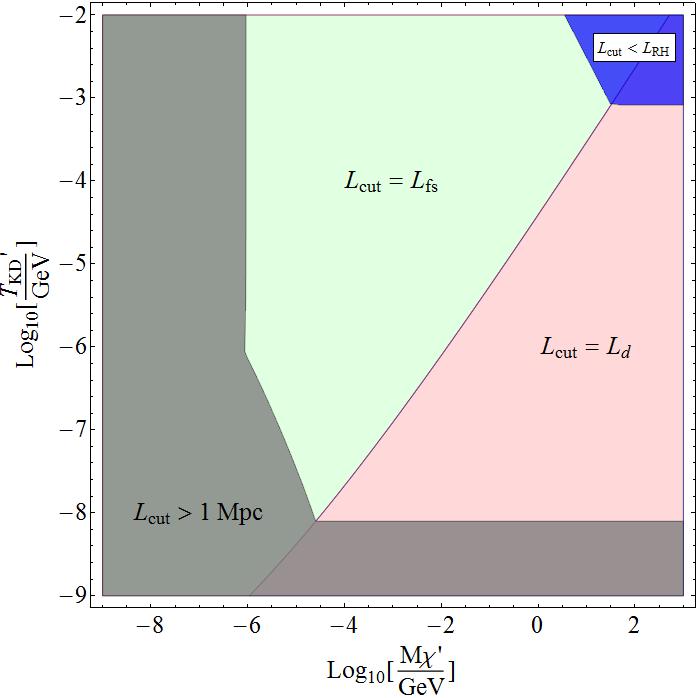}
\caption{\footnotesize{Hierarchies among the cosmological length scales $L_{fs}, L_{d}, L_{RH}$ shown in the $M_{X^\prime}-T_{KD}^\prime$ plane, assuming $T^\prime_{kd} < T^\prime_{D}$. The pink region corresponds to $L_{d} > L_{fs}$, the green region corresponds to $L_{fs} > L_d$, the blue region corresponds to $L_{RH} > L_{cut}$, and the brown region corresponds to $L_{cut} > 1$ Mpc. The other relevant parameters are set to their benchmark values, see (\ref{benchmark}).}\label{FS-Eff}}
\end{figure}

{\it \underline{Kinetic Decoupling During Modulus Domination}}: In this case, $X^\prime$ kinetically decouples before the beginning of radiation domination such that $T^\prime_{kd} > T^\prime_{d}$. If kinetic decoupling occurs \emph{after} $X^\prime$ production, $L_{fs}$ is given by (\ref{fsmod}) with  $T^\prime(t_*) = T^\prime_{kd}$ and $p_* = \sqrt{3} T^\prime_{kd} \,\omega^{1/2}$. Note that this scenario requires $M_{X^\prime} > T^\prime_{D}$, as for $M_{X^\prime} < T^\prime_{D}$ DM production occurs predominantly when $T^\prime \lesssim T^\prime_{D}$ (see Section \ref{summary}). For the allowed parameter regions depicted in Figure \ref{Fig5}, one finds in this case that $ L_{d} < L_{RH} < L_{fs} \ll 1$ Mpc, assuming a single DM particle accounts for all of the dark matter.

If kinetic decoupling occurs \emph{before} $X^\prime$ production, $T^\prime(a_*)$ is the characteristic temperature at which $X^\prime$ production occurs, and $p_*$ depends on the mechanism for $X^\prime$ production. The production mechanisms which allow for $X^\prime$ to be produced out of kinetic equilibrium are (see Section \ref{parametrics}):\begin{itemize}
\item \emph{Inverse annihilation}: As discussed in Section \ref{IA}, DM production from inverse annihilations peaks at $T^\prime_* \approx T_{D}^\prime/1.75$ for $IA_r$ and $T^\prime_* \approx 0.28 M_{X^\prime}$ for $IA_{nr}$. We then take $p_* \approx \sqrt{3} T^\prime_*$ in computing $L_{fs}$.
\item \emph{Production from Modulus Decay}: If the DM abundance comes predominantly from modulus decay (i.e. $\Omega_{DM}h^2 \approx \Omega_{decay} h^2$ (\ref{moduliBR})), then $T^\prime_* \approx T^\prime_D$ and $p_* \approx m_{\phi}/2$ assuming 2-body modulus decays.
\end{itemize} To be precise, if $X^\prime$ kinetically decouples before $X^\prime$ production occurs, one must replace $X^\prime$ in the Boltzmann equations with an integral over the $X^\prime$ phase space distribution function. However for the inverse annihilation and modulus decay production mechanisms, terms involving $X^\prime$ can be neglected in the $X^\prime$ Boltzmann equation; thus the results in Section \ref{inefficient} are still valid despite this departure from kinetic equilibrium\footnote{One additional subtlety is that if $X^\prime$ is out of kinetic equilibrium, we are no longer justified in assuming $E_{X^\prime} \approx \sqrt{3 {T^\prime}^2 + {M_{X^\prime}}^2}$ in (\ref{boltzmann}). However if $B_{X^\prime} \lesssim 0.1$, this subtlety will not significantly effect our results.} Nonetheless, in order to properly compute $p_*$ and $L_{fs}$, a precise knowledge of the DM phase space distribution function at $T^\prime_*$ is required. In lieu of a more precise computation we will use the approximate values for $p_*$ quoted above, with the understanding that our results for $L_{fs}$ are meant to be qualitative.

Figure \ref{FS-Ineff} summarizes the cosmological length scales which can arise in the case where $X^\prime$ is produced out of kinetic equilibrium. Because $X^\prime$ is not coupled to the dark radiation bath when produced, there is no acoustic damping effect to consider; thus $L_{cut} = L_{fs}$. We see from Figure \ref{FS-Ineff} that for the $IA_{r}$ case, most of the parameter space easily avoids Lyman-$\alpha$ constraints. The $IA_{r}$ scenario can also naturally accomodate warm DM candidates, with $L_{fs} \sim 1 - 100$ kpc. Perhaps more interestingly, we see that for a majority of the $IA_{nr}$ parameter space, $L_{fs} < L_{RH}$. Thus, the linear growth of DM perturbations during modulus domination is \emph{not} washed out for a large portion of the $IA_{nr}$ parameter space, leading to potentially interesting effects as discussed above.

Finally, let us comment on the case where relic DM is produced from modulus decay. If DM particles in this scenario are kinetically decoupled at $T^\prime_{D}$, they will be highly boosted when produced from modulus decay. If the modulus decay contribution is the dominant contribution to the overall DM abundance, DM masses within the range $10^{-3}\,{\rm GeV} \lesssim M_{X^\prime} \lesssim {\rm few}$ GeV are at odds with Lyman-$\alpha$ bounds $L_{fs} \lesssim 1$ Mpc; this is evident from Figure \ref{FS-Ineff}. Thus if the relic DM is predominantly produced via modulus decays, Lyman-$\alpha$ constraints require $M_{X^\prime} \gtrsim \mathcal{O}(1)$ GeV; this in turn implies $B_{tot} \lesssim 10^{-3}$ as can be seen from (\ref{moduliBR}). 

\begin{figure}[h!]
\centering
\includegraphics[scale=0.75]{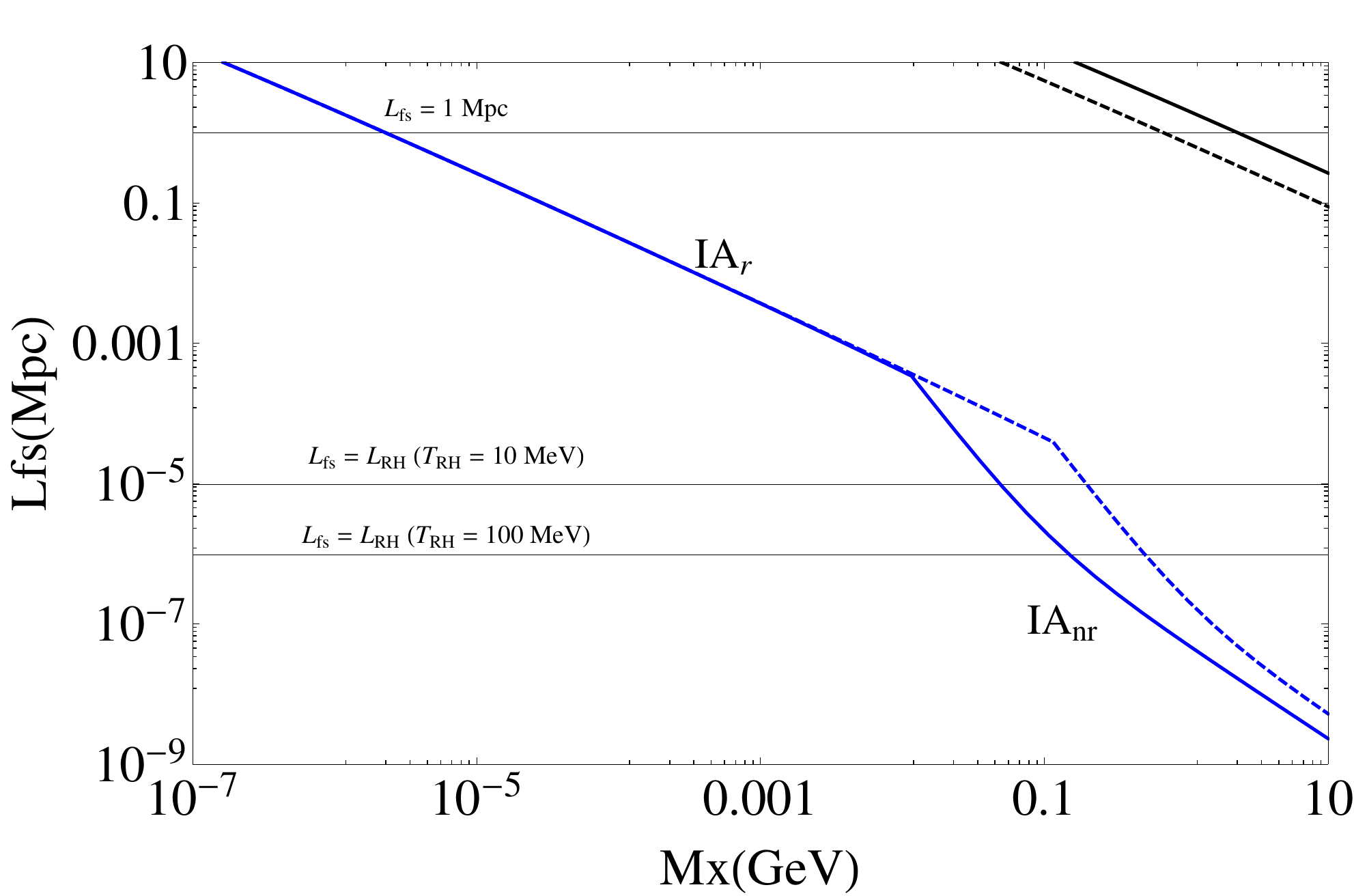}
\caption{\footnotesize{Plot of the free-streaming length $L_{fs}$ in Mpc. The blue lines correspond to DM produced predominantly via inverse annihilations, while the black lines correspond to DM produced predominantly via modulus decay. The solid lines were obtained for $T_{RH} = 10$ MeV, $m_{\phi} = 50$ TeV while the dashed lines were obtained for $T_{RH} = 100$ MeV, $m_{\phi} = 150$ TeV. The other relevant parameters are chosen as in (\ref{benchmark}).}\label{FS-Ineff}}
\end{figure}

To summarize, we find that there are various interesting possibilities for cosmological/astrophysical observables which can probe the framework considered, both in terms of providing constraints on the parameter space as well as by providing insights for potential signals. \emph{In particular, we find that there is sensitivity to the modulus domination era for a large portion of the $IA_{nr}$ parameter space}. This is in contrast to the result obtained in \cite{Fan:2014zua}, primarily because the framework considered here encompasses a wider variety of DM masses and couplings compared to the analysis in \cite{Fan:2014zua,Allahverdi:2013noa}. The results obtained in this section are largely qualitative. It would therefore be interesting to carry out a more detailed and comprehensive analysis of the constraints and potential observations which have been suggested in this section.

\section{Hidden/Extra Sectors in Explicit String Constructions}\label{models}

The system of Boltzmann equations studied in this paper are relevant for cases in which the the dark matter is located in a hidden sector, weakly coupled to the observable sector, and the universe undergoes a substantial period in which the total energy density is controlled by a single modulus field. In addition, we have chosen to study a benchmark case in which the number of relativistic degrees of freedom for the dark sector is similar to that of the Standard Model. How well motivated is this framework?

Certainly, the presence of hidden sectors is generic in string theory, as observed in the earliest days of model-building based on the $E_8 \times E_8$ heterotic string. Furthermore, every four-dimensional effective supergravity theory representing a string compactification has moduli. Barring some remarkable feat of engineering, therefore, a long period of modulus domination in the early universe is essentially guaranteed. These two components of our framework are therefore exceedingly well motivated from the point of view of string theory.

But this paper adds a third component: a portal that connects the hidden sector and observable sector. Our analysis generally assumes that the strength of this presumed coupling is greater than that of gravity-mediated operators suppressed by the Planck scale. Interactions between the dark sector and the observable sector complicate the system of Boltzmann equations, as we have explained at length. Measurements necessarily constrain the observable sector, but the presence of a portal of appreciable strength mean that these measurements also constrain the nature of the dark, hidden sector in a manner that would not exist if the coupling between the sectors was utterly negligible. This constraint is summarized in the very first equation presented in the paper, and it involves the number of relativistic degrees of freedom in the hidden sector. It is thus important to ask what, if any, statements can be made about the nature of hidden sectors in actual string constructions, and what sorts of interactions are observed to exist between these hidden sectors and the Standard Model.

\subsection{Heterotic Orbifolds}

Calculating the massless spectrum in a string compactification is easiest to perform in cases where conformal field theory tools are available. This tends to restrict explicit calculations to orbifolds and their orientifold analogues. These techniques have been used extensively in weakly-coupled heterotic string theory, but also in Type~II string theory. The latter have generally been conducted in the context of Type~IIA theory compactified on orientifolds with intersecting $D_6$-branes.

String phenomenologists tend to be concerned primarily with the observable sector, and ensuring that three generations of fields charged under the Standard Model gauge group emerge from the compactification. The hidden sector is often left undetermined, or only computed years later when they become necessary (for example) in guaranteeing global consistency conditions. Therefore, meaningful examples in the literature are relatively sparse. An important early computation involved $Z_6$ asymmetric orbifolds of heterotic string theory, in which quasi-realistic GUT models were constructed at higher Kac-Moody level~\cite{Kakushadze:1996jm,Kakushadze:1997ne}. A search for constructions which yielded an $E_6$, $SO(10)$, $SU(6)$ or $SU(5)$ GUT model was conducted. Satisfactory cases were required to have three (net) families of GUT representations capable of realizing the SM, and an adjoint Higgs representation for breaking the group to the Standard Model. The three families in this case arise from demanding a $Z_3$ outer-automorphism.

In addition to the GUT gauge group, the hidden sector groups were identified, and the massless matter content for all sectors was computed. For $E_6$ and $SO(10)$ GUTs, the hidden sector consisted of at least one, and sometimes two, $SU(2)$ factors. In one class of constructions bifundamental representations between the SM and a hidden $SU(2)$ were identified, suggesting the possibility of a Higgs-like portal between the sectors. In addition, there were several (non-anomalous) $U(1)$ factors and the SM states typically carried charges under these `hidden' $U(1)$'s. For the $SU(5)$ and $SU(6)$ GUTs, the hidden sector gauge groups can again be $SU(2)$, but occasionally $SU(3)$ and even $SU(4)$ were observed. Again, some twisted sectors tend to contain states that are fundamentals of the $SU(N)$ GUT group, but also a doublet under a hidden $SU(2)$ factor. We note that in most of the cases studied, the $SU(2)$ factor remains weakly coupled to very low energies. This suggests that a relatively sparse dark sector that contains a WIMP, interacting with the observable sector via a Higgs portal or $U(1)$ portal, would not be unusual in this particular class of theories.

An example in which the hidden sector analysis followed that of the observable sector by nearly a decade, is the venerable case of heterotic $E_8\times E_8$ string theory compactified on a $Z_3$ orbifold. The original work of classifying all possible twist embeddings for the $Z_3$ orbifold, with two Wilson lines, that yield the Standard Model gauge group was performed in the late 1980's~\cite{Casas:1989wu}. But this analysis did not fully specify the embedding for the hidden sector $E_8$ factor. This completion was performed in~2000~\cite{Giedt:2000bi}. The classification required three generations of Standard Model matter. In practice it is the demand that three quark doublets be present that puts the most restrictions on the allowed Wilson lines. This, in turn, restricts the allowed hidden sector gauge group to the relatively small list of $SO(10)$, $SU(5)$, $SU(4)$, $SU(3)$ and $SU(2)$, plus additional $U(1)$ factors to fill the rank-eight product group (one of which will be anomalous). In a follow-up study~\cite{Giedt:2001zw}, the field content charged under the full rank-sixteen gauge group was computed, and it was common to find states in twisted sectors which were bifundamental between the SM $SU(2)_L$ and a hidden $SU(2)$. In addition, Standard Model fields and hidden sector fields were generally charged under any number of common $U(1)$ factors.

Years later, a much more exhaustive search was performed, this time in the context of the $Z_{6-II}$ orbifold of heterotic string theory. This so-called `mini-landscape' study~\cite{Lebedev:2006kn} required a gauge embedding of the orbifold action such that an intermediary $SO(10)$ or $E_6$ GUT structure emerges. The authors then scan over all possible completions of the embedding with up to two Wilson lines such that intermediate GUT gauge group ultimately breaks to the SM gauge group. Further requirements included the demand of three (net) generations of SM fields and a hypercharge candidate which is non-anomalous. Unfortunately, only a single explicit example of a hidden sector was given, that of an $SO(8) \times SU(2)$ hidden sector. Again, the massless spectrum contained states which were bifundamental under the $SU(2)_L$ of the Standard Model and a hidden $SU(2)$, though these states were vector-like with respect to the overall gauge group and might therefore receive large masses if an appropriate set of singlet vacuum expectation values were to arise. 
A follow-up study relaxed the restriction to intermediate $SO(10)$ and $E_6$ structures, and allowed up to three Wilson lines~\cite{Lebedev:2008un}. Once again, however, a single example of a hidden sector was illustrated, containing both an $SU(3)$ and an $SU(5)$ factor. Interestingly, this example had states which were fundamentals under the hidden $SU(3)$ and yet were charged under various parts of the Standard Model gauge group.
It is hard to assess just how generic such portals are in this promising class of constructions, as the raw data was not presented in the papers. However, a later paper by Goodsell et al.~\cite{Goodsell:2011wn} investigated this same data set, analyzing the prospects for kinetic mixing between $U(1)_Y$ and hidden sector $U(1)$ factors. The authors found that over 95\% of the models allow for such mixing, and some explicit dark sectors and dark forces were constructed. Their conclusion was that such sectors and portals were indeed `generic' in this class.

It would be interesting to know if such properties were also common in smooth Calabi-Yau compactifications of the heterotic string, away from the orbifold point in moduli space. A systematic investigation of hidden sectors in this context has yet to be performed. An initial foray into the subject was presented in~\cite{Braun:2013wr}, in which a search was conducted for consistent vector bundle configurations of the $E_8 \times E_8$ hidden sector, given a holomorphic observable sector bundle with structure group $SU(4)$, which was shown to allow for the three-generation Standard Model field content~\cite{Braun:2005nv}. 
From this, two examples were presented, with gauge groups $SO(12)$ and $E_7$, and neither case seemed to contain a portal between the two sectors. But we note that this paper was meant as a proof-of-concept, not an exhaustive survey.

\subsection{Local Models in a Global Embedding}

Recent years has seen an explosion in model building in the context of Type~II string theory compactified on orientifolds. For the most part, this model building has occurred in the form of `local' models: the study of $D$-branes at singularities in the Calabi-Yau manifold, in which other effects (including the presence of possible hidden sectors) can be safely neglected. Models are constructed using representative quiver gauge theories, or the related techniques of dimer diagrams/brane tilings.

Unfortunately, these studies tend to focus exclusively on the Standard Model field content. Hidden sectors only emerge when an effort is made to embed these local constructions in a global Calabi-Yau context. To a first approximation, such global embeddings amount to the imposition of certain consistency conditions, including the requirement of $N=1$ supersymmetry, Ramond-Ramond tadpole cancellation and various anomaly constraints. These additional requirements generally necessitate sectors beyond the local Standard Model quiver (i.e. hidden sectors), as the Standard Model theory generally does not satisfy them on its own. In fact, in the context of quiver gauge theories, these consistency conditions will generally {\em require} that matter charged under hidden sector `nodes' are binfundamental with the nodes of the SM gauge group. 

An early example involved Type~IIA orientifolds on $T^6/\mathbb{Z}_2 \times \mathbb{Z}_2$ with intersectiong $D_6$-branes~\cite{Cvetic:2004ui}. These models achieve the three-generation Standard Model via the Pati-Salam gauge group $SU(4) \times SU(2) \times SU(2)$. 
A scan was performed over all possible brane configurations and wrapping numbers consistent with the Standard Model field content (via the Pati-Salam symmetry) and global consistency conditions. This yielded explicit hidden sectors which could then be classified. Typical hidden sectors involved the $USp(4)$ and $USp(2)$ symplectic groups. In some cases, the field content charged under these gauge groups allowed for confinement of the $USp(n)$ gauge group, and the authors speculate as to the appearance of various `mesonic' and `baryonic' bound states. It is noteworthy that such composites would generally carry charges under the various residual $U(1)$ symmetries, including that of the Standard Model. Typically, the number of such objects in the massless spectrum was of order ten, consistent with the number of degrees of freedom in the Standard Model below the QCD confinement scale.

A more expansive survey was conducted some time later~\cite{Cvetic:2012kj}. In this case the 
survey began with the original three-node `Madrid' quiver~\cite{Ibanez:2001nd} and all its three-node generalizations. These quivers represent gauge theories which contain the field content of the Standard Model. From this, additional nodes (i.e. gauge groups) were added to the quiver until all global embedding conditions were satisfied. All of the quiver extensions considered in this paper had an anomalous $U(1)$ factor, under which the newly introduced `hidden' states are chiral, not vector-like. As a result, there is a mixed anomaly between this $U(1)$ factor and any other $U(1)$ under which these states are charged. In particular, when the hidden sector states carry hypercharge, a mixed anomaly between hypercharge and the anomalous $U(1)$ provides a portal between the two sectors, with a potentially light mediating $Z'$ boson. For the phenomenology, and some toy models, see~\cite{Feng:2014cla}. 

In Type~IIB string theory, much of the recent work has focused on singularities of toric del Pezzo surfaces (dP$_n$ surfaces). The advantage here is that toric surfaces afford a certain `modularity' in constructing models, in which one can work in a bottom-up approach, beginning with various phenomenological demands~\cite{Balasubramanian:2009tv}. Another benefit is access to the large Kreuzer-Skarke database of reflexive polytopes~\cite{Kreuzer:2000xy}, which generate these toric ambient spaces, and the Calabi-Yau manifold realized as a hypersurface within these ambient spaces~\cite{Altman:2014bfa}.

Local models with promising phenomenological features were constructed in this context in recent years~\cite{Krippendorf:2010hj,Dolan:2011qu}. While these early efforts concentrated almost exclusively on the observable (Standard Model) sector, some attempts at embedding these into a global Calabi-Yau context have been made, by enforcing consistency conditions such as the Ramond-Ramond tadpole conditions and vanishing of K-theory torsion charges~\cite{Cicoli:2012vw,Cicoli:2013mpa}. In these papers, some rudimentary hidden sectors were constructed. The authors chose flux parameters in such a way as to avoid chiral matter in the hidden sector, resulting in a pure $SO(8)$ or $SU(4)$ gauge theory with no portal to the observable sector. However, this was again a proof-of-principle and not an exhaustive scan over all possible hidden sector configurations.

To our knowledge, no such survey has been conducted within the Type~IIB context in analogy with the above-mentioned work in Type~IIA. However, some interesting examples of non-trivial hidden, or `dark' sectors, were constructed using the `toric Lego' approach of~\cite{Balasubramanian:2009tv}. One such example involved the construction of a ``dark sector'' which mimics the MSSM (visible) sector. This was a toy model designed to exhibit the power of the modular approach. The model was based on two dP$_0$ singularities and a dP$_1$ singularity -- the former pair for the visible and dark sectors, the last for the SUSY breaking sector. The dark and visible sectors were patterned on the phenomenological model of~\cite{ArkaniHamed:2008qp}, in which kinetic mixing between $U(1)$ factors in the observable and dark sectors provide the portal. The global embedding was identified some time later~\cite{Balasubramanian:2012wd}, by identifying those reflexive polytopes from the Kreuzer-Skarke database with the appropriate singularity structure in one of their two-dimensional faces to give rise to this trio of sectors. Remarkably, nearly 300,000 such polytopes were shown to exist, implying at least as many (and perhaps many more) Calabi-Yau manifolds which would generate this model upon compactification.

Analysis of hidden sectors in {\it bona fide} string constructions -- at the level needed to describe early universe dynamics -- is still in its earliest stages, lagging the construction of viable observable sectors in many respects. The areas that have been investigated were those that were identified as being phenomenologically interesting from the point of view of observable sector physics, and may not be a representative sample of heterotic or Type~II string theory, let alone the entire string theory landscape. Nevertheless, the basic elements that are needed for our cosmological framework are often present.

\section{Summary and Future Directions}\label{conclude}
 
In this work, we have provided a general classification of dark matter models in a Universe which undergoes a  phase of pressure-less matter (modulus) domination. Such non-thermal cosmological histories are predicted in a wide class of UV completions to the Standard Model (e.g. compactified string theories), and are also phenomenologically viable provided that the matter dominated phase ends before BBN. Our analysis generalizes previous works by going far beyond the standard WIMP paradigm. In particular: 
\begin{itemize}

 \item We consider DM masses and annihilation cross sections which span several orders of magnitude above and below the electroweak scale.

 \item We allow the possibility that DM in thermal equilibrium with a `dark sector', whose temperature need not be the same as that of the visible sector. 
 \end{itemize}
Upon analyzing the relevant Boltzmann equations, we classify the mechanisms by which relic DM can be produced. We find four distinct mechanisms ($QSE_{{\rm nr}}, FO^{{\rm mod}}_{{\rm nr}}, IA_{\{{\rm r, nr}\}}$ and $FO^{{\rm rad}}_{\{{\rm r, nr}\}}$), each of which have different parametrics for $\Omega_{DM} h^2$. {\it The first three mechanisms are different from standard thermal freeze-out}. We derive semi-analytic approximations for these various production mechanisms, and discuss their regimes of validity. For the convenience of the reader, these results are summarized in Section~\ref{summary}.

Our results have interesting implications for supersymmetric theories containing moduli fields. As discussed in Sections \ref{overview} and \ref{solutions}, $\Omega_{DM} h^2$ does not depend on the masses or couplings of the (unstable) lightest visible sector superpartner (LOSP), provided the LOSP decays before the end of modulus domination\footnote{The contrary case is briefly considered in Appendix \ref{longlivedX}.}. Once the modulus mass and couplings are fixed and the dark relativistic degrees of freedom $g^\prime_*(T^\prime)$ are specified, $\Omega_{DM} h^2$ depends  only on $M_{X^\prime}$ and $\left<\sigma v \right>^\prime$. In Section \ref{viable}, we fixed the modulus mass and couplings by considering models with gravity mediated SUSY breaking in which $m_{\phi}$ is of order the gravitino mass. We mapped out the parameter space of these models by scanning over $M_{X^\prime}$, $\left<\sigma v \right>^\prime$ for various values of $B_{tot}$, see Figure \ref{Fig5}. Here $B_{tot}$ is the branching ratio of the modulus decay into DM, including contributions from intermediate states. For $B_{tot} \sim \mathcal{O}(0.1)$, the viable DM parameter space splits into two seperate regions: large annihilation cross section $\left<\sigma v \right>^\prime \gtrsim 10^{-9}$ GeV$^{-2}$, or small annihilation cross section $\left<\sigma v \right>^\prime \lesssim 10^{-17}$ GeV$^{-2}$. Intermediate values of $\left<\sigma v \right>^\prime$ result in DM overproduction. Moreover in the $\left<\sigma v \right>^\prime \lesssim 10^{-17}$ GeV$^{-2}$ region, the DM mass must be $\lesssim 100$ MeV to avoid being overproduced by moduli decay. If however the modulus branching ratio to DM is suppressed i.e. $B_{tot} \ll 1$, much more of the DM parameter space becomes available. These features can easily be inferred from Figure \ref{Fig5}.

We have also briefly discussed potential experimental signatures for the theoretical framework considered here. Since $\Omega_{DM} h^2$ is insensitive to the portal couplings between the visible and dark sectors for the models considered, the ``standard" DM signals in direct detection, indirect detection and collider experiments, which crucially depend on portal couplings between the visible and dark sectors, can cover a wide range of possibilities are rather model-dependent. On the other hand, observables which involve couplings within the dark sector yield more robust predictions, as these couplings are correlated with the DM relic abundance. One such set of observables involves the power spectrum of DM density perturbations. If the DM kinetically decouples during the radiation dominated era after BBN, the sensitivity of DM density perturbations to the modulus dominated phase is maintained only for a very small region of parameter space, as shown in Figure \ref{FS-Eff}. On the other hand, when DM kinetically decouples during modulus domination, the power spectrum of DM density perturbations depends on the mechanism by which relic DM is produced:\begin{itemize}

\item If DM is produced by annihilation of thermal bath particles while the DM is non-relativistic (we call this case $IA_{\rm nr}$, see Section~\ref{summary}), the free-streaming length is smaller than the comoving horizon at $T_{RH}$. The linear growth of DM density perturbations during modulus domination is \emph{not} washed out, leading to potentially interesting astrophysical signatures as discussed in \cite{Erickcek:2011us}.

\item If DM is produced by annihilation of thermal bath particles while the DM is relativistic (we call this case $IA_{\rm r}$, see Section~\ref{summary}), the free-streaming length is larger than the comoving horizon at $T_{RH}$. Even though the growth of DM perturbations during modulus domination is erased, a large region of parameter space yields $L_{\rm fs} \sim 1 - 100$ Kpc which leads to signatures similar to warm DM. 

\item If DM is dominantly produced by modulus decay, then the DM has large free streaming lengths $L_{\rm fs} \gtrsim 1$ Mpc, which is in tension with constraints on warm dark matter from Lyman-$\alpha$ measurements.
\end{itemize} 

There are many opportunities for future research. From the point of view of the Boltzmann equations, including $n \rightarrow 2$ annihilation processes where $n \ge 3$ would be worth understanding in this framework (this would be the non-thermal analog of \cite{Hochberg:2014dra}). From the point of view of model-building, it would be worthwhile to study explicit models of DM candidates and portal interactions within the general framework so that detailed predictions for ``standard" DM signals (e.g. direct and indirect detection) could be made. From a string theory perspective, although there is some existing work on dark sectors and portal interactions as described in Section \ref{models}, clearly much more needs to be done. Finally, our discussion in Section \ref{conseq} of the astrophysical/cosmological effects of DM interactions within its own sector has been largely qualitative. A more precise analysis would involve solving for the DM phase space distribution at kinetic decoupling in order to determine the appropriate transfer function relevant for the power spectrum of DM density fluctuations. We hope that future studies in these directions will help shed important light on the nature of dark matter.

\section*{Acknowledgments}

We would like to sincerely thank Bobby Acharya for numerous enlightening discussions. The work of GK and BZ is supported by DoE grant DE-FG-02-95ER40899 and by the Michigan Center for Theoretical Physics (MCTP). The work of PK is supported by DoE grant DE-FG-02-92ER40704, while that of BDN is supported by the National Science Foundation under grant PHY-0757959. BDN and PK would like to thank the MCTP for hospitality where part of the work was completed. 

\begin{appendices}

\section{Justifying Approximations for $R$ and $R^\prime$}\label{justifyingapprox}

In Section~\ref{approx}, analytic approximations for $R$ and $R^\prime$ were obtained assuming that all other terms aside from the modulus decay term can be neglected in $dR^\prime/dA$ and $dR/dA$ if $M_{X^\prime}, M_{X} \ll m_{\phi}$. In this appendix, we will justify this approximation. Note from~(\ref{boltzmann}) that the modulus decay terms in  $dR^\prime/dA$ and $dR/dA$ grow like $A^{3/2}$ during the modulus domination phase, and peak when $T \sim T_{D}$. Thus in determining whether or not certain terms in $dR^\prime/dA$ and $dR/dA$ are negligible compared to the modulus decay term, it is sufficient to focus on the Boltzmann equations at temperatures near $T_{D}$.

First, consider the $X \rightarrow X^\prime + ...$ decay term in $dR^\prime/dA$ and $dR/dA$. At $T \gtrsim T_{D}$, $X$ has already reached QSE, assuming $\left<\Gamma_{X}\right> > \Gamma_{\phi}$. Taking $X = X_{\rm QSE}$ with $b \approx 1$, the $X^\prime \rightarrow X + ...$ decay terms are given by:\begin{equation}
\widetilde{H}\frac{d R}{dA } =B_{X} B_{X \rightarrow X^\prime R}  \,c_{\rho}^{1/2}\left( \frac{E_{X} - E_{X^\prime}}{m_{\phi}}\right) A^{3/2} \Phi + ... \hspace{4mm} \widetilde{H} \frac{d R^\prime}{dA } =B_{X} B_{X \rightarrow X^\prime R^\prime}  \,c_{\rho}^{1/2}\left( \frac{E_{X} - E_{X^\prime}}{m_{\phi}}\right) A^{3/2} \Phi + ...
\end{equation} Here $B_{X \rightarrow X^\prime R} $ and $B_{X \rightarrow X^\prime R^\prime} $ are the branching fractions of $X$ into $X^\prime R$ and $X^\prime R^\prime$. Thus we see that the $X \rightarrow X^\prime + ...$ decay terms are suppressed with respect to the modulus decay term by a factor of $(E_{X} - E_{X^\prime})/m_{\phi}$; a similar conclusion holds if $X$ does not decay to $X^\prime$. Next, consider the annihilation terms. For the $\left<\sigma v \right>$ term in $d R/ dA$, $X_{\rm eq} \approx 0$ and $X \approx X_{\rm QSE}$ for $T \sim T_{D}$. Thus for temperature-independent $\left<\sigma v \right>$, the annihilation term in $dR/dA$ falls like $A^{-3/2}$ for $T \gtrsim T_{D}$, and will be numerically insignificant at $T_D$ due to suppression by negative powers of the scale factor. 

The argument for the $\left<\sigma v \right>^\prime$ term in $dR^\prime/dA$ is less straightforward. First, consider the case where $\left<\sigma v \right>^\prime > \left<\sigma v \right>^\prime_c$ such that $X^\prime$ reaches QSE at $T^\prime \sim T_{D}^\prime$. If $X^\prime_{\rm QSE} \approx X^\prime_{\rm eq}$, the annihilation term vanishes and is trivially negligible. If instead $X^\prime_{\rm eq}$ is negligible in $X^\prime_{\rm QSE}$ (\ref{xpqse}) at $T^\prime \sim T_{D}^\prime$ (as is the case for $QSE_{\rm nr}$), we can take $X^\prime \approx X^\prime_{QSE}$ and write the $\left<\sigma v \right>^\prime$ term as:\begin{equation}
\widetilde{H} \frac{d R^\prime}{dA } = c_{\rho}^{1/2} B_{\rm tot} \left(\frac{2 E_{X^\prime}}{m_{\phi}}\right) A^{3/2} \Phi+ ...
\end{equation} which is suppressed with respect to the modulus decay term by a factor of $E_{X^\prime}/ m_{\phi}$. Now consider the case where $\left<\sigma v \right>^\prime < \left<\sigma v \right>^\prime_c$ such that $X^\prime$ is not in QSE at $T_{D}^\prime$. As discussed in Section~\ref{inefficient}, we can write $X^\prime$ at $T^\prime \gtrsim T_{D}^\prime$ as $X^\prime = X^\prime_{\rm mod} + X^\prime_{\rm ann}$, where $X^\prime_{\rm mod}$ comes from integrating the modulus decay term:
\begin{equation}X^\prime_{\rm mod} = \frac{2}{3} {c_{\rho}}^{1/2} \frac{T_{RH} B_{\rm tot}}{m_{\phi}} A^{3/2} \Phi^{1/2}\end{equation} and $X^\prime_{\rm ann}$ is determined by the $\left<\sigma v \right>^\prime$ term. In the case where $M_{X^\prime} > T_{D^\prime}$, $X^\prime_{\rm ann}$ is negligible compared to $X^\prime_{\rm mod}$ unless $B_{\rm tot} \ll 1$ (see Sections~\ref{nrFO} and~\ref{IA}). Taking $X^\prime \approx X^\prime_{\rm mod}$, the $\left<\sigma v \right>^\prime$ term in $dR^\prime/dA$ can be written as:
\begin{equation}
\widetilde{H}\frac{d R^\prime}{d A} = \frac{4}{9} {c_{\rho}}^{1/2} \left(2 \frac{B_{\rm tot} E_{X^\prime}}{m_{\phi}}\right) \frac{\left<\sigma v \right>^\prime}{\left<\sigma v \right>^\prime_c} A^{3/2} \Phi+ ...
\end{equation} Thus in the case where $\left<\sigma v \right>^\prime < \left<\sigma v \right>^\prime_c$ and $M_{X^\prime} > T_{D}^\prime$, the $\left<\sigma v \right>^\prime$ term in $dR^\prime/dA$ is suppressed by at least a factor of $E_{X^\prime}/m_{\phi}$ with respect to the modulus decay term.

Finally, consider the case where $\left<\sigma v \right>^\prime < \left<\sigma v \right>^\prime_c$ and $M_{X^\prime} < T_{D}^\prime$, corresponding to the $I A_{\rm r}$ scenario (see Section~\ref{IA}). In this case $X^\prime \ll X^\prime_{eq}$ and we can write the $\left<\sigma v \right>^\prime$ annihilation term as:
\begin{equation}\label{drprimeda}
\widetilde{H} \frac{d R^\prime}{d A} \approx    c_{1}^{1/2}\,M_{\rm pl}\, \eta^2\, \frac{48\, g_*(T_{RH}) \,{c_\xi}^2 \,\zeta(3)^2 E_{X^\prime} \left<\sigma v \right>^\prime T_{RH}^2}{5 \pi^6 g^\prime_*(T^\prime)^2\,{T^\prime}^2} \Phi_I A^{3/2} + ...
\end{equation} where we have used (\ref{tempa})-(\ref{Tratio}) to relate $A$ and $T^\prime$. Evaluating~(\ref{drprimeda}) at $T^\prime = T_{D}^\prime$, we obtain:\begin{align}
\widetilde{H} \frac{d R^\prime}{d A}\Big|_{T^\prime = T_{D}^\prime} &  \approx c_{1}^{1/2} \,M_{\rm pl} \,T_{RH}\,  \eta \left(\frac{\eta \,g_*(T_{RH})}{g^\prime_*(T_{D}^\prime)}\right)^{3/4}\, \left(\frac{48\sqrt{3}\, {c_\xi}^2  \zeta(3)^2 \left<\sigma v \right>^\prime}{5 \pi^6 \,g^\prime_*(T^\prime)}\right) \Phi_I A^{3/2} + ...
\\& \approx c_{\rho}^{1/2} \,\eta \left(\frac{0.16\, c_{\xi} \,\zeta(3)}{g^\prime_*(T_D^\prime)}\right) \left(\frac{\left<\sigma v \right>^\prime}{\left<\sigma v \right>^\prime_c}\right)  \Phi_I A^{3/2}+ ...
\end{align} Thus the $\left<\sigma v \right>^\prime$ term in $dR^\prime/dA$ is suppressed with respect to the modulus decay term by a factor of the order of $0.1 \left<\sigma v \right>^\prime/ \left<\sigma v \right>^\prime_c$.

To summarize, the above arguments show that the approximations made in solving the equations for $R$ and $R^\prime$ in solving~(\ref{boltzmann}) are justified, as can also be confirmed by the agreement of the approximate and exact solutions in Appendix~\ref{accuracy}.

\section{A Very Long-lived $X$ Particle ($\Gamma_{X} \lesssim \Gamma_{\phi}$)}\label{longlivedX}

For most of this work, we have assumed $\Gamma_{X} > {\cal O}(1)\,\Gamma_{\phi}$ such that $X$ decays are efficient before the end of modulus domination. This assumption is well-motivated from both theoretical and phenomenological points of view. To see this, note that the modulus decays through Planck suppressed operators such that the decay width is parametrically given by: $\Gamma_{\phi} \sim m_{\phi}^3/M_{\rm pl}^2 \sim 10^{-24}\, \mathrm{GeV}$ for $m_{\phi} \sim 50$ TeV. Thus, as long as the visible and dark sectors are coupled by larger than gravitational strength interactions,  one expects $\Gamma_{X} \gg \Gamma_{\phi}$ for a wide class of dark sector models. This is also true if the coupling between the two sectors arises by integrating out Kaluza-Klein (KK) modes of the extra dimensions or heavy GUT multiplets of some underlying GUT model, as even these mediators are lighter than the Planck scale. In addition, from a phenomenological point of view, $X$ decays to visible sector particles can spoil the successful predictions of BBN if $\Gamma_{X} < H(T_{BBN}) \sim T_{BBN}^2/M_{pl}$ where $T_{BBN} \sim 1$ MeV~\cite{Jedamzik:2006xz}. To avoid these constraints, for $\Gamma_X \lesssim \Gamma_{\phi}$, $\Gamma_X$ should lie in a narrow window: \ba T_{RH}^2 \simeq \Gamma_{\phi}M_{\rm pl}  \gtrsim \; \Gamma_{X}M_{\rm pl}\;\gtrsim T_{BBN}^2.\ea

Despite these considerations, for completeness we briefly discuss in this appendix the case where $\Gamma_{X} \lesssim \Gamma_{\phi}$. In this case, $X$ is effectively stable during modulus domination (as $H > \Gamma_{\phi} > \Gamma_{X}$). Thus for $H > \Gamma_{X}$ we can treat $X$ as a stable relic. If $X$ is a WIMP, its comoving abundance will become fixed at $T \sim T_{D}$ via the $QSE_{\rm nr}$ mechanism, which is the precise generalization of the non-thermal WIMP miracle~\cite{Moroi:1999zb, Acharya:2009zt}. Once the Hubble parameter drops below $\Gamma_{X}$ during radiation domination, the remaining $X$ abundance will decay to yield $X^\prime$ particles. The dynamics of such a process was studied in detail in~\cite{Cheung:2010gj}.  From the results of~\cite{Cheung:2010gj}, we see that there are three possibilities for the resulting parametrics of $\Omega_{X^\prime}h^2$:
\begin{itemize}
\item $X^\prime$ is in equilibrium when $H = \Gamma_{X}$ (which is only possible for $FO^{\rm rad}_{\rm nr}$ and $FO^{\rm rad}_{\rm r}$). $X^\prime$ will continue to track its equilibrium abundance until freeze-out. In this case $\Omega_{DM}\,h^2$ is completely insensitive to $X$ decays.

\item $X^\prime$ is out of equilibrium when $H = \Gamma_{X}$, and $X$ decays yield an $X^\prime$ abundance which is less than the critical abundance required for $X^\prime$ annihilations. This gives rise to the {\it freezeout $\&$ decay} ($FO\&D$) mechanism described in~\cite{Cheung:2010gj}. In terms of dimensionless comoving variables, $X_{\rm QSE}(A_c) < X^\prime_{\rm crit}\big|_{H = \Gamma_{X}}$, where $X_{\rm QSE}(A_c)$ is given by the $QSE_{\rm nr}$ mechanism as described in Section~\ref{nr-qse} and $X^\prime_{\rm crit}$ is defined in~(\ref{Xcrit}). The resulting contribution to the $X^\prime$ comoving abundance is insensitive to $\Gamma_{X}$, and is given simply by $\Delta X^\prime \approx X_{\rm QSE}(A_c)$. This contribution must be added to the $X^\prime$ abundance which results from the production mechanisms described in Section \ref{parametrics}. 

\item $X^\prime$ is out of equilibrium when $H = \Gamma_{X}$, and $X$ decays yield an $X^\prime$ abundance which exceeds the critical abundance required for $X^\prime$ annihilations. In terms of dimensionless comoving variables this occurs if $X_{\rm QSE}(A_c) > X^\prime_{\rm crit}\big|_{H = \Gamma_{X}}$. The $X^\prime$ particles produced from $X$ decays will then annihilate until $X^\prime \approx X^\prime_{\rm crit}\big|_{H = \Gamma_{X}}$. This was referred to as the ``{\it freezeout $\&$ decay and re-annihilation}" ($FO\&D_{\rm r}$) in~\cite{Cheung:2010gj}; the resulting $X^\prime$ relic abundance scales like $\Omega_{DM}\,h^2 \propto \frac{1}{\Gamma^{1/2}_{X}\left<\sigma v \right>^\prime}$.
\end{itemize}

Before concluding this appendix, we remark that the `{\it freeze-in}' mechanisms ($FI$ and $FI_r$) described in \cite{Hall:2009bx,Cheung:2010gj} are not important for the models considered here. Recall that $FI$ is due to $X \rightarrow X^\prime + ...$ decays which occur during the radiation domination era when $X$ is still relativistic and in equilibrium. However, it turns out that freeze-in due to $X$ decays is negligible during the modulus dominated era. To see this, consider the $X$ decay term in $d X^\prime/d A$. We saw in section \ref{X-sol} that $X$ attains QSE at some scale factor $A$ (say $A_X$) before $A_D$ if $\Gamma_X > {\cal O}(1)\,\Gamma_{\phi}$. For $A< A_X$, $X$ is given by $X \approx X_{\rm eq}$, while for $A_X < A \lesssim A_D$, $X$ is given by $X \approx X_{\rm QSE}$. In the analysis in Section \ref{X-sol}, the effect of $X$ decays when $1 < A \leq A_X$ and $X \approx X_{eq}$, which corresponds to freeze-in effects from $X$ decays, was neglected. To see that it is justified to do so, note that the integration of the decay term gives (up to overall constants):\begin{align}\label{freezein}
\int^{A_D}_1 dA\, X A^{1/2} & \approx \int^{A_X}_1 dA \, X_{\rm eq} A^{1/2} + \int^{A_D}_{A_X} d A \, X_{\rm QSE} A^{1/2}\notag\\
& \approx \frac{c_{\xi}}{\pi^2 T_{RH}^3} \int^{A_X}_1 dA \, A^{7/2} T^3 + \frac{2}{3} {A_D}^{3/2}\left(g_{X} B_{X} \frac{\Gamma_{\phi} T_{RH}}{\Gamma_{X} m_{\phi}}\right)\Phi
\end{align} where $A_{X}$ corresponds to the scale factor at which either $X$ becomes non-relativistic or $X$ enters QSE (whichever occurs first). Comparing the first and second terms in (\ref{freezein}), we find:\begin{equation}\label{ratio}
\frac{\int^{A_X}_1 dA \, X_{\rm eq} A^{1/2} }{\int^{A_D}_{A_X} d A \, X_{\rm QSE} A^{1/2}} \sim \left(\frac{T_{D}}{T_X}\right)^{4}\left(\frac{\Gamma_{X} m_{\phi}}{B_{X} \Gamma_{\phi} T_{RH}^4}\right)\frac{\left(\kappa T_{\rm max}\right)^8}{{T_X}^5 \Phi_I}\sim \left(\frac{T_{RH}^6 \Gamma_{X} m_{\phi} M_{\rm pl}}{{T_X}^9 B_{X}}\right)\end{equation} where we have used $T \approx \kappa T_{\rm max} A^{-3/8}$ and $(\kappa T_{\rm max})^8/\Phi_I \sim {T_{RH}}^8$ (see (\ref{tmax})). There are now two possibilities for $T_{X}$. If $X$ enters QSE before $X$ becomes non-relativistic, then $T_{X} \sim (\Gamma_{X} M_{\rm pl} {T_{RH}}^2)^{1/4} > M_{X}$. If instead $X$ becomes non-relativistic before QSE is reached, then $T_{X} \sim M_{X}$ and $\Gamma_{X} \lesssim \frac{M_{X}^4}{{T_{RH}}^2 M_{\rm pl}}$. Since $T_X$ is smaller in the latter case, the ratio (\ref{ratio}) is maximized for $T_X \sim M_X$, and one gets:\begin{equation}
\frac{\int^{A_X}_1 dA \, X_{\rm eq} A^{1/2} }{\int^{A_D}_{A_X} d A \, X_{\rm QSE} A^{1/2}} \lesssim \frac{T_{RH}^4 \,m_{\phi}}{B_{X}\,M_{X}^5} \simeq  \frac{10^{-13}}{B_X}\left(\frac{T_{RH}}{10\,\mathrm{MeV}}\right)^4 \left(\frac{m_{\phi}}{100\,\mathrm{TeV}}\right) \left(\frac{100 \,\mathrm{GeV}}{M_{X}}\right)^5\, .
\end{equation} Thus the freeze-in production of $X^\prime$ from $X$ decays can be neglected for reasonable choices of parameters, provided $B_{X}$ is not extremely tiny.

\section{Accuracy of Approximate Solutions}
\label{accuracy}

\begin{figure}[h!]
\includegraphics[scale=0.48]{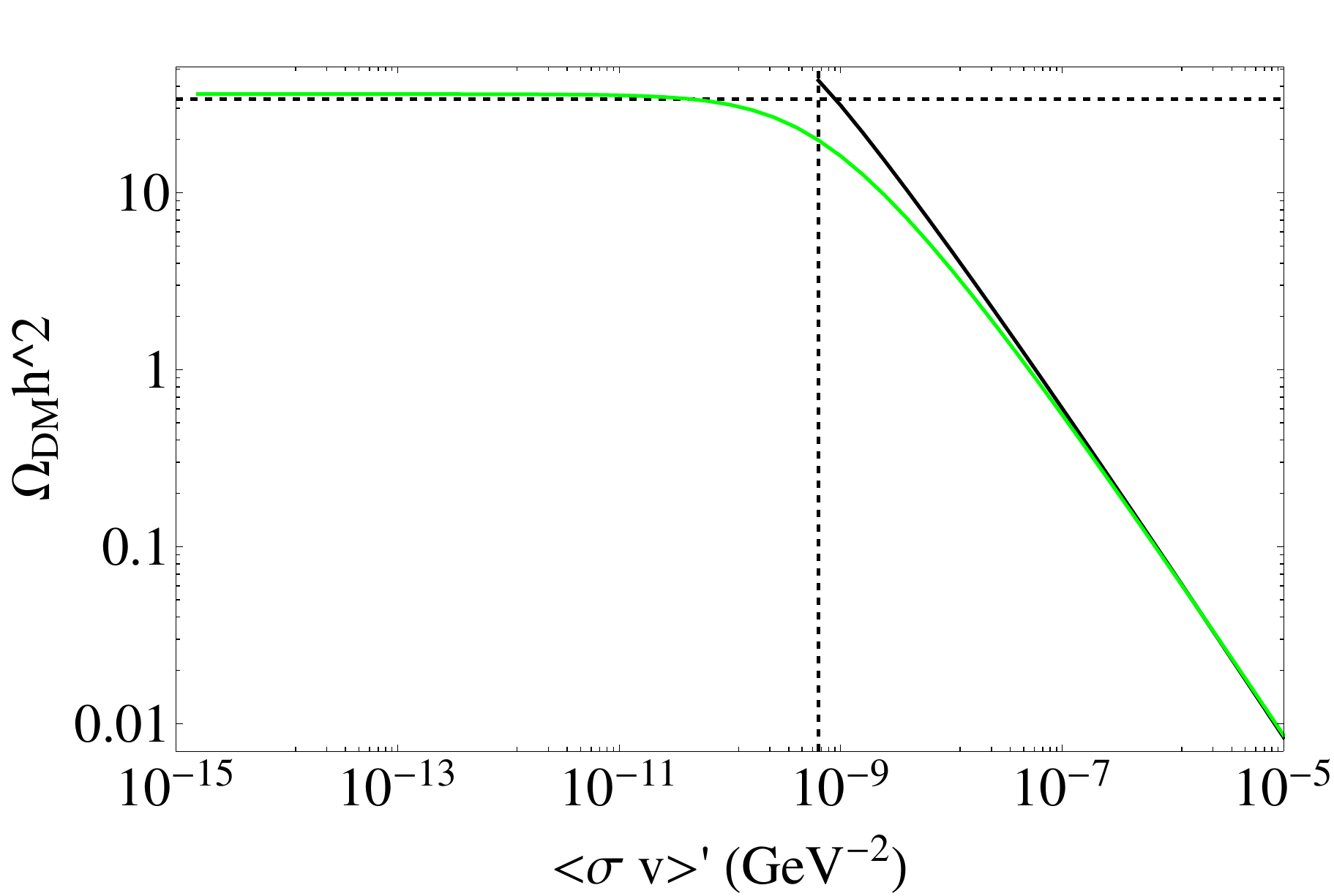}
\includegraphics[scale=0.48]{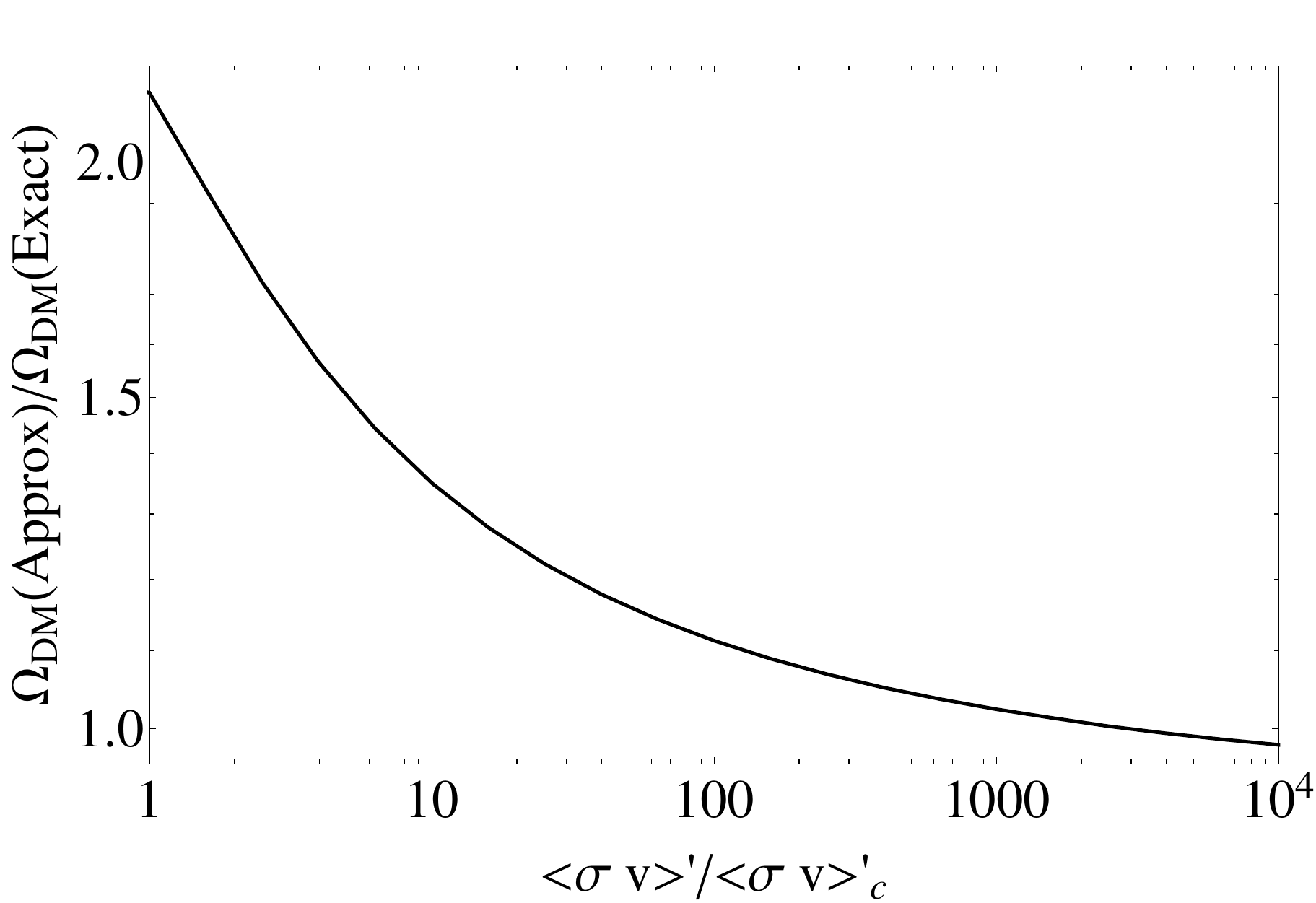}
\caption{\footnotesize{Left: $\Omega_{DM} h^2$ as a function of $\left<\sigma v \right>^\prime$ for $M_{X^\prime} = 10$ GeV and $B_{\rm tot} = 0.1$. The green curve shows the numerical solution, while the black curve shows the approximate  $QSE_{\rm nr}$  solution~(\ref{critnr}). The vertical dashed line represents $\left<\sigma v \right>^\prime = \left<\sigma v \right>^\prime_c$ as defined in~(\ref{critnr}), while the horizontal dashed line shows represents the modulus decay contribution given in~(\ref{moduliBR}), which is valid for $\left<\sigma v \right>^\prime < \left<\sigma v \right>^\prime_c$. Right: the ratio of the approximate result for $QSE_{\rm nr}$ to the exact result.}\label{Fig2}}
\end{figure}

\begin{figure}[h!]
\centering
\includegraphics[scale=0.48]{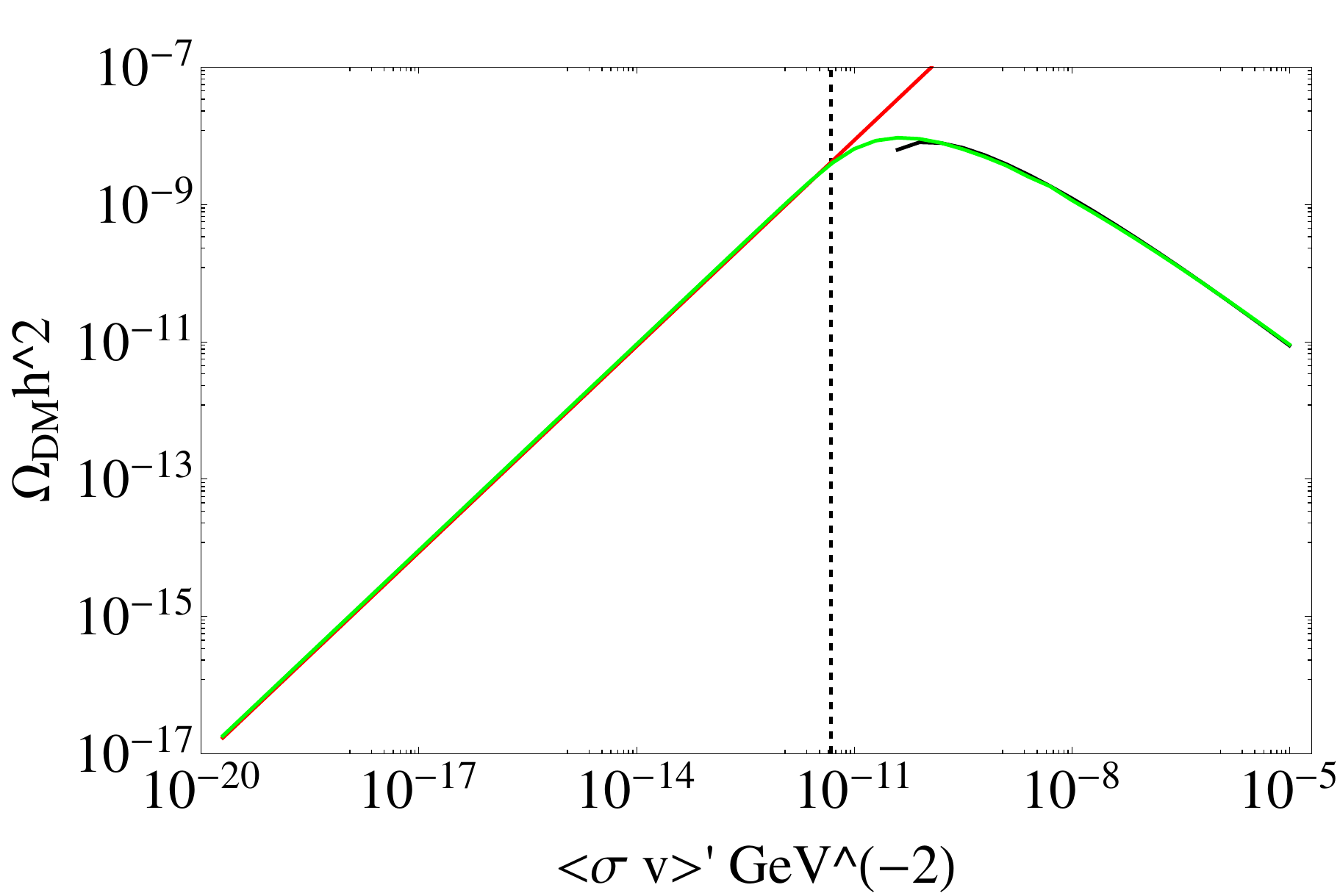}
\includegraphics[scale=0.48]{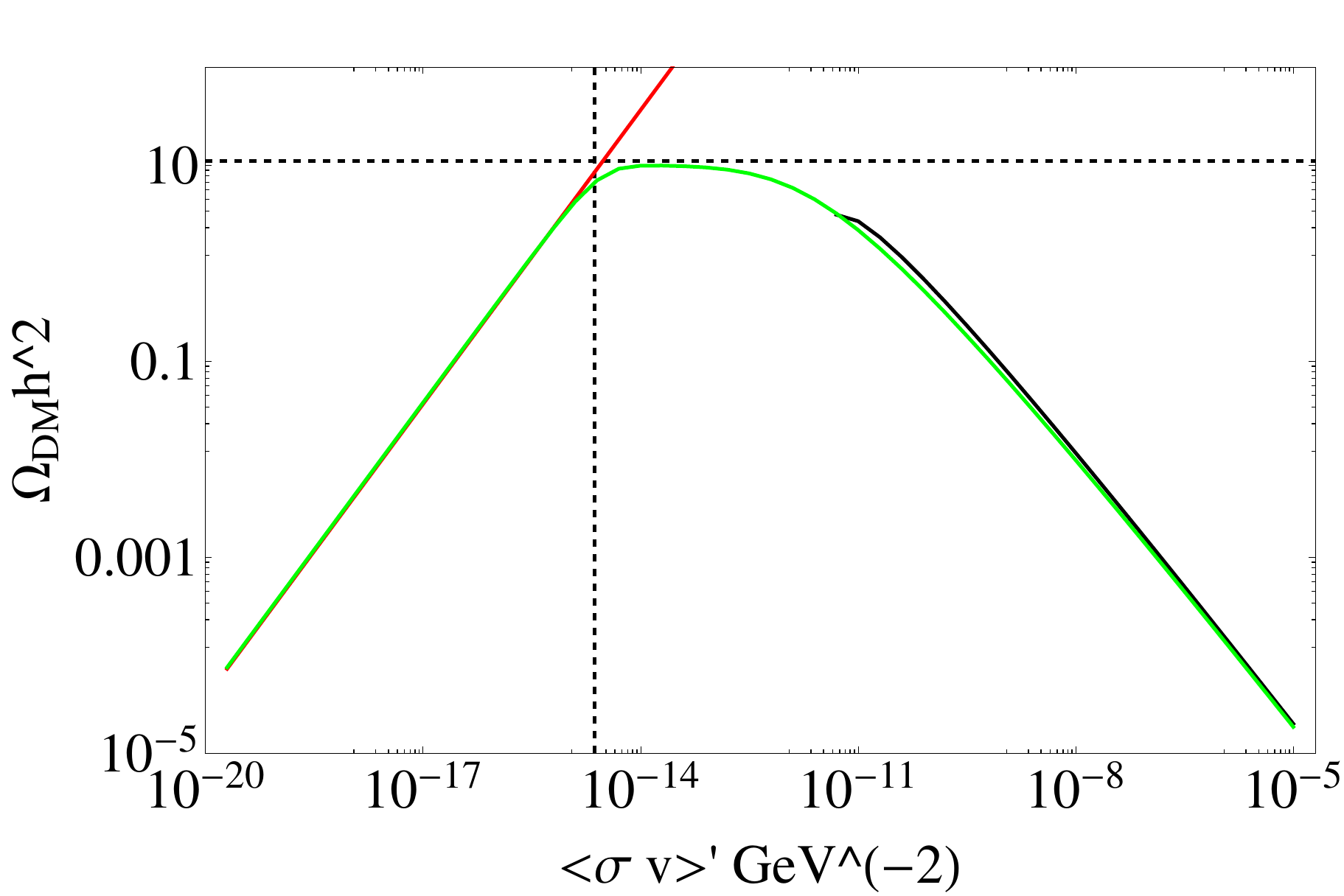}
\caption{\footnotesize{Left: $\Omega_{DM}h^2$ as a function of $\left<\sigma v \right>^\prime$ for $M_{X^{\prime}} = 10$ GeV and $B_{\rm tot} = 0$. Right: similar plot for $M_{X^{\prime}} = 10^{-6}$ GeV and $B_{\rm tot} = 0$. The green curves show the numerical solution, the red curve shows the approximation for $I A_{\rm nr}$ (left) and $I A_{\rm r}$ (right), while the black curve shows the approximation for $FO^{\rm mod}_{\rm nr}$ (left) and $FO^{\rm rad}_{\rm nr}$ (right). In the left plot the vertical dashed line represents $\left<\sigma v \right>^\prime = \left<\sigma v \right>^\prime_0$, defined in~(\ref{cond})}, while in the right plot the vertical line represents $\left<\sigma v \right>^\prime = \left<\sigma v \right>^\prime_c$ in the case where $M_{X^\prime} < T^\prime_{D}$ (see ~(\ref{critnr})). The dashed horizontal line in the right plot shows the approximate solution for $FO^{\rm rad}_{\rm r}$. \label{Fig3}}
\end{figure}

In this section, we compare the semi-analytic approximations obtained above with the full numerical solution to~(\ref{boltzmann}). The accuracy of these approximations is depicted in Figures~\ref{Fig2} and~\ref{Fig3}. In these plots we use the benchmark values of parameters as in~(\ref{benchmark}); however we take $g_*^\prime(T^\prime) = 20 \neq g_*(T)$ to ensure that the $g^\prime_*(T^\prime)$ dependence has been properly captured. Figure~\ref{Fig2} shows the accuracy of the approximate solutions for  $M_{X^\prime} = 10$ GeV and $B_{\rm tot} = 0.1$. In the left plot, the green curve shows the numerical solution, while the black curve in the left plot shows the approximate expression for DM production through $QSE_{nr}$ (\ref{omegadm1}). The right-hand plot shows the ratio of the approximate $QSE_{\rm nr}$ result to the exact result, which is close to unity if $\left<\sigma v \right>^\prime \gg \left<\sigma v \right>^\prime_c$. 

Figure \ref{Fig3} shows the accuracy of  the approximate solutions for $B_{\rm tot} = 0$; note that in this case $\left<\sigma v \right>^\prime_{c}$ becomes effectively infinite for $M_{X^\prime} > T^\prime_{D}$, see~(\ref{critnr}). The left plot shows $\Omega_{DM} h^2$ as a function of $\left<\sigma v \right>^\prime$ for $M_{X^\prime} = 10$ GeV. For these parameters, DM production occurs either via $FO^{\rm mod}_{\rm nr}$ for $\left<\sigma v \right>^\prime > \left<\sigma v \right>^\prime_0$ or via $I A_{\rm nr}$ for $\left<\sigma v \right>^\prime < \left<\sigma v \right>^\prime_0$ where $\left<\sigma v \right>^\prime_0$ is defined in~(\ref{cond}). The green curve shows the numerical solution; the red curve shows the approximation for $I A_{\rm nr}$~(\ref{invannNR}); and the black curve shows the approximation for $FO^{\rm mod}_{\rm nr}$~(\ref{nrfo}). The right plot shows a similar plot with $M_{X^\prime} = 10^{-6}$ GeV. In this case DM production occurs via $I A_{\rm r}$ for $\left<\sigma v \right>^\prime < \left<\sigma v \right>^\prime_c$ and via thermal freeze-out  ($FO^{\rm rad}_{\rm r}$ and $FO^{\rm rad}_{\rm nr}$) for  $\left<\sigma v \right>^\prime > \left<\sigma v \right>^\prime_c$. The green curve shows the numerical solution, the red curve shows the approximate expression for $I A_{\rm r}$~(\ref{invannR}), while the black curve shows the approximate expression for $FO^{\rm rad}_{\rm nr}$~(\ref{radnr}). Within their respective regimes of validity, (\ref{invannNR}), (\ref{invannR})~and~(\ref{nrfo}) are accurate to within $\sim 5 \%$, while (\ref{radr})~and~(\ref{radnr}) are accurate to within $\sim 15 \%$.

\section{Temperature Dependence of $\left<\sigma v \right>^\prime$}\label{tempsigmavd}

In Section \ref{parametrics}, (semi)-analytic expressions for $\Omega_{DM}\,h^2$ were obtained assuming that $\left<\sigma v \right>^\prime$ is temperature dependent. In this section, we generalize the results of Section~\ref{parametrics} for temperature dependent $\left<\sigma v \right>^\prime$. For scenarios where the contribution to $\Omega_{DM}\,h^2$ is determined by non-relativistic $X^\prime$ annihilation ($QSE^{\rm nr}$, $FO^{\rm rad}_{\rm nr}$, $FO^{\rm mod}_{\rm nr}$ and $IA_{\rm nr}$) we will consider p-wave annihilations where $\left<\sigma v \right>^\prime = T^\prime/\Lambda^3$. For scenarios where the contribution to $\Omega_{DM} h^2$ is determined by relativistic annihilations ($IA_{\rm r}$) we consider the case where $\left<\sigma v \right>^\prime = T^2/\Lambda^4$, corresponding to annihilation through a heavy bosonic mediator. Note that for $FO^{\rm rad}_{\rm r}$, $\Omega_{DM} h^2$  is independent of $\left<\sigma v \right>^\prime$ so~(\ref{radr}) holds regardless of the temperature dependence of $\left<\sigma v \right>^\prime$. 

\begin{itemize}

\item For $QSE_{\rm nr}$, (\ref{acrit})~and~(\ref{omegadm1}) are still valid for p-wave annihilation, provided the annihilation cross section is parameterized as:\begin{equation} \left<\sigma v \right>^\prime = \frac{T^\prime}{\Lambda^3} = \frac{T_{D}^\prime}{\Lambda^3}\left( \frac{\tilde{A}_D}{\tilde{A}_c}\right)\, , \end{equation} where in the second equality we have assumed $T^\prime \propto A^{-1}$ as in radiation domination. In order to match the numerical result, we instead use $\kappa = 1.8$ in~(\ref{acrit}) and~(\ref{omegadm1}).

\item For $FO^{\rm rad}_{\rm nr}$, the expression for $\hat{x}_{F}^\prime$ is given by:\begin{equation}\label{xfrad2}
\hat{x}_F^{\prime}\equiv \frac{M_{X^{\prime}}}{T^{\prime}_{FO}} = \log\left( \frac{3}{8 \pi^3} \sqrt{\frac{10\,\eta}{g^\prime_*(T^\prime_{FO})}}  g^\prime \left(\frac{{M_{X^\prime}}^2 M_{\rm pl}}{\Lambda^3}\right) \kappa_{p}^{\rm rad} (\hat{x}_F^{\prime})^{-1/2}\right)
\end{equation} while $\Omega_{DM} h^2$ is given by:\ba
\Omega\,h^2\;[FO^{\rm rad}_{\rm nr}] &\approx& \left[\frac{4 \sqrt{5}}{\sqrt{\pi}}\right] \left[\frac{\eta^{1/4}} {\left(1-\eta\right)^{3/4}}\right] \left[\frac{1}{g_*(T_{FO}){g^\prime_{*}(T^\prime_{FO})}}\right]^{1/4} \left[\frac{\kappa_p^{\rm mod} \left(\hat{x}_{F}^{\prime}\right)^2 \Lambda^3}{{M_{X^{\prime}}}^2\,M_{\rm pl}}\right]\left[\frac{M_{X^{\prime}}}{T_{\rm now}}\right]\,[\Omega_{R}\,h^2] \ea Here $\kappa_{p}^{\rm rad} = 2$ is a constant which is chosen to match the full numerical result. 

\item For $FO^{\rm mod}_{\rm nr}$, the expression for $x^\prime_{F}$ is given by:\begin{equation}
x_{F}^{\prime} = \ln \left[\left(\frac{3}{2 \sqrt{10} \pi^3}\right) \left(\frac{g^\prime g_*(T_{RH})^{1/2}}{g^\prime_*(T^\prime_{FO})}\right)\left(\frac{M_{\rm pl} T_{RH}^2}{\Lambda^3}\right) \kappa_{p}^{\rm mod} \,\eta\, {x_{F}^{\prime}}^{3/2}\right]
\end{equation} while $\Omega_{\rm ann} h^2$ is given by:\begin{equation}\label{nrfo2}\Omega_{\rm ann}\,h^2\;[FO^{\rm mod}_{\rm nr}] \approx \left[\frac{8\, \eta}{\sqrt{5\pi}\,L^{3/4}}\right] 
 \left[\frac{g_*(T_{RH})^{1/2}}{g^\prime_{*}(T^\prime_{FO})}\right] \left[\frac{T_{RH}}{M_{X^\prime}}\right]^3
\left[\frac{{\kappa_{p}^{\rm mod} x_F^{\prime}}^5 \Lambda^3}{{M_{X^{\prime}}}^2\,M_{\rm pl}}\right]\left[\frac{M_{X^{\prime}}}{T_{\rm now}}\right]\;[\Omega_{R}\,h^2]\end{equation} 

Here $\kappa_{p}^{\rm mod} = 5/4$ is a constant which is chosen to match the full numerical result (see also \cite{Giudice:2000ex}).

\item For $IA_{\rm nr}$, it is straightforward to show that for $\left<\sigma v \right>^\prime = T^\prime/\tilde{M}^3$, the expression analogous to~(\ref{invannNR}) is given by:\begin{equation}
\Omega_{\rm ann}\, h^2\;[IA_{\rm nr}] \approx  \left[\frac{48\,{g^{\prime}}^2\,\chi_{p}\, \eta^3}{125^{1/2}\pi^{15/2}\,L^{3/4}}\right]\left[\frac{{g_{\star}}^{3/2}(T_{RH})}{{g^{\prime}_{\star}}^3(T_{\star}^{\prime})}\right]\,\left[\left(\frac{T_{RH}}{M_{X^{\prime}}}\right)^7\,\left(\frac{M_{\rm pl} {M_{X^\prime}}^2}{\Lambda^3}\right)\right]\left[\frac{M_{X^\prime}}{T_{\rm now}}\right]\,[\Omega_{R}\,h^2] 
\end{equation} where $\chi_p$ is given by:\begin{equation} \chi_p \equiv \int_{\frac{M_{X^\prime}}{T^\prime_{\rm max}}}^{\frac{M_{X^\prime}}{T^\prime_D}} dx^\prime {x^\prime}^8 K_2(x^\prime)^2\end{equation}The integrand peaks at $T^\prime_* \approx 0.33 M_{X^\prime}$; if $T^\prime_* \ll T^\prime_{\rm max}$ and $T^\prime_* \gg T^\prime_{D}$, $\chi_p \approx 80$.

\item For $IA_{\rm r}$, we are interested in the case where $\left<\sigma v \right>^\prime = {T^\prime}^2/\Lambda^4$ (see above). The expression analgous to~(\ref{invannR}) is given by:\begin{equation}\label{invannRpwave}
\Omega_{\rm ann}h^2 = \left[\frac{48 {c_{\xi}}^2 \zeta(3)^2}{125^{1/2} \pi^{15/2} L^{3/4}}\left(\frac{T^\prime_D}{T^\prime_*}\right)^4\right] \left[\frac{\eta^2 g_*(T_{RH})^{1/2} }{g^\prime_*(T^\prime_D)^{2}}\right] \left[\frac{{T_{RH}}^3 M_{X^\prime} M_{\rm pl}}{T_{\rm now} \Lambda^4}\right] \Omega_{R} h^2 \end{equation} In the above, $T^\prime_* \approx T^\prime_{D}/1.35$ is chosen to match the numerical result, and is related to the temperature at which the integrand of $\int dA \, {T^\prime}^2 A^{7/2} {n^\prime_{\rm eq}}^2 \widetilde{H}^{-1}$ peaks. Note that we can recover~(\ref{invannRpwave}) from~(\ref{invannR}) by making the replacement:\begin{equation}
\left<\sigma v \right>^\prime \rightarrow 0.17 \times \left(\frac{\eta^{1/2} g_*(T_{RH})^{1/2}}{\Lambda^4 g^\prime_*(T_{D}^\prime)^{1/2}}\right) T_{RH}^2
\end{equation}
\end{itemize} 

\end{appendices}

\bibliographystyle{utphys}
\bibliography{refs}

\end{document}